\documentclass[nopubldata,hyper]{lpor2014}

\graphicspath{{figs/}{}}

\usepackage{amsmath}
\usepackage{amssymb}
\usepackage{IEEEtrantools}

\hypersetup{pdfstartview={FitH},pdfpagemode={UseNone},
            colorlinks,linkcolor=blue, citecolor=blue, urlcolor=blue,
            bookmarks=true, bookmarksopen=true, pdfnewwindow=true}

\newcommand\hrefBib[3][]{\href{#2}{#3}} %links in references
\newcommand\hrefEmpty[3][]{#3} %links in references

\newcommand\hrefBibPDF[3][]{}

\newcommand\sectpath{.}
\newcommand\pictc[5]{\begin{figure}[#2]
                       \centerline{
                       \includegraphics[width=#1\columnwidth,height=0.7\textheight,keepaspectratio]{#3}}
                   \protect\caption{\protect\label{fig:\sectpath/#4} #5}
                    \end{figure}            }
\newcommand\pict[4][1]{\pictc{#1}{!tb}{#2}{#3}{#4}}
\newcommand\pictcW[5]{\begin{figure*}[#2]
                       \centerline{
                       \includegraphics[width=#1\textwidth,height=0.7\textheight,keepaspectratio]{#3}}
                   \protect\caption{\protect\label{fig:\sectpath/#4} #5}
                    \end{figure*}            }
\newcommand\pictW[4][1]{\pictcW{#1}{!tb}{#2}{#3}{#4}}
\newcommand\rpict[2][\sectpath]{\ref{fig:#1/#2}}

\newcommand\leqt[1]{\protect\label{eq:\sectpath/#1}}
\newcommand\reqtn[2][\sectpath]{\ref{eq:#1/#2}}
\newcommand\reqt[2][\sectpath]{(\reqtn[#1]{#2})}

\newcommand\lsect[1]{\protect\label{sect:\sectpath/#1}}
\newcommand\rsect[2][\sectpath]{\ref{sect:#1/#2}}

\newcommand\REMOVE[1]{}

\newcommand\PT[0]{$\mathcal{PT}$}

\DeclareMathOperator*{\sech}{sech}
\newcommand\e[1]{\mathrm{e}^{#1}}
\newcommand\dd[0]{\mathrm{d}}
\makeatletter
\newcommand{\showfontsize}{\f@size{} pt}
\makeatother

\begin{document}
\sloppy

\titlefigure{TitFig}

\abstract{
One of the challenges of the modern photonics is to develop all-optical devices enabling
increased speed and energy efficiency for transmitting and processing information on an optical chip. It is believed that the recently suggested Parity-Time (PT) symmetric photonic systems with alternating regions of gain and loss can bring novel functionalities. In such systems, losses are as important as gain and, depending on the structural parameters, gain compensates losses. Generally, PT systems demonstrate nontrivial non-conservative wave interactions and phase transitions, which can be employed for signal filtering and switching, opening new prospects for active control of light. In this review, we discuss a broad range of problems involving nonlinear PT-symmetric photonic systems with an intensity-dependent refractive index. Nonlinearity in such PT symmetric systems provides a basis for many effects such as the formation of localized modes, nonlinearly-induced PT-symmetry breaking, and all-optical switching. Nonlinear PT-symmetric systems can serve as powerful building blocks for the development of novel photonic devices targeting an active light control.
}

\title{Nonlinear switching and solitons in PT-symmetric photonic systems}
\titlerunning{Nonlinear PT-symmetric photonic systems}
\author{
Sergey V. Suchkov\inst{1}, % sergey.v.suchkov@gmail.com
Andrey A. Sukhorukov\inst{1,*}, % ans124@gmail.com
Jiahao Huang\inst{2,3}, % eqjiahao@gmail.com
Sergey V. Dmitriev\inst{4,5}, % dmitriev.sergey.v@gmail.com
Chaohong Lee\inst{2,3}, % lichaoh2@mail.sysu.edu.cn
Yuri S. Kivshar\inst{1} % ysk@internode.on.net
}
\authorrunning{Sergey V. Suchkov et al.}
\institute{%
    Nonlinear Physics Centre, Research School of Physics and Engineering, The Australian National University, Canberra, ACT 2601 Australia
\and
   Institute of Astronomy and Space Science, Sun Yat-Sen University, Guangzhou 510275, China
\and
    State Key Laboratory of Optoelectronic Materials and Technologies, School of Physics and Engineering, Sun Yat-Sen University, Guangzhou 510275, China
\and
    Institute for Metals Superplasticity Problems of RAS, Ufa 450001, Russia
\and
    %National Research Tomsk State University, 36 Lenin Prospekt, Tomsk 634050, Russia
    Research Laboratory for Mechanics of New Nanomaterials, Peter the Great St. Petersburg Polytechnical University, St. Petersburg 195251, Russia
}
\mail{\email{ans124@physics.anu.edu.au}}
\maketitle

{\protect\renewcommand\sectpath{Introduction}
%-------------------------------------------
\section{Introduction} \lsect{}
%-------------------------------------------

The concept of Parity-Time (PT) symmetry first emerged in the studies of quantum operators and their spectra~\cite{Bender:1998-5243:PRL, Bender:2002-270401:PRL, Bender:2007-947:RPP}. Whereas it is well known that Hermitian operators have real spectra, Carl Bender and collaborators posed a question whether there exist non-Hermitian Hamiltonians which still have real spectra. They found that this can happen for a broad class of Hamiltonians which are symmetric with respect to the so-called PT transformations, defined by the parity operator ${\mathbf{P}}$ as $\mathbf{P}\psi ({\bf r},t) = \psi (-{\bf r}, t)$, and time reversal operator $\mathbf{T}$, as $\mathbf{T} \psi ({\bf r},t)=\psi^*({\bf r},-t)$, where $\psi ({\bf r},t)$ is a wave function in quantum mechanics. When a simple one-dimensional system is described by the Schr\"odinger-type Hamiltonian with a complex potential $U(x)$,
\begin{equation}
\label{PT}
H = - \frac{\dd^2}{\dd x^2} + U(x), \;\;\;  U(x) = V(x) + i W(x)
\end{equation}
where $V(x)$ and $W(x)$ are both real, the system (\ref{PT}) is called PT-symmetric provided $U^*(x) = U(-x)$, i.e.
$V(x)=V(-x)$ and $W(x) = -W(-x)$. Although initially the analysis was motivated by the development of rather abstract quantum theories~\cite{Bender:1998-5243:PRL, Bender:2002-270401:PRL, Bender:2007-947:RPP}, it soon emerged that the concept of PT symmetry is relevant to various physical systems.

In optics, the propagation of light can often be described theoretically as an evolution governed by an effective Hamiltonian~(\ref{PT}) in the so-called paraxial approximation. In this case variation of the medium refractive index is described by the potential $U(x)$. Hermitian Hamiltonians correspond to the cases where the optical energy is conserved and $U(x)$ is real. On the other hand, the presence of {\em losses} or {\em gain} in optical structures maps to non-conservative non-Hermitian operators (\ref{PT}) with a complex $U(x)$. However, if such an operator possesses PT-symmetry and has a real spectrum, then the amplitudes of optical modes would be preserved, corresponding to an effective compensation of the effects of loss and gain, which can be important for a range of practical applications. First theoretical proposals for optical PT systems were formulated a decade ago~\cite{Ruschhaupt:2005-L171:JPA, El-Ganainy:2007-2632:OL}, and it was shown that in waveguiding structures, $\mathbf{P}$ operator performs to spatial reflection, whereas $\mathbf{T}$ reverses the propagation direction. Accordingly, conventional optical PT-symmetric structures have symmetrically positions regions of loss and gain, corresponding to the same real but opposite imaginary parts of the optical refractive index. The PT symmetry definition can be generalized to include the gauge transformations enabling applications to practical structures without exact balance of gain and loss~\cite{Ruter:2010-192:NPHYS}, and including the structures with pure loss~\cite{Guo:2009-93902:PRL}.

The spectrum of PT operators is not always real, and is inherently connected to the symmetry of the eigenmodes. If an eigenmode profile is invariant with respect to the PT transformation, then it has real eigenvalue. However, the modes with PT-symmetry-broken profiles generally have complex eigenvalues. The PT-symmetry breaking transition and the associated change from real to complex eigenvalues can be generally observed when the amount of gain/loss is increased~\cite{Guo:2009-93902:PRL, Ruter:2010-192:NPHYS}. Such {\em a phase transition} can in particular facilitate single-mode lasing~\cite{Feng:2014-972:SCI, Hodaei:2014-975:SCI} and find various other applications (see the recent review paper~\cite{Zyablovsky:2014-1063:PUS} and references therein).

Even before the PT symmetry concept was introduced, it was suggested that optical systems with gain and loss can improve the operation of nonlinear all-optical switching devices, including the lowering of the switching power~\cite{Chen:1992-239:IQE, Malomed:1996-330:OL}. The development of PT concept brought a new perspective and stimulated further active investigations of nonlinear effects in PT-symmetric  systems. In addition to switching, the topic of nonlinear self-focusing and the formation of self-localized solitons received a strong attention. Optical solitons in the conservative systems or the systems with low losses originate from a balance of nonlinearity and dispersion and usually form families of localized solutions~\cite{Kivshar:2003:OpticalSolitons}. In nonlinear dissipative systems, 
localized modes appear due to an additional balance of gain and loss at some special values of parameters and the existence of families of localized modes is very rare~\cite{Rosanov:2002:SpatialHysteresis,Akhmediev:2005:DissipativeSolitons}. While the PT-symmetric systems 
can be viewed as a special class of dissipative systems, they are quite unique, combining many features of conservative 
and dissipative systems, being able to support families of localized and periodic modes. 

In this paper, we review the studies of nonlinear phenomena in optical PT symmetric systems. We discuss both theoretical concepts and predictions, as well as a relatively small yet quickly growing number of experimental demonstrations. We start with the discussion of the simplest configuration of PT-symmetric systems in optics in the form of a pair of coupled waveguides,
an optical coupler, in Sec.~\rsect[PT-couplers]{}. Then, in Sec.~\rsect[Oligomers]{} we extend the discussion to multiple coupled waveguides, or oligomers. Further in Sec.~\rsect[Solitons_couplers_periodic]{} we analyse the properties of solitons in couplers and extended periodic systems. In Sec.~\rsect[Scattering]{}, scattering of waves incident on nonlinear PT structures is considered. In Sec.~\rsect[Modulated]{}, we discuss effects in PT structures, where gain and loss are modulated along the propagation direction.
Finally, in Sec.~\rsect[Conclusion]{} we present conclusion and outlook.

{\protect\renewcommand\sectpath{PT-couplers}
%-------------------------------------------
\section{PT-symmetric couplers} \lsect{}
%-------------------------------------------

A pair of coupled waveguides with gain and loss represents the simplest configuration of PT-symmetric optical system. However, such a system can already showcase rich physics and phenomena associated with the interplay of nonlinearity and PT-symmetry. In the following, we first overview the properties of linear couplers, and then discuss the effect of nonlinearity on the optical modes and the phenomenon of nonlinearly-induced PT symmetry breaking, as well as some generalizations.

\subsection{Linear properties} \lsect{PT-couplers-linear}

%\subsection{Linear properties of PT-symmetric couplers} \lsect{PT-couplers-linear}

Directional couplers composed of waveguides with gain and loss regions can be used to realize PT-symmetric optical structures. In the pioneering papers~\cite{Ruschhaupt:2005-L171:JPA, El-Ganainy:2007-2632:OL}, different designs of optical structures were suggested, where optical beams would exhibit an effective PT-symmetric potential. %PT-symmetry in optics requires that
In such structures, the real part of the optical refractive index should be symmetric, and the imaginary part - antisymmetric, i.e.
\begin{equation}\label{PTrefractiveIndex}
  n(x,y) = n^\ast(-x,y).
\end{equation}
This means that the absolute amounts of gain and loss should be the same, and gain/loss regions should have mirror configurations with respect to the central symmetry point.
%Such requirements can be satisfied in different ways, and we reproduce the first suggested configuration based on coupled waveguides in Fig.~\rpict{PTcouplerOrig}(a), where each waveguide combines regions with gain and loss.

%---------------------------------------------------------------------------------------------------
\pictW{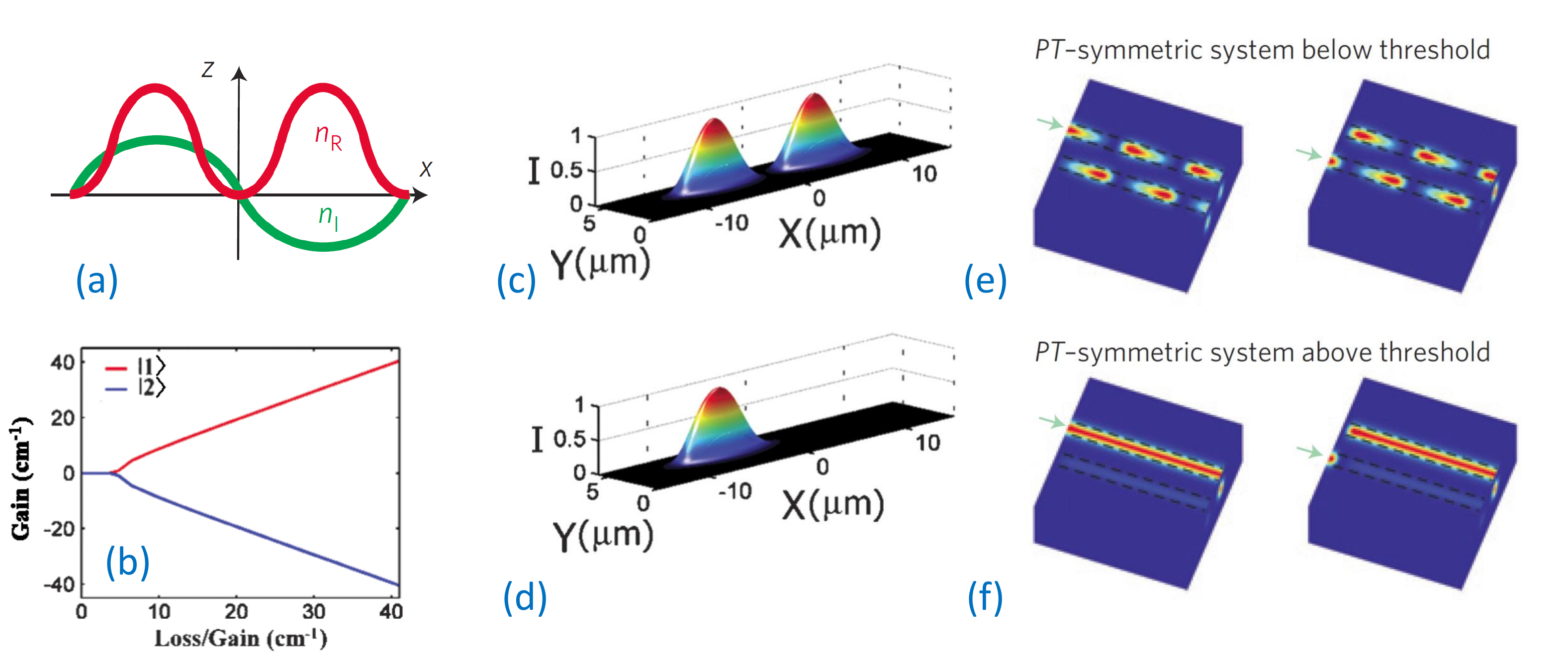}{PTcouplerLinear}{
Linear PT couplers. (a)~Characteristic real ($n_R$, red line) and imaginary ($n_I$, green line) parts of the complex refractive-index distribution. (b)~Imaginary eigenvalue part for optical supermodes characterizing their gain or loss vs. the gain/loss coefficients in individual waveguides.
(c,e)~Calculated optical supermodes and dynamics below the critical gain/loss value: (c)~the supermode intensity is equally divided between the two waveguides and (e)~periodic beating of light between two waveguides, with zero average gain or loss, when the system is excited at either channel 1 or channel 2.
(d,f)~Regime above the critical gain/loss value: (d)~a supermode is isolated at one waveguide and (f)~light localizes in waveguide with gain and exhibits amplification.
Figures (a,e,f) after Ref.~\cite{Ruter:2010-192:NPHYS}; (b,c,d) after Ref.~\cite{Guo:2009-93902:PRL}.
}
%---------------------------------------------------------------------------------------------------

For experimental realization of optical PT couplers, directional couplers with different values of gain or loss in each of the waveguides were considered~\cite{Guo:2009-93902:PRL, Ruter:2010-192:NPHYS}, see Fig.~\rpict{PTcouplerLinear}(a). Beam propagation in such structures can be described by a pair of coupled equations for the amplitudes of modes in each of the waveguides, which can be represented in a Hamiltonian form as
\begin{equation} \leqt{CouplerCM}
i\frac{d \mathbf{a}}{d z} = \mathbf{H}\mathbf{a}, \,\,
\mathbf{a}(z)=\left( \begin{array}{cc}
a_{1}(z) \\
a_{2}(z)  \end{array} \right) , \,\,
\mathbf{H} = \left( \begin{array}{cc}
i \rho_1, & - C \\
- C, & - i \rho_2  \end{array} \right),
\end{equation}
%
%\begin{equation} \leqt{CouplerCM}
%   \begin{array}{l} {\displaystyle
%       i \frac{d a_1}{d z} + i \rho_1 a_1 + C a_2 = 0, \quad
%   } \\*[9pt] {\displaystyle
%       i \frac{d a_2}{d z} - i \rho_2 a_2 + C a_1 = 0,
%   } \end{array}
%\end{equation}
%
where
$z$ is the propagation distance, $a_1$ and $a_2$ are the mode amplitudes in the first and second waveguide, respectively, $\rho_{1,2}$ define the rates of gain in the first waveguide and loss in the second waveguide, and $C$ is the coupling coefficient between the modes of two waveguides.
%In the following

The Hamiltonian possesses {\em PT symmetry} when applied together with Gauge transformation~\cite{Guo:2009-93902:PRL},
\begin{equation} \leqt{CouplerHPT}
\mathbf{P}\mathbf{T} (\mathbf{H} - \bar{\rho} \mathbf{I}) = (\mathbf{H} - \bar{\rho} \mathbf{I}) \mathbf{P}\mathbf{T} ,
%
%\mathbf{P}_{1,+}\mathbf{P}_{2,+}\mathbf{T}\mathbf{H}=-\mathbf{H}\mathbf{P}_{1,+}\mathbf{P}_{2,+}\mathbf{T} , \quad
%\mathbf{P}_{j,\pm}=\{ a_{sj}\leftrightarrow \pm a^{*}_{ij}, a_{ij}\leftrightarrow \pm a^{*}_{sj} \},
\end{equation}
where parity operator swaps the mode amplitudes between the waveguides, $\mathbf{P}=(0, 1; 1, 0)$,
$\mathbf{T}$ is a time-reversal operator which changes $z \rightarrow -z$ and performs a complex conjugation, $\mathbf{I}$ is an identity matrix, and $\bar{\rho} = (\rho_1 - \rho_2)/2$ defines the average gain or loss between the two waveguides.

Important features of PT couplers become apparent by studying the properties of optical supermodes. These solutions have the form $\mathbf{a}^{(m)}(z) = \mathbf{a}^{(m)}(0) \exp(i \beta_{m} z)$, where $m=1,2$ is the mode index. The mode propagation constants are $\beta_m = \beta_0 + \tilde{\beta}_m$, $\tilde{\beta}_m = C \cos(\varphi_{\pm})$ and the profiles are $a_2^{(m)} = \exp(-i \varphi_{\pm}) a_1^{(m)}$, where $\sin(\varphi_\pm) = \rho / C$, $\rho = (\rho_1+\rho_2)/2$, $\beta_0 = i (\rho_1-\rho_2)/2$.

When $|\rho / C| \le 1$, i.e. when the amount of gain and loss is below a critical value
%defined by the coupling coefficient
, then $\varphi_{\pm}$ are real, accordingly  ${\rm Im}\tilde{\beta}_m = 0$, and both supermodes exhibit the same amount of gain/loss as ${\rm Im}\beta_1 \equiv {\rm Im}\beta_2$, see Fig.~\rpict{PTcouplerLinear}(b). In this regime the supermode profiles are PT-symmetric, i.e. $\mathbf{P}\mathbf{T} \mathbf{a}^{(m)}(0) =  \exp[i \varphi_{\pm} - 2 i \arg(a_1^{(m)})] \mathbf{a}^{(m)}(0)$. Accordingly, the mode intensity is symmetrically distributed between the waveguides, i.e. $|a_1| = |a_2|$, see Fig.~\rpict{PTcouplerLinear}(c).

However, when the gain/loss exceeds a critical value, $|\rho / C| > 1$, then $\varphi_m$ become complex, supermodes exhibit different gain/loss [Fig.~\rpict{PTcouplerLinear}(b)], and their profiles become asymmetric and no longer conform to PT symmetry [Fig.~\rpict{PTcouplerLinear}(d)]. Accordingly, this regime is called broken PT-symmetry, referring to the symmetry of the supermodes.

%and their profiles are asymmetric
%
An input optical beam in general excites a superposition of the supermodes, which properties thereby determine the beam dynamics. Below the threshold, the beating of two supermodes can lead to a periodic switching of light between the coupled waveguides, see Fig.~\rpict{PTcouplerLinear}(e). Such behavior is similar to a conventional conservative coupler, and indeed the power is conserved on average after every oscillation between the lossy and amplifying waveguides. The beating period is $z_p = 2 \pi / |\beta_+ - \beta_-| = \pi (C^2-\rho^2)^{-1/2}$, and it grows monotonously as the gain/loss is increased, approaching infinity close to the critical threshold.
Above the threshold, one supermode decays, and the remaining supermode primarily localized in the waveguide with gain determines amplification irrespective of the input condition, see Fig.~\rpict{PTcouplerLinear}(f).

%, which are found as eigen-solutions of Eq.~\reqt{CouplerCM} of the form $a_j(z) = a_j(0) \exp(i \beta z)$.

\subsection{Nonlinear coupler and symmetry breaking} \lsect{PT-couplers-breaking}

The effect of nonlinearity on the beam dynamics in directional couplers composed of waveguides with gain and loss was originally described theoretically already two decades ago in Refs.~\cite{Chen:1992-239:IQE, Malomed:1996-330:OL}. It was predicted that such structures can offer benefits for all-optical switching in the nonlinear regime, lowering the switching power and attaining sharper switching transition. In the last decade, the renewed interest in such structures as realizations of ${\cal PT}$ symmetric optical systems has led to a series of extensive theoretical and more recently experimental studies.

%However the renewd however the recently identified analogy with ${\cal PT}$ symmetry property has stimulating
%extensive theoretical~\cite{Berry:2008-244007:JPA, Makris:2008-103904:PRL, Longhi:2009-123601:PRL,
%Bendix:2009-30402:PRL, West:2010-54102:PRL, Longhi:2010-22102:PRA, Ramezani:1005.5189:ARXIV} and experimental~\cite{Guo:2009-93902:PRL, Ruter:2010-192:NPHYS} studies.

%As was demonstrated in the original study of directional couplers with gain and loss~\cite{Chen:1992-239:IQE}, such structures can offer benefits for all-optical switching in the nonlinear regime, lowering the switching power and attaining sharper switching transition. Recently, these conclusions were complimented by the prediction of unidirectional switching and exact analytical solution describing the switching dynamics in nonlinear ${\cal PT}$ symmetric couplers~\cite{Ramezani:1005.5189:ARXIV}.

We first consider the effect of Kerr-type nonlinearity, which can modify the refractive index in each of the waveguides, depending on the optical intensity~\cite{Chen:1992-239:IQE, Malomed:1996-330:OL, Ramezani:2010-43803:PRA, Sukhorukov:2010-43818:PRA, Kevrekidis:2013-365201:JPA, Barashenkov:2014-45802:PRA}. This effect can be modeled by including nonlinear terms in the coupled-mode Eq.~\reqt{CouplerCM} as follows~\cite{Sukhorukov:2010-43818:PRA},
\begin{equation} \leqt{DNLS}
%       i \frac{d a_j}{d z} + i \rho_j a_j + C a_{3-j} + G(|a_j|^2) a_j = 0, \quad
   \begin{array}{l} {\displaystyle
       i \frac{d a_1}{d z} + i \rho a_1 + C a_2 + G(|a_1|^2) a_1 = 0, \quad
   } \\*[9pt] {\displaystyle
       i \frac{d a_2}{d z} - i \rho a_2 + C a_1 + G(|a_2|^2) a_2 = 0,
   } \end{array}
\end{equation}
where the function $G$ characterizes the nonlinear response (we assume it to be real-valued), and $\rho > 0$ is the loss/gain coefficient in the first and second waveguides. We will consider here the regime below the linear ${\cal PT}$-symmetry breaking threshold, when $\rho < C$.
%We now analyze the effect of nonlinearity

It was found that nonlinear solutions belong to two classes~\cite{Sukhorukov:2010-43818:PRA}: (i)~periodic solutions, where the intensities and relative phases in two waveguides are exactly restored after each period ($z\rightarrow z+z_p$) [Fig.~\rpict{PTcouplerNonlinear}(a)], or (ii)~solutions where the total intensity grows without bound due to nonlinearly-induced symmetry breaking [Fig.~\rpict{PTcouplerNonlinear}(b)]. This classification is valid for arbitrary Kerr-type nonlinearities with smooth response functions $G(I)$.
Quite remarkably, in case of cubic nonlinear response with $G(I) = \gamma I$, Eqs.~\reqt{DNLS} were shown to be integrable and their solutions can be formulated analytically~\cite{Ramezani:2010-43803:PRA, Kevrekidis:2013-365201:JPA, Barashenkov:2013-53817:PRA, Pickton:2013-63840:PRA, Barashenkov:2014-45802:PRA, Barashenkov:2015-325201:JPA}.
%, which agree with the general classification formulated above for arbitrary Kerr-type nonlinearities.
%
The dependence of the minimal input intensity ($I_{\rm cr}$) required for nonlinear switching on gain/loss coefficient is shown in Fig.~\rpict{PTcouplerNonlinear}(c), which illustrates that the threshold for nonlinear switching is drastically reduced for larger gain/loss coefficients.

%In order to analyze nonlinear dynamics, i
It is convenient to represent the mode amplitudes in the following form,
\begin{equation} \leqt{a12}
   \begin{array}{l} {\displaystyle
       a_{1}=\sqrt{I(z)} \cos[\theta(z)] \exp[+i \varphi(z)/2] \exp[i \beta(z)],
   } \\*[9pt] {\displaystyle
       a_{2}=\sqrt{I(z)} \sin[\theta(z)] \exp[-i \varphi(z)/2] \exp[i \beta(z)],
   } \end{array}
\end{equation}
where $I$ is the total intensity,
%$0 < \theta < \pi/2$ and $-\pi < \varphi < \pi$
$\theta$ and $\varphi$ define the relative intensities and phases between the two input waveguides, and $\beta$ is the overall phase.
Then, the initial conditions corresponding to different solution types (oscillating or growing) can be conveniently visualized on a phase plane, see an example in Fig.~\rpict{PTcouplerNonlinear}(d). This plot illustrates that even for high intensity above the threshold, the type of dynamics depends on the relative amplitudes and phases in the two waveguides.
%
%dependence of solution dynamics
%on the relative amplitudes and phases in the two waveguides
%
%Even for high intensities above the threshold, the type of dynamics depends on the relative amplitudes and phases in the two waveguides, see a characteristic example in Fig.~\rpict{PTcouplerNonlinear}(d).
%
Interestingly, the type of nonlinear dynamics remains the same if we swap the intensities between the two waveguides, which corresponds to a transformation $\theta \rightarrow \pi/2 - \theta$. In particular, we could couple light at the input just to the waveguide with loss, or to the waveguide with gain, and the type of dynamics would be the same. This is a counter-intuitive result, since in the first case the total intensity will initially decrease, whereas in the second case the total intensity will be growing. However, in both cases the type of dynamics will be determined only by the initial intensity level. This is a highly nontrivial consequence of linear PT-symmetry in the strongly nonlinear regime.

%It can be further demonstrated that for certain input conditions
Whereas in general solutions are oscillating or growing, for particular initial conditions wave intensities can remain constant. Such regime corresponds to the excitation of stationary nonlinear modes, which can be viewed as a generalization of linear supermodes. They are positioned at the phase space points $(I=I_0,\theta = \pi/4,\varphi=\varphi_\pm)$ and $\beta = \beta_\pm z$, where $\beta_\pm = G(I_0 / 2) + C \cos(\varphi_\pm)$, $\sin(\varphi_\pm) = \rho / C$, and $\cos(\varphi_\pm) = \mp [ 1 - (\rho/C)^2 ]^{1/2}$. The fixed point in phase space is a stable center if (i)~$\tilde{\gamma} < \tilde{\gamma}_{\rm cr}$ and $\varphi = \varphi_-$ or (ii)~$\tilde{\gamma} > - \tilde{\gamma}_{\rm cr}$ and $\varphi = \varphi_+$, where $\tilde{\gamma}_{\rm cr} = |\cos(\varphi_\pm)|$ and $\tilde{\gamma} = G'(I_0 / 2) I_0 / ( 2 C)$ (prime stands for the derivative). If these conditions are not satisfied, then the fixed point is a saddle, which indicates an instability.
In case of self-focusing Kerr nonlinearity, $G(I) = \gamma I$ with $\gamma > 0$, the stationary point at $\varphi_+$ is always stable at arbitrarily high intensities, whereas at $\varphi_-$ the instability appears for $I_0 > I_{\rm 0cr} = 2 [ 1 - \rho^2 ]^{1/2}$. The stable point at $\varphi_+$ is indicated in Fig.~\rpict{PTcouplerNonlinear}(d) by a star, and unstable at $\varphi_-$~-- by a triangle. We note that the unstable point is located at a boundary between the domains of periodic or growing solutions.

%---------------------------------------------------------------------------------------------------
\pict{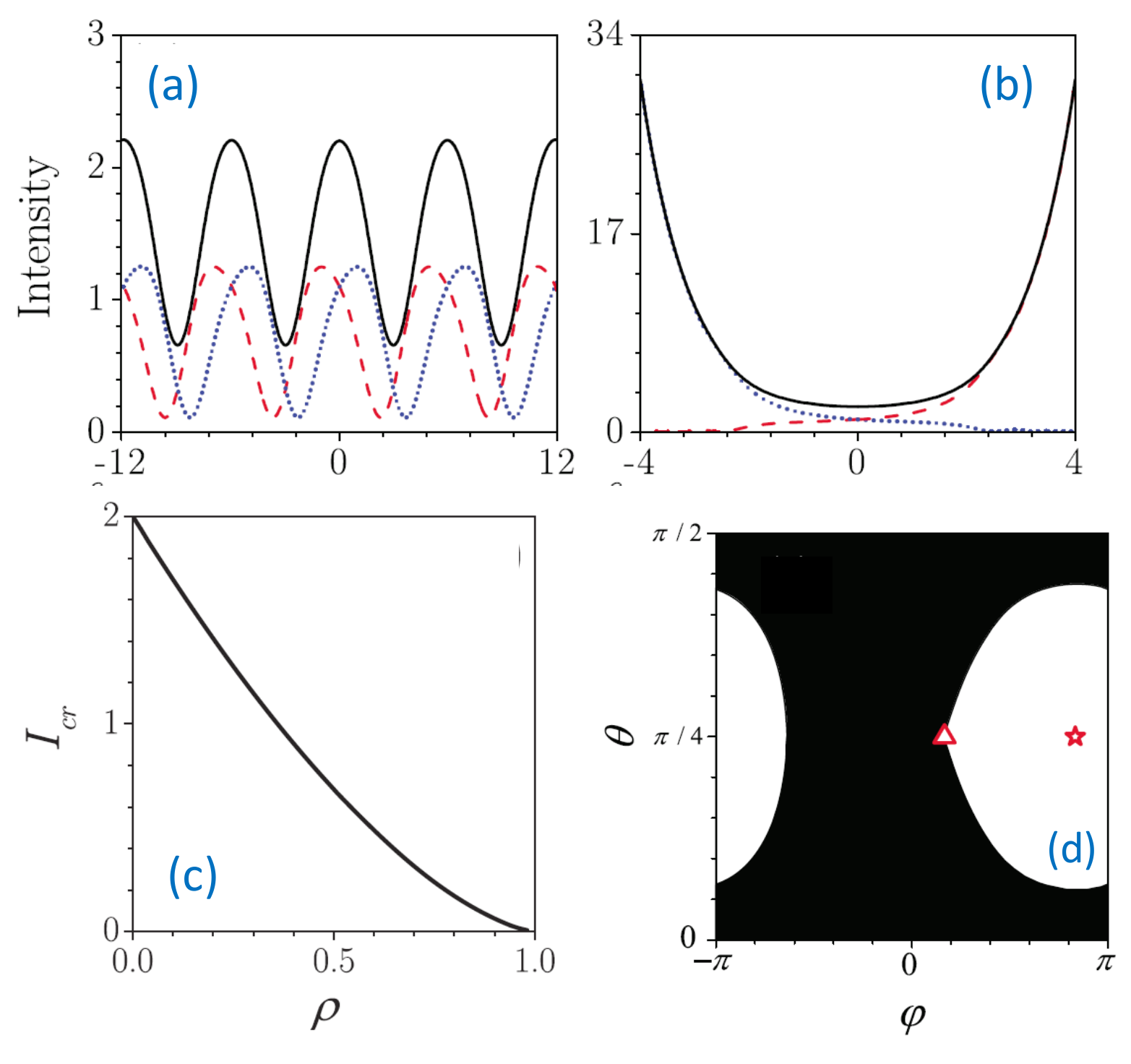}{PTcouplerNonlinear}{
Nonlinear PT coupler.
(a,b)~Intensity dependencies on propagation distance in the first (dotted line) and second (dashed line) waveguides; the solid line shows the sum of individual intensities. Shown are the regimes of (a)~periodic oscillations and (b)~nonlinear localization and amplification in the waveguide with gain.
(c)~Minimum critical intensity required for nonlinear PT-symmetry breaking vs the gain/loss coefficient.
(d)~Regions of PT symmetry (white shading) and symmetry breaking with nonlinear switching (black shading) in the plane of initial conditions. Stars and triangles mark stable and unstable stationary solutions, respectively.
Parameters are $G(I) = \gamma I$, $\gamma=1$, $C=1$, (a,b,d)~$\rho=0.5$ and $I = 2.2$, (a)~$\varphi=\pi/6-\pi/20$, (b)~$\varphi=\pi/6+\pi/20$.
Figures after Ref.~\cite{Sukhorukov:2010-43818:PRA}.
}
%---------------------------------------------------------------------------------------------------

Nonlinear PT couplers can have applications for efficient all-optical signal manipulation with reduced switching power and the possibility to perform intensity-dependent amplification. Additionally, such couplers can operate as unidirectional optical valves~\cite{Ramezani:2010-43803:PRA} when nonlinearity breaks PT symmetry, as in this case the output beam becomes localized in the gain channel, irrespective of initial conditions.
%Optical switching in PT couplers can also be realized based on quadratic, or $\chi^{(2)}$ nonlinearity~\cite{Li:2013-53820:PRA}.
It was suggested that switching performance can be enhanced with application of Bragg gratings, involving specially engineered modulation of the real and imaginary part of the optical refractive index~\cite{Lupu:2014-305:PNFA}.
These concepts can be extended to other physical systems beyond optics, including electrical circuits~\cite{Schindler:2011-40101:PRA, Cuevas:2013-32108:PRA}.

A PT-symmetric coupler, with additional gain and loss proportional to nonlinear terms has been studied in Refs.~\cite{Duanmu:2013-20120171:PTRSA, Miroshnichenko:2011-12123:PRA, Zezyulin:2011-64003:EPL}. The system can be described by the following equations,
\begin{eqnarray}
  i\dot{u}_1 &=& - u_2 +(\chi-i\gamma_{non})|u_1|^2 u_1 -i\gamma u_1,\\
  i\dot{u}_2 &=& - u_1 +(\chi+i\gamma_{non})|u_2|^2 u_2 +i\gamma u_2.
\end{eqnarray}
where $\chi$ determines the nonlinearity strength and $\gamma_{non}$ accounts for the PT-symmetric nonlinear loss and gain. The symmetric and asymmetric eigenstates of the system are found analytically.
The additional nonlinear gain and loss result in the emergence of asymmetric solutions which not only are involved in symmetry-breaking bifurcations, but also produce asymmetric linearization matrices with spectral properties that reflect this asymmetry. If such nonlinear PT-symmetric coupler implemented into a linear chain, it may give raise to new types of nonlinear Fano resonances, with entirely suppressed or greatly amplified transmission~\cite{Miroshnichenko:2011-12123:PRA}.

%\REMOVE{
%\subsection{Nonlinear nonreciprocity in PT coupled ring resonators} \lsect{PT-couplers-ring}

\subsection{Actively coupled waveguides} \lsect{PT-couplers-active}

Nonlinearity can provide an important mechanism to balance gain and loss on average~\cite{Alexeeva:2014-13848:PRA, Barashenkov:2014-282001:JPA, Barashenkov:2015-325201:JPA}.
%PT symmetry can also appear
Such situation can arise even when linear PT symmetry is absent, in particular when the coupling between the waveguides is non-conservative and provides gain or loss depending on the relative phase between the guided modes~\cite{Alexeeva:2014-13848:PRA}. Physically, this can be realized when a pair of nonlinear optical waveguides with absorption are placed in a medium with power gain.
%, see Fig.~\rpict{PTcouplerActive}(a).
The evolution of mode amplitudes in two waveguides is governed by the following normalized equations,
\begin{equation} \leqt{CouplerActive}
   \begin{array}{l} {\displaystyle
        i\frac{d \psi_1}{d z} + i \gamma \psi_1 + |\psi_1|^2 \psi_1 + (1 - i a) \psi_2 = 0,
   } \\*[9pt] {\displaystyle
        i\frac{d \psi_2}{d z} + i \gamma \psi_2 + |\psi_2|^2 \psi_2 + (1 - i a) \psi_1 = 0,
   } \end{array}
\end{equation}
where $\gamma > 0$ is the net loss rate and $a$ is the active coupling coefficient.
Note that these equations are different from Eqs.~\reqt{DNLS}, and it was found that nonlinear dynamics demonstrates distinct regimes including stationary, periodic, and chaotic oscillations as outlined in Fig.~\rpict{PTcouplerActive}(a).

The reason for the appearance of oscillations is the following. For $a > \gamma$, symmetric supermodes with $\psi_1 = \psi_2$ exhibit a linear amplification rate $(\gamma - a)$. As the amplitude increased beyond a certain level, nonlinearity results in symmetry breaking and localization of light in one waveguides, and then the mode exhibits linear loss with rate approaching $\gamma$. Therefore, real nonlinearity acts as an effective nonlinear absorber.
Examples of the resulting oscillations are shown in Figs.~\rpict{PTcouplerActive}(b,c).
An important advantage of the actively coupled nonlinear PT structure is its structural
stability. For the given loss rate, the system supports stationary
and periodic light beams in a wide range of gain coefficients. This is a fundamental
distinction from the conventional PT-symmetric coupler discussed above, where one has to tune the gain to match the loss exactly to obtain periodic oscillations.

%---------------------------------------------------------------------------------------------------
\pict{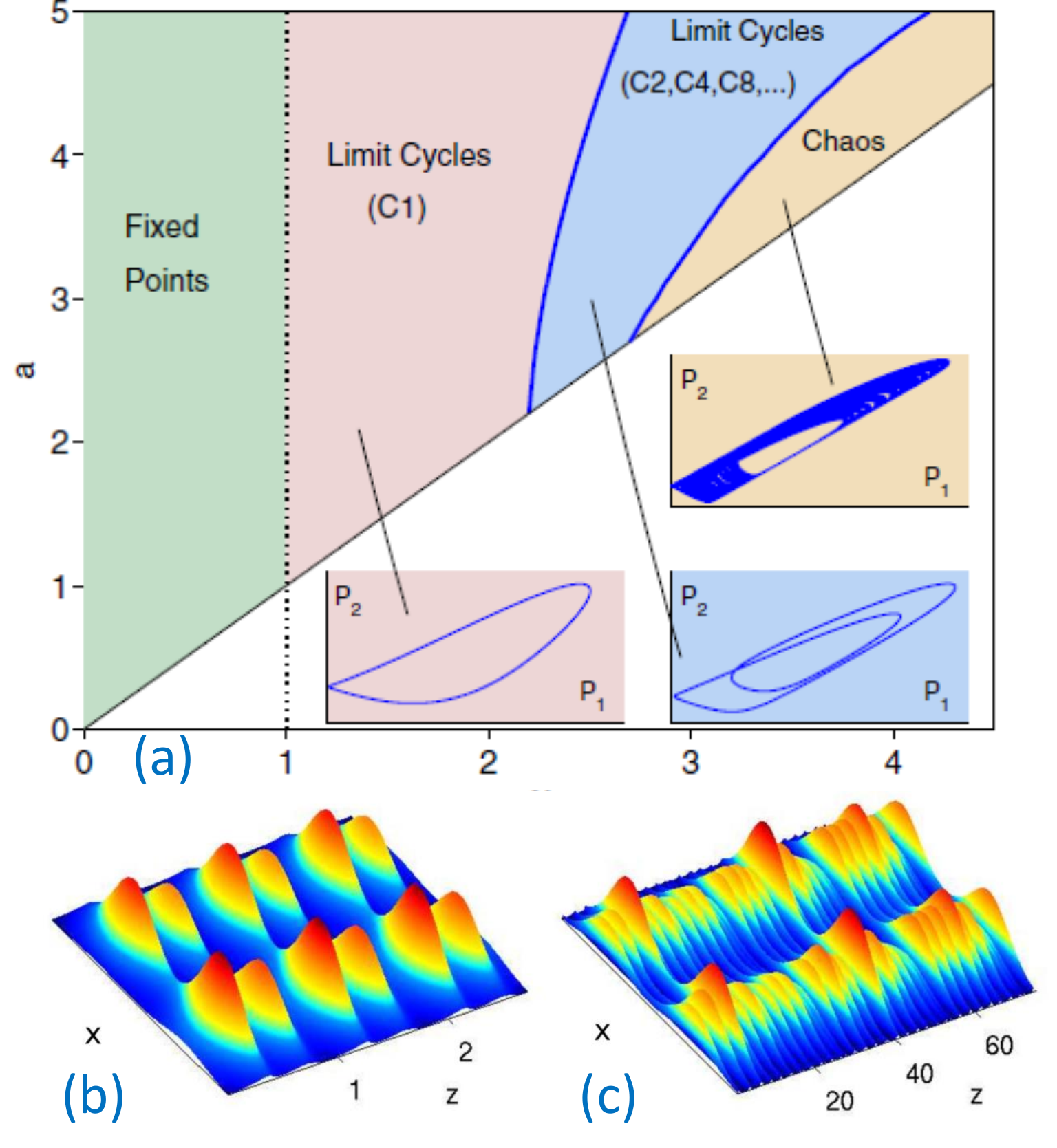}{PTcouplerActive}{
Actively coupled PT system: two lossy waveguides placed in an active medium with gain.
%(a)~Illustration of two coupled lossy waveguides placed in an active medium with gain.
(a)~The chart of attractors in the plane of the normalized net loss rate ($\gamma$) and active coupling coefficient ($a$).
(b)~The limit cycle with two oscillations per period ($a = 9$, $\gamma = 4.2$).
(c)~The chaotic attractor ($a = 9$, $\gamma = 7.6$).
Figures after Ref.~\cite{Alexeeva:2014-13848:PRA}.
}
%---------------------------------------------------------------------------------------------------

Another type of PT symmetry can appear when active coupling between the modes is nonlinear~\cite{Sarma:2014-52918:PRE}. This can be potentially realized in a two waveguide coupler in the presence of a nonlocal nonlinearity, which respects PT symmetry.
The corresponding model shows unique behavior, in particular admitting both bright and dark soliton solutions
at the same time.

\subsection{Anti-PT-Symmetry and parametric amplification} \lsect{PT-couplers-parametric}

Nonlinear modes in PT potentials can also be supported by quadratic ($\chi^{(2)}$) optical nonlinearity~\cite{Moreira:2012-53815:PRA, Li:2013-53820:PRA}. Compared to the case of Kerr-type nonlinearity discussed above, quadratic nonlinear response involves parametric coupling of the fundamental and second-harmonic optical waves. Such interactions give rise to a rich family of localized solitons.

Quadratic nonlinearity can also facilitate a fundamentally important regime of parametric amplification, which is efficiently realized through difference-frequency generation~\cite{Boyd:2008:NonlinearOptics}. Here the amplification rate is determined by the pump, enabling ultrafast all-optical tunability. Quadratic nonlinear PT-symmetric systems can, on one hand, realize ultrafast spatial signal switching through pump-controlled breaking of PT symmetry, and on the other hand enable spectrally-selective mode amplification~\cite{Antonosyan:1506.02143:ARXIV}.
The process of optical parametric amplification in PT coupler based on nonlinear mixing between a strong pump, and signal and idler waves is illustrated in Fig.~\rpict{PT-OPA}(a). The evolution of near-degenerate (with close similar frequencies) signal and idler waves in the undepleted pump regime
is modelled by coupled-mode equations, which can be represented in a Hamiltonian form:
% describing the PT-symmetric coupler with the modevector $\mathbf{a}$ and the Hamiltonian $\mathbf{H}$
%
\begin{equation} \leqt{HEM}
    i\frac{\partial\mathbf{a}}{\partial z}=\mathbf{{H}}\mathbf{a},
\end{equation}
where $\mathbf{a}(z)= \left(a_{s1}(z), a_{s2}(z), a^{\ast}_{i1}(z), a^{\ast}_{i2}(z)\right)^T$ is a vector of wave amplitudes in the two waveguides,
the subscripts stand for signal (`s') and idler (`i') waves in two waveguides (`1' and `2'),
\begin{equation}\leqt{HF}
    \mathbf{H}=\left( \begin{array}{cccc}
    \beta-i\gamma_{1} & -C & iA_{1} & 0 \\
    -C & \beta-i\gamma_{2} & 0 & iA_{2} \\
    iA^{\ast}_{1} & 0 & -\beta-i\gamma_{1} & C \\
    0 & iA^{\ast}_{2} & C & -\beta-i\gamma_{2} \end{array} \right) ,
\end{equation}
%
%$\beta = (\beta_p - \beta_s - \beta_i) / 2$
$\beta$ defines the phase mismatch of parametric wave mixing, $\gamma$ are the linear loss coefficients, $C$ is the mode coupling coefficient between the waveguides,
%and $\chi$ are effective quadratic nonlinear coefficients.
and $A$ are the normalized input pump amplitudes.

%---------------------------------------------------------------------------------------------------
\pict{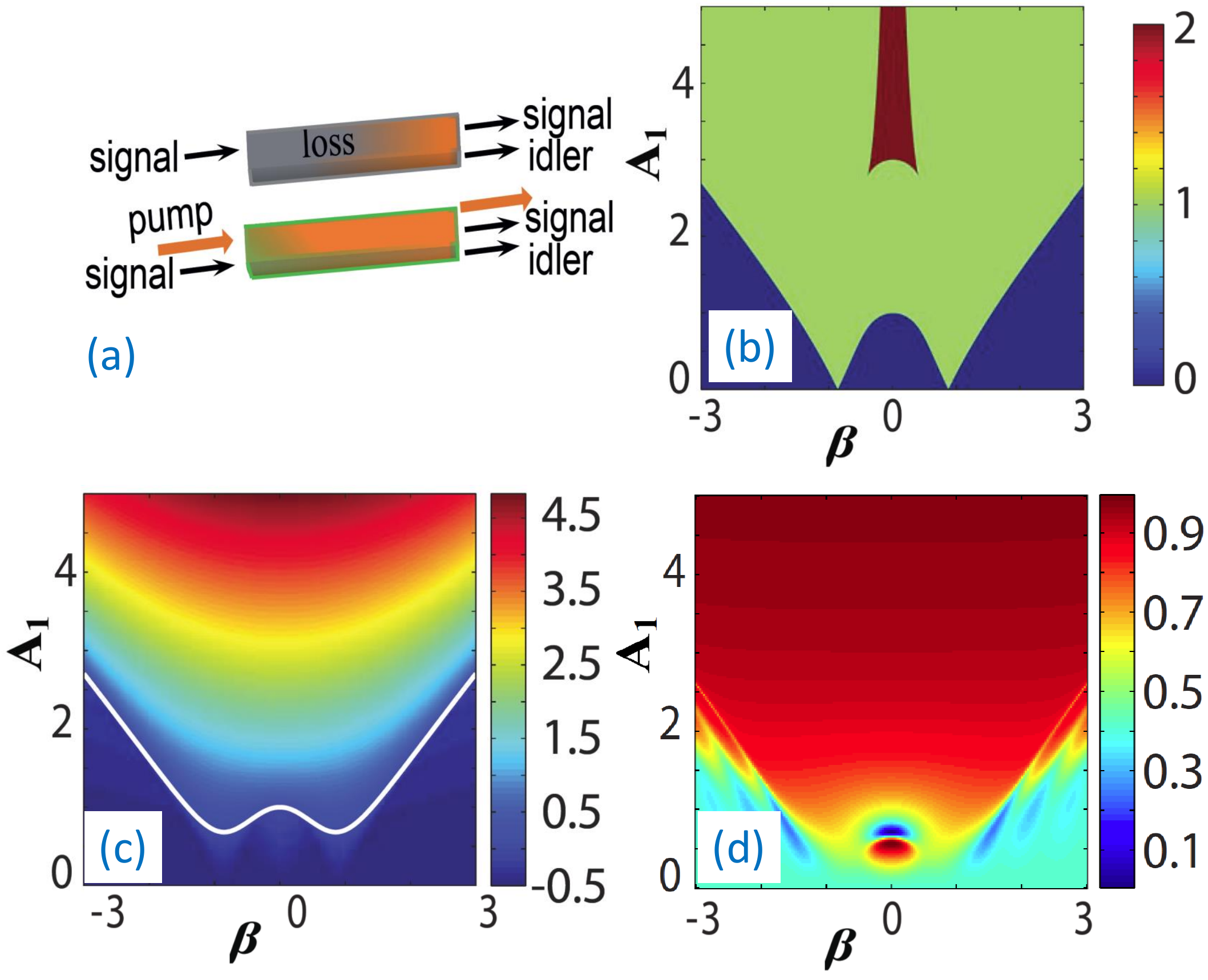}{PT-OPA}{
PT parametric amplifier:
(a)~scheme of wave mixing in a quadratic nonlinear coupler with linear absorption in one waveguide.
(b)~Number of spectrally anti-PT symmetric mode pairs,
(c)~the largest mode gain (white line marks zero),
(d)~fraction of signal intensity in the first waveguide at the output
vs. the input pump amplitude in the first waveguide and the phase-mismatch.
Figures after Ref.~\cite{Antonosyan:1506.02143:ARXIV}.
}
%---------------------------------------------------------------------------------------------------

Remarkably, the Hamiltonian possesses a {\em spectral anti-PT symmetry}, corresponding to a negative sign on the right-hand side of the following equality,
\begin{equation} \leqt{AntiSC}
    \mathbf{P}_{1,+}\mathbf{P}_{2,+}\mathbf{T}\mathbf{H}
        = -\mathbf{H}\mathbf{P}_{1,+}\mathbf{P}_{2,+}\mathbf{T}.
\end{equation}
Here $\mathbf{T}$ is a usual time-reversal operator which changes $z \rightarrow -z$ and performs a complex conjugation. The {\em parity operators operate in spectral domain}, interchanging the signal and idler waves,
\begin{equation}
 \mathbf{P}_{1,\pm}=\left\{ a_{s1}\leftrightarrow \pm a^{\ast}_{i1} \right\},
 \, \mathbf{P}_{2,\pm}=\left\{ a_{s2}\leftrightarrow \pm a^{\ast}_{i2} \right\}.
\end{equation}
The Hamiltonian is linear, and the properties of its solutions are defined by the spectrum of its four eigenmodes.
%We have found four eigenvalues corresponding to each of the four eigenmodes for the signal and idler modes in each waveguide, respectively. The behavior of the real and the imaginary parts of the eigenmodes depending on the loss coefficient shows the symmetry in the system.  Based on the fact that our system has four eigenvalues, it is easy to understand that
Overall, there can be three possible symmetry regimes: (i)~two mode pairs with broken spectral anti-PT symmetry, (ii)~one pair of PT-broken modes and a pair of anti-PT symmetric modes, or (iii)~two pairs of anti-PT symmetric modes. Characteristic dependance of the number of PT-symmetric mode pairs in depending on the pump amplitude in the first waveguide and the phase-mismatch is shown in Fig.~\rpict{PT-OPA}(b), and the largest mode gain is presented in Fig.~\rpict{PT-OPA}(c).
Remarkably, the established relations of mode symmetry and gain/loss
are reversed in comparison to previously discussed spatial PT-symmetry in directional couplers, due to the spectral {\em anti}-PT symmetry of parametric wave mixing.
Specifically, when all modes have broken spectral PT symmetry [blue shaded regions in Fig.~\rpict{PT-OPA}(b)], the pairs of eigenmodes exhibit the same amount of gain/loss, and effectively the amounts of gain and loss are averaged out between the eigenmodes. However, upon transition to the region with spectrally PT-symmetric modes [green and red shaded regions in Fig.~\rpict{PT-OPA}(b)], there appears an unequal redistribution of gain and loss between the modes. One PT-symmetric eigenmode exhibits gain much larger then all other modes, while the latter experience stronger loss. Such sensitivity of amplification to PT-breaking threshold could be used to discriminate between multiple spectral modes, analogous to the concept of PT-lasers~\cite{Feng:2014-972:SCI, Hodaei:2014-975:SCI}.

In case of perfect phase-matching, $\beta=0$, and pump amplitudes of the same phase ${\rm Im}(A_{1}) = {\rm Im}(A_{2}) = 0$, the Hamiltonian also features {\em spatial PT symmetry}. Both the spatial and spectral PT-symmetry breaking occurs simultaneously at the threshold $|\gamma_1-\gamma_2 - \eta (A_1 + A_2)| = 2 C$. However, the spatial and spectral symmetries are opposite: a mode pair is spatially PT-symmetric and has spectrally broken symmetry below threshold, whereas the situation is reversed above the threshold. Numerical simulations reveal a strong connection between spectral symmetry and spatial dynamics even for non-zero phase mismatches. Indeed, an increase of pump amplitude can control the period of mode coupling between the waveguides, while the oscillations get suppressed close to the spectral PT threshold, see Fig.~\rpict{PT-OPA}(d). It is expected that due to the universality of parametric amplification processes,
these concepts can be extended to different physical mechanisms including four wave mixing in media with Kerr-type optical nonlinearity.

\subsection{Nonlinear modes} \lsect{PT-couplers-modes}

Analysis in previous sections was based on the consideration of coupling between the fundamental guided modes. However, new effects can appear when PT systems support multiple modes, including bound and extended (non-localized) states. Various types of double-well potentials have been investigated~\cite{Tsoy:2012-3441:OC, Cartarius:2012-444008:JPA, Rodrigues:2013-5:ROMRP}.
It was confirmed that discrete coupled equations can accurately describe the full continuous dynamics in a broad range of parameters.
%~\cite{Tsoy:2012-3441:OC, Rodrigues:2013-5:ROMRP}
However, continuous models can be used to predict a range of important effects beyond two-mode approximation, such as the appearance of radiation losses in curved PT waveguides~\cite{Tsoy:2012-3441:OC}. On the other hand, it was predicted that a PT structure placed in a nonlinear medium can trap soliton~\cite{Mayteevarunyoo:2013-22919:PRE}. Furthermore, soliton scattering by a localized mode of PT coupler can exhibit nonreciprocal and non-conservative scattering, which is strongly dependent on the relative phase between the mode and the soliton~\cite{Suchkov:2014-443:APA}.

Properties of nonlinear modes supported by a harmonic PT-symmetric potential also show a number of distinguishing features~\cite{Zezyulin:2012-43840:PRA}. In particular, the modes bifurcating from different eigenstates of the underlying linear problem can belong to the same family of nonlinear modes. By adjusting the potential profile, it is possible to enhance the stability of small-amplitude and strongly nonlinear modes compared
to the case of conservative real harmonic potential.

{\protect\renewcommand\sectpath{Oligomers}
\section{Discrete PT-symmetric oligomers}\lsect{}

PT-symmetric discrete oligomers have recently attracted a lot of attention, motivated by possibilities of their experimental realisation~\cite{Ruter:2010-192:NPHYS, Ramezani:2010-43803:PRA}. In particular, many scientific works address the questions of  existence and stability of nonlinear states in PT-symmetric oligomers, which may drastically  differ from eigenmodes of underlying linear systems. The nonlinear effects in PT-symmetric systems can be utilized for an efficient control of light including all-optical low-threshold switching and unidirectional invisibility~\cite{Ramezani:2010-43803:PRA, Lin:2011-213901:PRL}. The possibility to engineer PT-symmetric oligomers, which may include nonlinearity, triggers a broad variety of studies on both the few-site systems and entire PT-symmetric lattices, including one-dimensional PT-symmetric dimer~\cite{Li:2011-66608:PRE, Miroshnichenko:2011-12123:PRA}, trimer~\cite{DAmbroise:2012-444012:JPA, Li:2011-66608:PRE}, quadrimer~\cite{Li:2011-66608:PRE, Zezyulin:2012-213906:PRL}, 2D PT-symmetric plaquettes~\cite{Li:2012-444021:JPA, Zezyulin:2012-213906:PRL}, and PT-symmetric finite/infinite chains~\cite{Bendix:2009-30402:PRL, Pelinovsky:2014-85204:JPA} and necklaces~\cite{Barashenkov:2013-33819:PRA} and multicore fibers~\cite{Martinez:2015-23822:PRA}.

In this section we discuss PT-symmetric configurations consisting of three or more coupled waveguides. We first formulate the general model of finite discrete PT-symmetric oligomer and next consider some special configurations of PT-symmetric trimers and quadrimers. We also discuss finite open and periodic few-site geometries.

Consider an array of $N$ waveguides coupled to each other. We assume that each waveguide experiences some gain or loss determined by parameter $\gamma_n$, where $n$ is a waveguide number. Taking into account conservative Kerr nonlinear response, the light propagation through this system can be modeled by a system of discrete nonlinear Schr\"{o}dinger equations (DNLSE),
%-------------------------------------------------------------------------------------------------
\begin{equation}\label{DNLSE}
    i\dot{u}_n=-\sum_{m=1}^N K_{n m} u_m -|u_n|^2 u_n -i\gamma_n u_n,\quad n=\overline{1:N},
\end{equation}
%-------------------------------------------------------------------------------------------------
where $u_n$ are mode amplitudes, $\dot{u}_n\equiv\dd u_n/\dd z$, $z$ is the propagation distance, $\gamma_n>0$ ($\gamma_n<0$) corresponds to loss (gain), and $K_{n m}$ are the coupling coefficients between sites $n$ and $m$. We assume that $\textbf{K}=\{K_{nm}\}$ is a real symmetric matrix, i.e. $K_{nm}=K_{mn}=K_{nm}^{*}$. The system~\eqref{DNLSE} can be rewritten in matrix form as
%--------------------------------------------------------------------------------------------------
\begin{IEEEeqnarray}{l}\label{DNLSE1}
    i\dot{\textbf{u}}=-[\mathbf{H}+\mathbf{F}(\mathbf{u})]\mathbf{u}, \quad \mathbf{H}\equiv\mathbf{K}+i\mathbf{G},\\
    \mbox{where } \textbf{G}=\mathrm{diag}(\gamma_1,\gamma_2,...,\gamma_N)\nonumber\\
    \mbox{and } \textbf{F}(\textbf{u})=\mathrm{diag}(|u_1|^2,|u_2|^2,...,|u_N|^2).
\end{IEEEeqnarray}
%-------------------------------------------------------------------------------
Note that if $\textbf{H}$ commutes with the $\mathbf{PT}$ operator, i.e. $[\mathbf{PT},\textbf{H}]=0$, then $\mathbf{H}$ is a PT-symmetric matrix.

Stationary nonlinear modes of Eq.~\eqref{DNLSE1} are sought in the form $\textbf{u}(z)=\exp(ibz)\textbf{w}$, with propagation constant (eigenvalue) $b$ and $\textbf{w}$ satisfying the following equation
%-------------------------------------------------------------------------------
\begin{equation}\label{DNLSE2}
    b\textbf{w}=[\textbf{H}+\textbf{F}(\textbf{w})]\textbf{w}.
\end{equation}
%-------------------------------------------------------------------------------
Hereafter we assume that $\sum\limits_{n=1}^N\gamma_n=0$, which is a necessary condition for the system spectrum to be real~\cite{Zezyulin:2012-213906:PRL}.

In the following, we consider in detail two PT-symmetric geometries with the nearest-neighbor interaction and homogeneous coupling strength between the sites. Then the coupling coefficients can be expressed as $K_{nm}=k\delta_{|n-m|,1}$.

\subsection{PT-symmetric trimer}

%--------------------------------------------------------------------------------------------------------
\pict{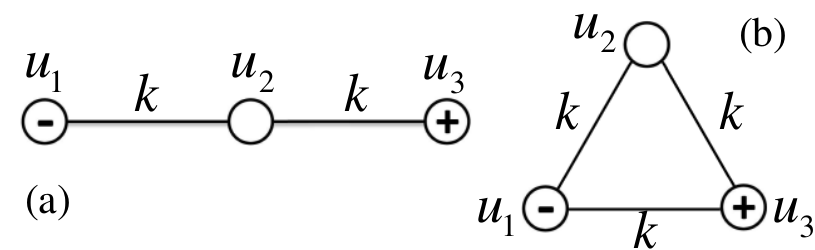}{F4}{
Schematic of PT-symmetric trimer for (a) open and (b) periodic boundary conditions. Signs "-" and "+" denote loss and gain sites,
respectively, $k$ - coupling coefficient.
}
%--------------------------------------------------------------------------------------------------------

PT-symmetric trimer, first discussed in~\cite{Li:2011-66608:PRE} and revisited in~\cite{Li:2013-375304:JPA,Duanmu:2013-20120171:PTRSA} is schematically shown in Fig.~\rpict{F4}. It is composed of three waveguides: active, lossy and conservative ones. In the linear regime there is only one possible PT-symmetric interposition of these waveguides with non-zero PT-symmetry breaking threshold: equal gain and loss parameters, i.e. $\gamma_1=-\gamma_3\equiv\gamma$, with conservative waveguide in the middle. The PT-symmetry breaking threshold is defined by the relation $\gamma_{\rm crit}=\sqrt{2}k$.

In the nonlinear regime, the dynamical equations for the open trimer [open boundary conditions, Fig.~\rpict{F4}(a)] are
%---------------------------------------------------------------------------------------------------------
\begin{eqnarray}\label{TrimerPl}
  i\dot{u}_1=-k u_2 -|u_1|^2 u_1 -i\gamma u_1,\nonumber\\
  i\dot{u}_2=-k(u_1+u_3) -|u_2|^2 u_2,\\
  i\dot{u}_3=-k u_2 -|u_3|^2 u_3 +i\gamma u_3\nonumber.
\end{eqnarray}
Here we are interested in the interplay of nonlinearity with PT-symmetry and its effect on stationary modes and their stability. We search for the stationary solutions of Eqs.~\eqref{TrimerPl} in the form $u_1=A\exp[ibz+\phi_a]$, $u_2=B\exp[ibz+\phi_b]$, and $u_3=C\exp[ibz+\phi_c]$. Substituting these ansatz into Eqs.~\eqref{TrimerPl} we find relations on the soliton parameters,
%-----------------------------------------------------------------------------------------------------------------------------------------------------------------
\begin{IEEEeqnarray}{l}
  \label{TrimerCoefConstrain}
  A^2\left[\gamma^2+(b-A^2)^2\right]^2- k^2 b\left[\gamma^2+(b-A^2)^2\right]-\nonumber \\
 2k^4A^2+2k^4b=0, \nonumber\\
  B^2 = \frac{b\pm\sqrt{b^2-8A^2(b-A^2)}}{2},\\
  C = A,\nonumber\\
  \sin(\phi_b-\phi_a)=-\sin(\phi_b-\phi_c)=-\frac{\gamma A}{k B}, \nonumber \\
  \cos(\phi_b-\phi_a)=\cos(\phi_b-\phi_c)=\frac{b A- A^3}{k B}\nonumber
\end{IEEEeqnarray}
%-----------------------------------------------------------------------------------------------------------------------------------------------------------------
Note that without loss of generality we may put $\phi_b=0$, which yields $A=C$, $\phi_a=-\phi_c$ and thus $$u_3=u_1^*.$$ Thereby, all stationary modes of the PT-symmetric trimer possess equally distributed intensity between loss and gain waveguides.

Analysis of Eqs.~\eqref{TrimerCoefConstrain} shows that there are up to five (at least one) branches of stationary solutions of Eqs.~\eqref{TrimerPl} depending on the particular values of $b$ and $k$. The dependence of these branches on gain/loss parameter $\gamma$ for particular case $b=k=1$ has been discussed in detail in~\cite{Li:2011-66608:PRE, Li:2013-375304:JPA}. Three different branches were found: first branch is mostly unstable (except small region that splits unstable region into two domains), while the second one is chiefly stable. These branches collide in a saddle-center bifurcation at $\gamma=1.043$ and disappear. Interestingly, the third branch bifurcates from zero for $\gamma>\sqrt{2-b^2}$ and persists for larger values of $\gamma$ even beyond the PT-symmetry breaking threshold $\gamma_{\rm crit }=\sqrt{2}$, although this branch is unstable for $\gamma>1.13$. In Fig.~\rpict{F5} the system dynamics for the initial conditions belonging to the first branch is presented for two different values of gain/loss parameter $\gamma$. In Fig.~\rpict{F5}(a) for $\gamma=0.5$ (first unstable region), the nonlinear modes are unstable and the growth of intensity is observed in the active site $u_3$, while two other modes $u_1$ and $u_2$ decay. In Fig.~\rpict{F5}(b) for $\gamma=1.04$ is taken from the second unstable region, but this time intensity growth is observed not only in the active waveguide but also in the conservative one. Both these types of dynamics are also observed for the second branch.
%--------------------------------------------------------------------------------------------------------
\pict{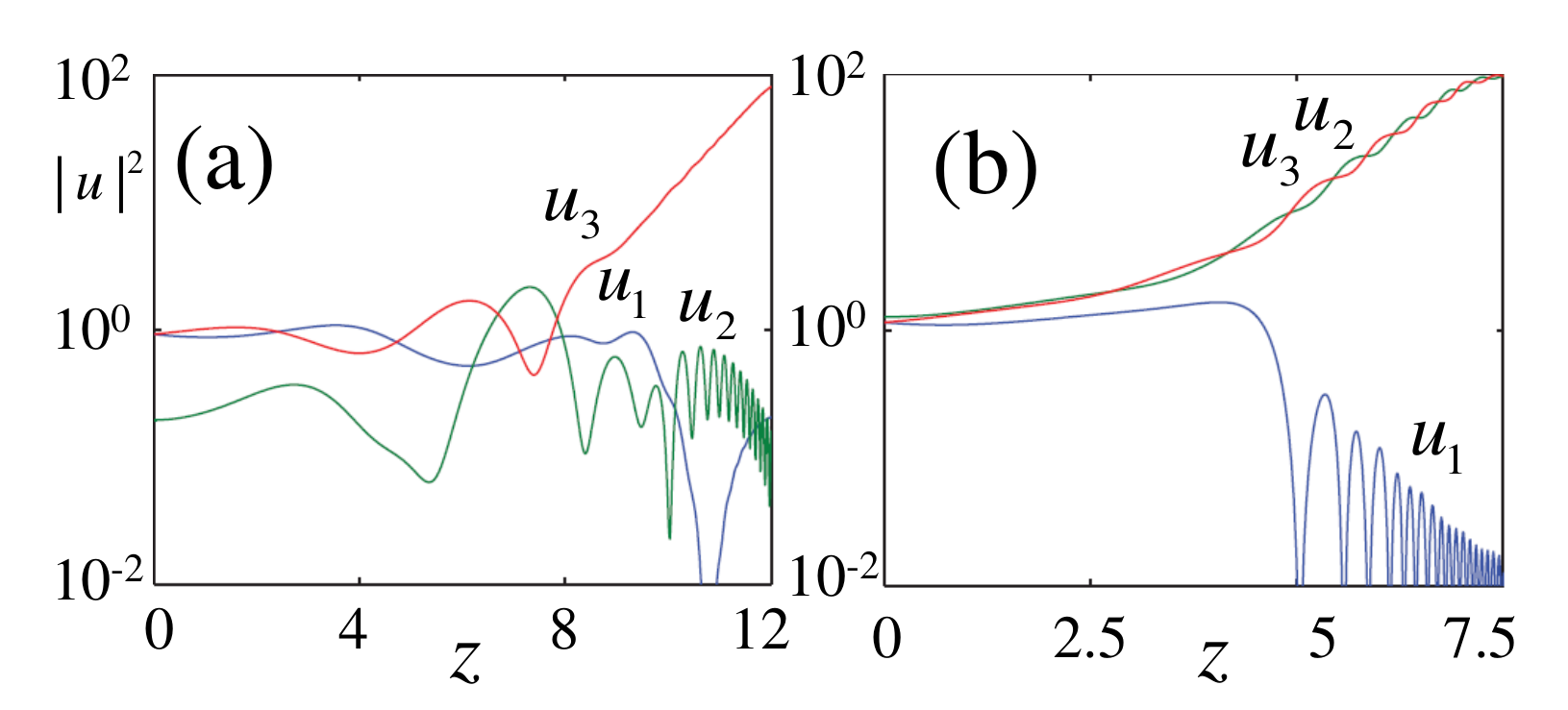}{F5}{
Dynamical evolution of initial data belonging to the first branch of the PT-symmetric trimer. (a) $\gamma=0.5$ (first unstable region), (b) $\gamma=1.04$ (second unstable region). Red, blue and green lines denote $u_3$, $u_2$, and $u_1$, respectively. Adopted from~\cite{Li:2011-66608:PRE}.
}
%--------------------------------------------------------------------------------------------------------

Note that in the considered case it is possible to find two additional branches of solutions of Eqs.~\eqref{TrimerCoefConstrain} assuming $b$ to be complex-valued, although modes with such propagation constants will not be longer solutions of the original system~\eqref{TrimerPl}. Such so-called ghost states can predict evolution of the systems~\eqref{TrimerPl} at short propagation distances~\cite{Li:2013-375304:JPA}.

The second configuration of the PT-symmetric trimer schematically depicted in Fig.~\rpict{F4}(b) has also been investigated in Ref.~\cite{Li:2011-66608:PRE} (and also discussed in~\cite{Leykam:2013-371:OL}). Although the PT-symmetry breaking threshold for underlying linear system is $\gamma_{\rm crit}=0$, the nonlinear branches of solutions still exist for nonzero values of gain/loss parameter $\gamma$. Thus, nonlinearity provides stable stationary nontrivial solutions of the system, the linear spectrum of which has complex eigenvalues.

More complicated structure of PT-symmetric trimer including competing nonlinear gain with linear loss in the first waveguide and vise versa in the third waveguide has also been studied in~\cite{Duanmu:2013-20120171:PTRSA}. The introduced nonlinear gain and loss give rise to rich features of asymmetric modes not observed in linear gain/loss profile systems.

It is worth to note here that experimental realization of dissipative trimer having "gain-loss-gain" profile has been done in~\cite{Li:2013-375304:JPA}. Although this trimer does not possess PT-symmetry, this experimental setup can be considered as a first step to realization of PT-symmetric trimer.

\subsection{PT-symmetric quadrumer}

In contrast to the PT-symmetric trimers considered above the four-site geometry admits  several PT-symmetric configuration. Taking into account PT-symmetry requirements (symmetrical distribution of gain and loss sites) the general model of a one-dimensional PT-symmetric quadrimer with nearest-neighbor interaction [see Fig.~\rpict{F3}] can be described by the following equations,
%-----------------------------------------------------------------------------------------------------------------------
\begin{eqnarray}
  \label{quadr}
  i\dot{u_1} &=& -k u_2 -|u_1|^2 u_1 -i\gamma_1 u_1,\nonumber\\
  i\dot{u_2} &=& -k(u_1+u_3) -|u_2|^2 u_2-i\gamma_2 u_2,\\
  i\dot{u_3} &=& -k(u_2+u_4) -|u_3|^2 u_3 +i\gamma_2 u_3,\nonumber\\
  i\dot{u_4} &=& -ku_3 -|u_4|^2 u_4 +i\gamma_1 u_4.\nonumber
\end{eqnarray}
%-----------------------------------------------------------------------------------------------------------------------
%-----------------------------------------------------------------------------------------------------------------------
\pict{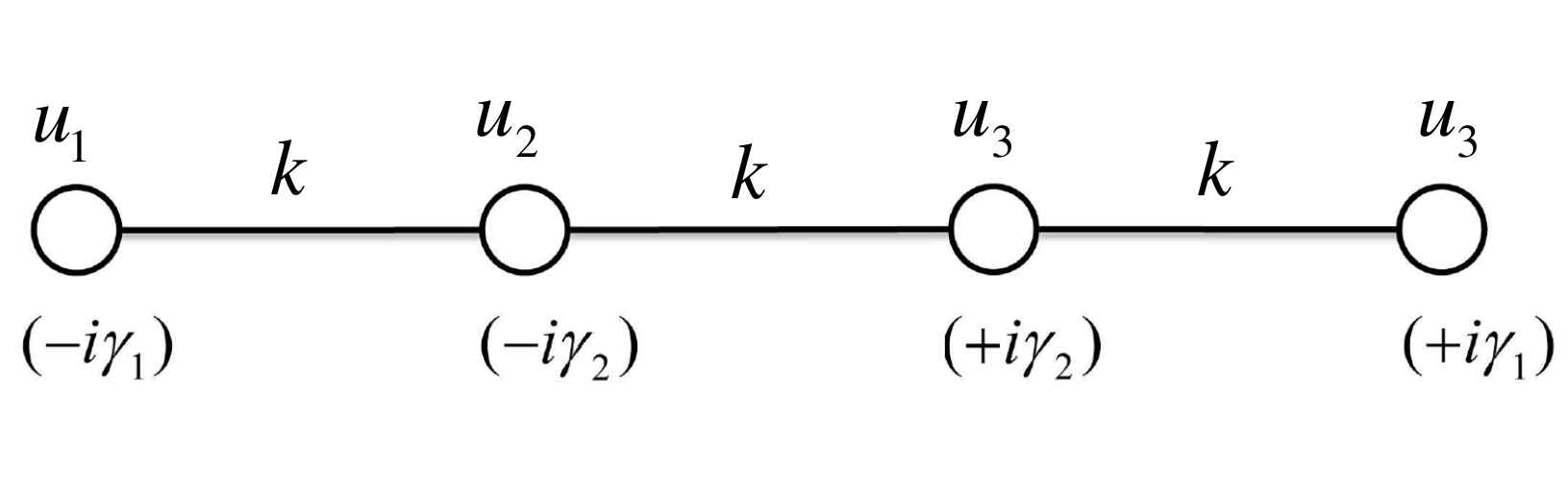}{F3}{
Schematic of PT-symmetric quadrimer with nearest-neighbor interaction with equal coupling strength.
}
%-----------------------------------------------------------------------------------------------------------------------

Similar to the trimer case, the existence and stability of stationary nonlinear modes for $\gamma_1=\gamma_2$ has been investigated in~\cite{Li:2011-66608:PRE}. Such structure additionally possesses completely asymmetric solutions and the symmetry-breaking bifurcation structure is more complicated.

Let us consider in detail general case of bi-parameter gain and loss, i.e. $\gamma_1\neq\gamma_2$~\cite{Zezyulin:2012-213906:PRL}. The analysis of the underlying linear system of~\eqref{quadr} shows that there are three types of regions in the parameter plane $\gamma_1 \gamma_2$ [see Fig~\rpict{Fig3} left top panel]: (i) unbroken PT-symmetry, with all eigenvalues real; (ii) broken PT symmetry with two real and two complex conjugated eigenvalues (this case is only possible for $\gamma_1\neq\gamma_2$); (iii) broken PT-symmetry with all eigenvalues complex. Note that there are "triple" points in left top panel Fig.~\rpict{Fig3}~ denoted $T_j$, which correspond to contact points of three regions. At these points, all eigenvalues $\tilde{b}_n=0$, hereafter "tilde" indicates eigenstates of the underlying linear system.
%-------------------------------------------------------------------------------------------------
\pict{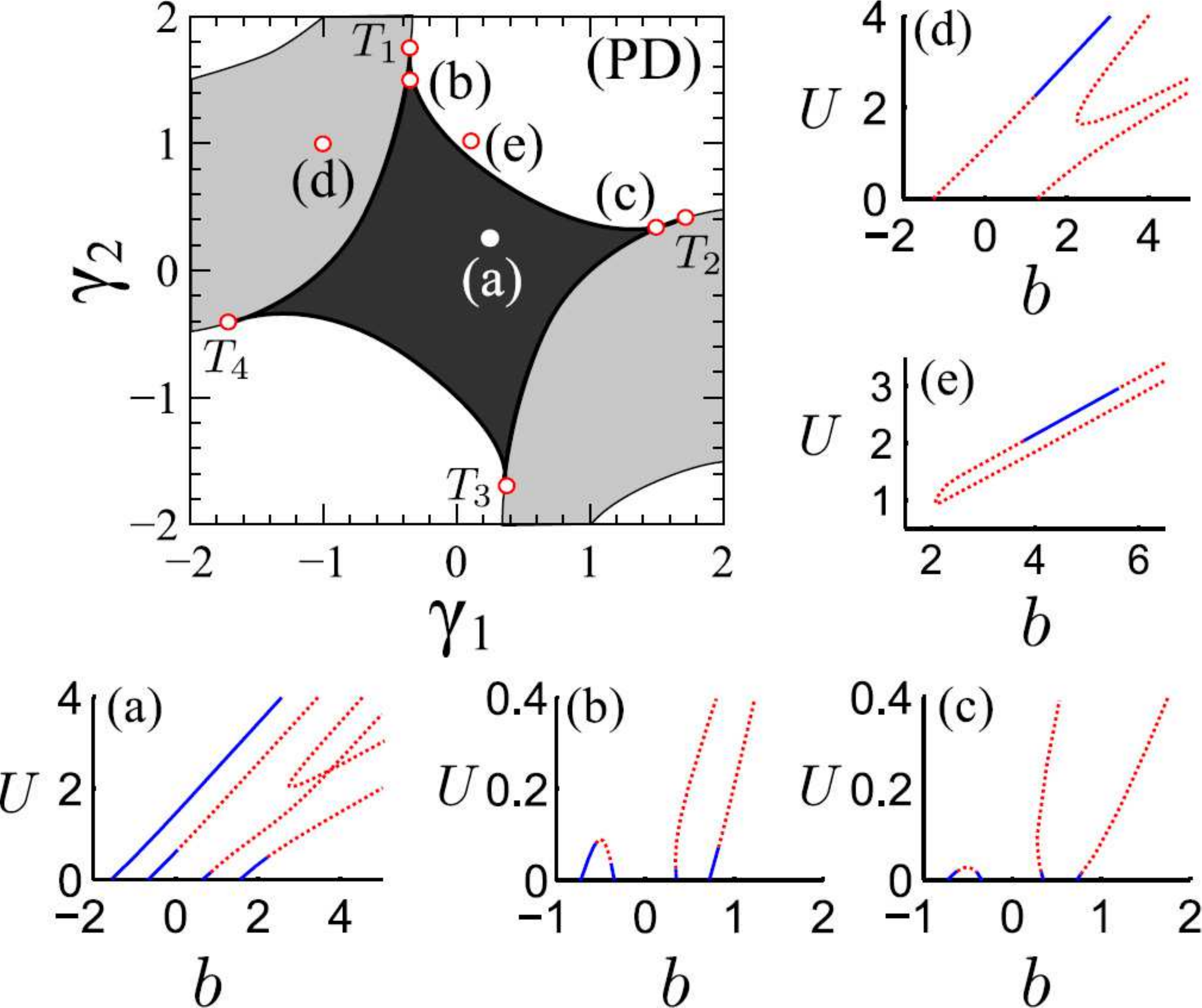}{Fig3}{
Regions of stability underlying linear system of Eqs.~\eqref{quadr}. The dark diamond-shaped region corresponds to unbroken PT-symmetry phase; the light-grey regions represent parameter ranges where two of eigenvalues are real and two are complex; and the white regions correspond to all eigenvalues are complex. Curves in (a)-(e) correspond to the branches of  nonlinear modes, where $U$ is a total energy flow determined by~\eqref{powflow}. The solid blue and dashed red lines indicate the stable and unstable branches of modes, respectively. Panels (a)-(e) correspond to the points (a)-(e) in the Phase Diagram. Adopted from Ref.~\cite{Zezyulin:2012-213906:PRL}.
}
%-------------------------------------------------------------------------------------------------

Next, we discuss an effect of nonlinearity on stationary modes, considering unbroken PT phase where eigenstates $\tilde{\mathbf{w}}$ of the matrix $\mathbf{H}$ of the system~\eqref{quadr} are simultaneously the eigenstates of PT operator, i.e $P_0T\tilde{\mathbf{w}}=\tilde{\mathbf{w}}$ up to an irrelevant phase shift. Here $P_0$ is parity operator of the form
%-------------------------------------------------------------------------------------------------
\begin{equation}\label{parityoperator}
  P_0=\left(
  \begin{array}{c c}
  \mathbf{0} & \sigma_1\\
  \sigma_1 & \mathbf{0}
  \end{array}
  \right).
\end{equation}
%------------------------------------------------------------------------------------------------
Hereafter $\sigma_{1,2,3}$ are Pauli matrices and $\mathbf{0}$ is $2\times2$ zero matrix. It is reasonable to look for nonlinear modes possessing the same property  $\mathcal{P}_0\mathcal{T}\mathbf{w}=\mathbf{w}$. Thus, we may put $w_1=w_4^*$ and $w_2=w_3^*$ which leads to the set of equations
%-----------------------------------------------------------------------------------------------------------------------------
\begin{eqnarray}
% \nonumber % Remove numbering (before each equation)
  b w_1&=&w_2+|w_1|^2w_1+i\gamma_1 w_1,\label{FirstEq}\\
  b w_2&=&w_1+w_2^*+|w_2|^2w_2+i\gamma_2 w_2. \label{SecEq}
\end{eqnarray}
%-----------------------------------------------------------------------------------------------------------------------------
Substituting $w_1=W_1\exp(i\Phi)$ and $w_2=W_2\exp(i\Phi)$  with real $W_1$ and $\Phi$ into Eq.~\eqref{FirstEq}, we get $W_2=W_1(b-W_1^2-i\gamma)$, while from Eq.~\eqref{SecEq} we have $\exp(-2i\Phi)=f(W_1)\equiv (b W_2-W_1-|W_2|^2W_2-i\gamma W_2)/W_2^*$. The roots of the equation $|f(W_1)|^2=1$ immediately give us $w_1$ and $w_2$. The latter equation can be reduced to the $P_8(W_1^2)=0$, where $P_8$ is a eighth-degree polynomial with real coefficients, each real root of which corresponds to a nonlinear branch of modes. Since the roots depend on $b$ continuously, we have continuous families of nonlinear modes for fixed system parameters. Note that these families of modes are additional to one found in~\cite{Li:2011-66608:PRE} for the case $\gamma_1=\gamma_2$. The branches of modes can be characterized by the total energy flow
%---------------------------------------------------------------------------------------------------------------------------------------
\begin{equation}\label{powflow}
  U(b)=\frac14\sum\limits_{n=1}^{4}|w_n|^2.
\end{equation}
%---------------------------------------------------------------------------------------------------------------------------------------
In Fig.~\rpict{Fig3}(a)-(e) the branches of solutions are depicted for the gain/loss parameters $\gamma_1$ and $\gamma_2$ corresponding to the points (a)-(e) in Fig.~\rpict{Fig3} left top panel. It is found that: (i) for small values of gain/loss parameters $\gamma_{1,2}$ from the \textit{exact} PT-symmetry phase region, nonlinear modes bifurcate from linear ones; (ii) if $\gamma_{1,2}$ are chosen from an unbroken PT-symmetry region, but close to triple points $T_j$ as shown in Figs.~\rpict{Fig3}~(b),~(c), nonlinear modes bifurcate from linear limit, but rapidly loose stability and two of these branches cease to exist as $U$ reaches some critical value; (iii) the stable nonlinear modes can be found even in the region of broken PT symmetry [see Figs.~\rpict{Fig3}~(d),~(e)].

We now give a general perspective on the role of nonlinearity for PT-symmetric systems~\cite{Zezyulin:2012-213906:PRL}. It is known that linear PT-symmetric systems exhibit similar behaviour (e.g. real spectrum and energy conservation of eigenmodes) as conservative ones, if they operate below critical point. Thus, there exists a unitary equivalence between PT-symmetric and Hermitian systems~\cite{Mostafazadeh:2002-205:JMP,Mostafazadeh:2003-7081:JPA}. In terms of considered oligomers (we return to matrix representation~\eqref{DNLSE1}), it means that there is a unitary matrix $\mathbf{R}$ that transforms \textit{exactly} PT-symmetric matrix (in PT unbroken phase) $\mathbf{H}_0$ into Hermitian one with the same set of eigenvalues, i.e. $\mathbf{RH}_0\mathbf{R}^{-1}\equiv\mathbf{H}_H=\mathbf{H}_H^*$. This new matrix $\mathbf{H}_H$ will include coefficients $\gamma_n$, which will determine the coupling strength. Thus, the following question naturally arises: are there any significant differences between these PT-symmetric and Hermitian systems having the same eigenvalues in nonlinear regime?

To answer this question we introduce new system, described by the equation $i\dot{\mathbf{u}}=-[\mathbf{H}_H+\mathbf{F}(\mathbf{u})]\mathbf{u}$, where $\mathbf{H}_H$ can be constructed explicitly, following~\cite{Mostafazadeh:2002-205:JMP,Mostafazadeh:2003-7081:JPA}. Families of stationary nonlinear modes of this new system were obtained in~\cite{Zezyulin:2012-213906:PRL} and are presented in Fig.~\rpict{Fig7} for the same values of $\gamma_{1,2}$ as in Figs.~\rpict{F3}~(a)-(c). While for $\gamma_{1,2}$ taken from the "middle" of the unbroken PT phase the behaviour of mode branches for PT-symmetric and Hermitian systems looks similar (except their stability regions)[cf. Fig.~\rpict{Fig3}(a) and Fig.~\rpict{Fig7}(a)], for $\gamma_{1,2}$ placed near the border of the unbroken PT phase, the difference becomes significant [cf. Figs.~\rpict{Fig3}~(b),(c) and Figs.~\rpict{Fig7}~(b),(c)]. While for the Hermitian system the mode branches do not change qualitatively depending on $\gamma_{1,2}$ and there is always at least one stable branch, in the PT-symmetric case there is a bifurcation of two branches and loss of stability for even small values of the energy flow $U$.
%-------------------------------------------------------------------------------------------------
\pict{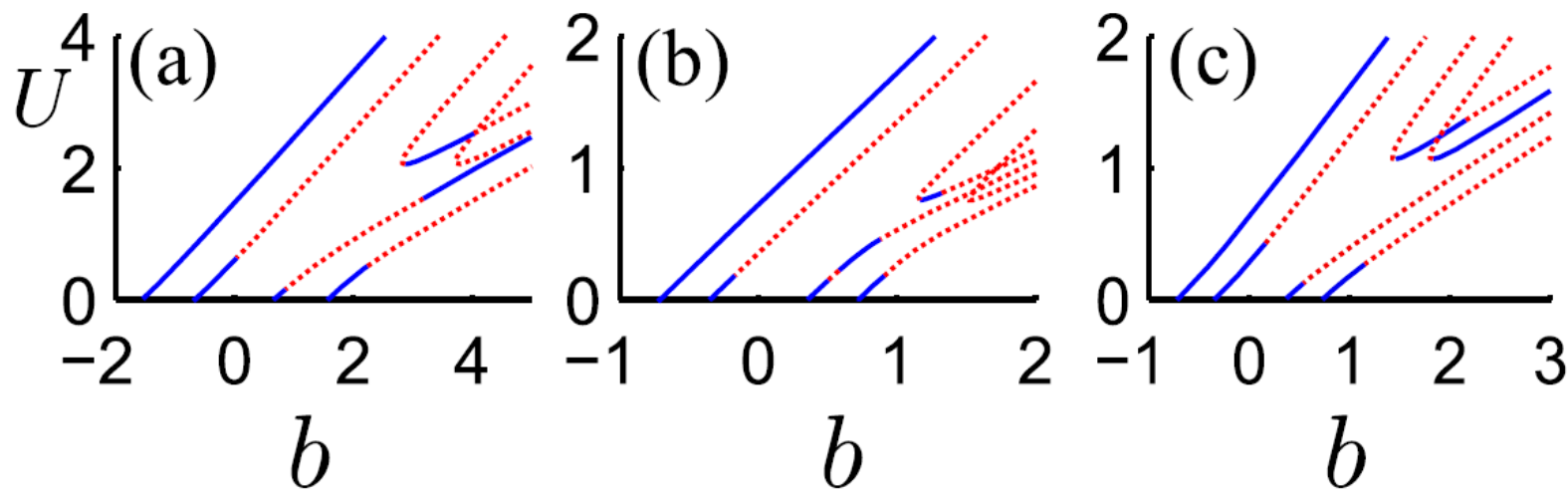}{Fig7}{
Families of nonlinear modes of the Hermitian quadrimer determined by $\mathbf{H}_H$. Parameters $\gamma_{1,2}$, which now affect on the coupling strength, are the same as in panels (a-c) of Fig.~\rpict{Fig3}. Adopted from~\cite{Zezyulin:2012-213906:PRL}.
}
%-------------------------------------------------------------------------------------------------

From the result above, there could be an impression that an existence of continuous families of the nonlinear modes is an attribute of PT-symmetric systems. To check this conjecture, we consider the case $\gamma_1=\gamma_2\equiv\gamma$, but this time take into account not only the nearest-neighbor interactions, according to the following matrix:
%-------------------------------------------------------------------------------------------
\begin{eqnarray}
  \mathbf{K}_0 =\left(
  \begin{array}{c c c c}
    0 & 1 & 0 & 0 \\
    1 & 0 & \cos\beta & -\sin\beta \\
    0 & \cos\beta & \sin2\beta & \cos2\beta \\
    0 & -\sin\beta & \cos2\beta & -\sin2\beta
  \end{array}
  \right),
\end{eqnarray}
%-------------------------------------------------------------------------------------------
where $\beta$ determines a coupling strength.  This structure is schematically shown in Fig.~\rpict{Fig1} in the left top corner. This configuration is symmetric with respect to the operator
%-------------------------------------------------------------------------------------------
\begin{eqnarray}
  \mathrm{P}_0(\beta) =\left(
  \begin{array}{cc}
    \mathbf{0} & \rho(\beta) \\
    \rho(\beta) & \mathbf{0}
  \end{array}
  \right),
\end{eqnarray}
where $\rho(\beta)=\cos\beta\sigma_1+\sin\beta\sigma_3$. Note that $\mathbf{H}_0(0)$ coincides with the case considered above for $\gamma_1=\gamma_2$.

It can be shown that $\mathbf{H}_0(\beta)$ admits continuous families of nonlinear modes only for discrete number of points $\beta_k=\pi k/2$~\cite{Zezyulin:2012-213906:PRL}. Thus, continuous families of nonlinear modes do not always exist for PT-symmetric system.

Interestingly, by using appropriate transformation, e.g.
%------------------------------------------------------------------------------------------------
\begin{eqnarray}
  \mathbf{M} =\left(
  \begin{array}{cc}
    \mu & \mathbf{0} \\
    \mathbf{0} & \mu
  \end{array}
  \right),
\end{eqnarray}
%------------------------------------------------------------------------------------------------
with $\mu=\sigma_3+i\sigma_1$, it is possible to obtain new PT-symmetric matrix $\mathbf{H}_1(\beta)=\mathbf{MH}_0(\beta)\mathbf{M}^{-1}$ with the same set of eigenvalues. Thus, the PT-symmetric quadrimer can be reconfigured to a PT-symmetric plaquette. In Fig.~\rpict{Fig1} the graph representations of both PT-symmetric quadrimer $\mathbf{H}_0(\beta)$ and PT-symmetric plaquette $\mathbf{H}_1(\beta)$ are shown for a generic case of arbitrary $\beta$ and a particular one with $\beta=\pi/4$. Here $k_{\pm}=\sqrt{2}/2\sin(\beta\pm\pi/4)$.
%-------------------------------------------------------------------------------------------------
\pict{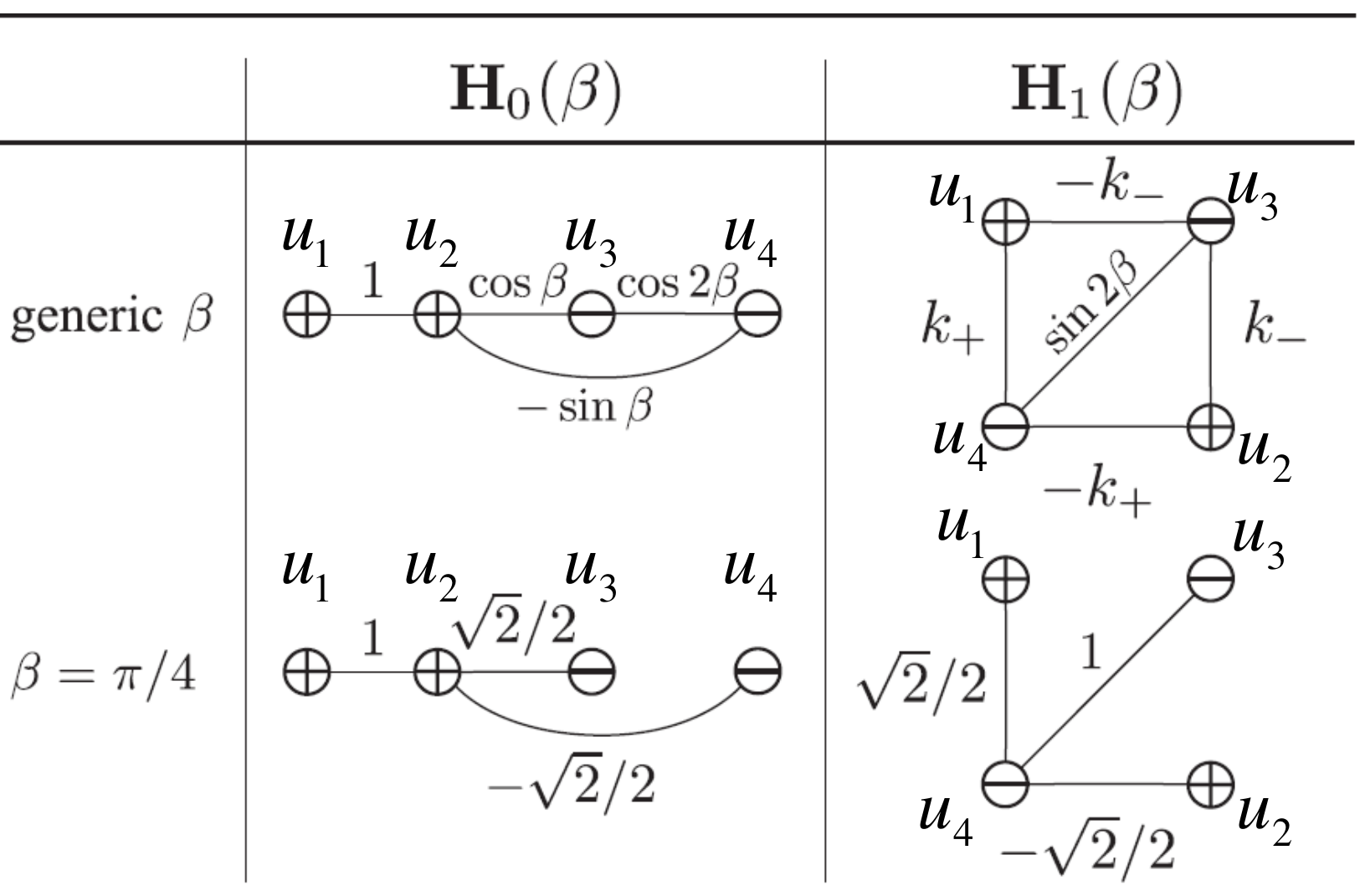}{Fig1}{
PT-symmetric quadrimer $\mathbf{H}_0(\beta)$ and the corresponding plaquette configuration $\mathbf{H}_1(\beta)$ obtained by means of $\mathbf{M}$ transformation of $\mathbf{H}_0(\beta)$. Adopted from~\cite{Zezyulin:2012-213906:PRL}.
}
%-------------------------------------------------------------------------------------------------

The PT-symmetric plaquettes can also be considered as blocks for construction of two-dimensional PT-symmetric lattices. Therefore, it is important to figure out what are the properties of the nonlinear modes in such structures. Four types of fundamental PT-symmetric plaquettes corresponding to different interpositions of active and lossy waveguides were studied in Ref.~\cite{Li:2012-444021:JPA} (one of possible realisation of such a plackets is presented in Fig.~\rpict{Fig5}(b)). Existence of solution branches past the critical point of the underlying linear system and the appearance of additional asymmetric modes were reported, although  most of the stationary modes were found to be unstable.

The plaquette models can describe real physical systems such as PT-symmetric multi-modes fibers~\cite{Li:2013-33812:PRA,Benisty:2015-53825:PRA}, including PT-symmetric coupler with birefringent arms was considered in~\cite{Li:2013-33812:PRA}. The schematic of such a coupler and its graph representation are shown in Fig.~\rpict{Fig5}. The coupling between elements includes both linear and nonlinear nature, which complicates the mode structure. Stationary modes with different polarisation were obtained and regions of their stability were identified.
%------------------------------------------------------------------------------------------------
\pict{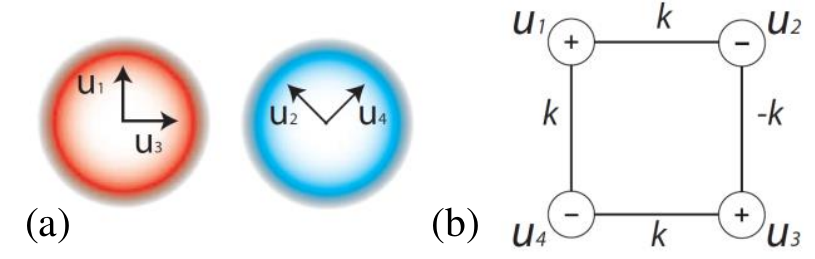}{Fig5}{
(a) PT-symmetric birefringent arm coupler with principal optical axes of the two waveguides are $\pi/4$ rotated with respect to each other, and (b) its graph representation in form of PT-symmetric plaquette. Red ("+") and blue ("-") colors denote active and lossy waveguides, respectively. Adopted from~\cite{Li:2013-33812:PRA}.
}

\subsection{Multicore fibers}

Being driven by rapid development of multicore optical fibres, the theory of coherent propagation and power transfer in low-dimensional array of coupled nonlinear waveguides has been suggested~\cite{Turitsyn:2012-31804:PRA, Rubenchik:2013-4232:OL}. This offers an opportunity to explore the multicore waveguide systems with PT-symmetry in the presence of gain and loss. Existence, stability, and dynamics of both linear and nonlinear stationary modes propagating in radially symmetric multicore waveguides with balanced gain and loss have recently been studied~\cite{Martinez:2015-23822:PRA}. A multicore waveguide array of $N$ identical waveguides arranged in a circular geometry with the central one being gain (loss) and all peripheral ones being loss (gain), as shown in Fig.~\rpict{Fig-sec6-F2}(a)-(b), has been investigated. Taking into account only  nearest-neighbor interaction, the evolution equations for the mode amplitudes read as,
%--------------------------------------------------------------------------------------------------------------
\begin{IEEEeqnarray}{l}
    -i\frac{d A}{d z}=(\epsilon_0+i\gamma_0)A + C_0\sum^{N}_{j=1}B_j +G |A|^2 A,\label{core-eq}\\
    i\frac{d B_j}{d z}=(\epsilon_1+i\gamma_1)B_j + C_1(B_{j+1}+B_{j-1})+C_0 A + G|B_j|^2 B_j,\nonumber
\end{IEEEeqnarray}
%--------------------------------------------------------------------------------------------------------------
where $A$, $\epsilon_0$ and $\gamma_0$ denote the mode amplitude, refractive index, and gain/loss parameter in the central core waveguide, respectively, while $B_j$, $\epsilon_1$ and $\gamma_1$ are the same but in the $j$-th waveguide on the ring. Here $B_0=B_N$ and $B_{N+1}=B_1$ and the nonlinear coefficients for all waveguides are identical and characterized by $G$. $C_0$ is the coupling coefficient between the core and other waveguides on the ring, while $C_1$ stands for the coupling between two nearest-neighbors on the ring.

\pict{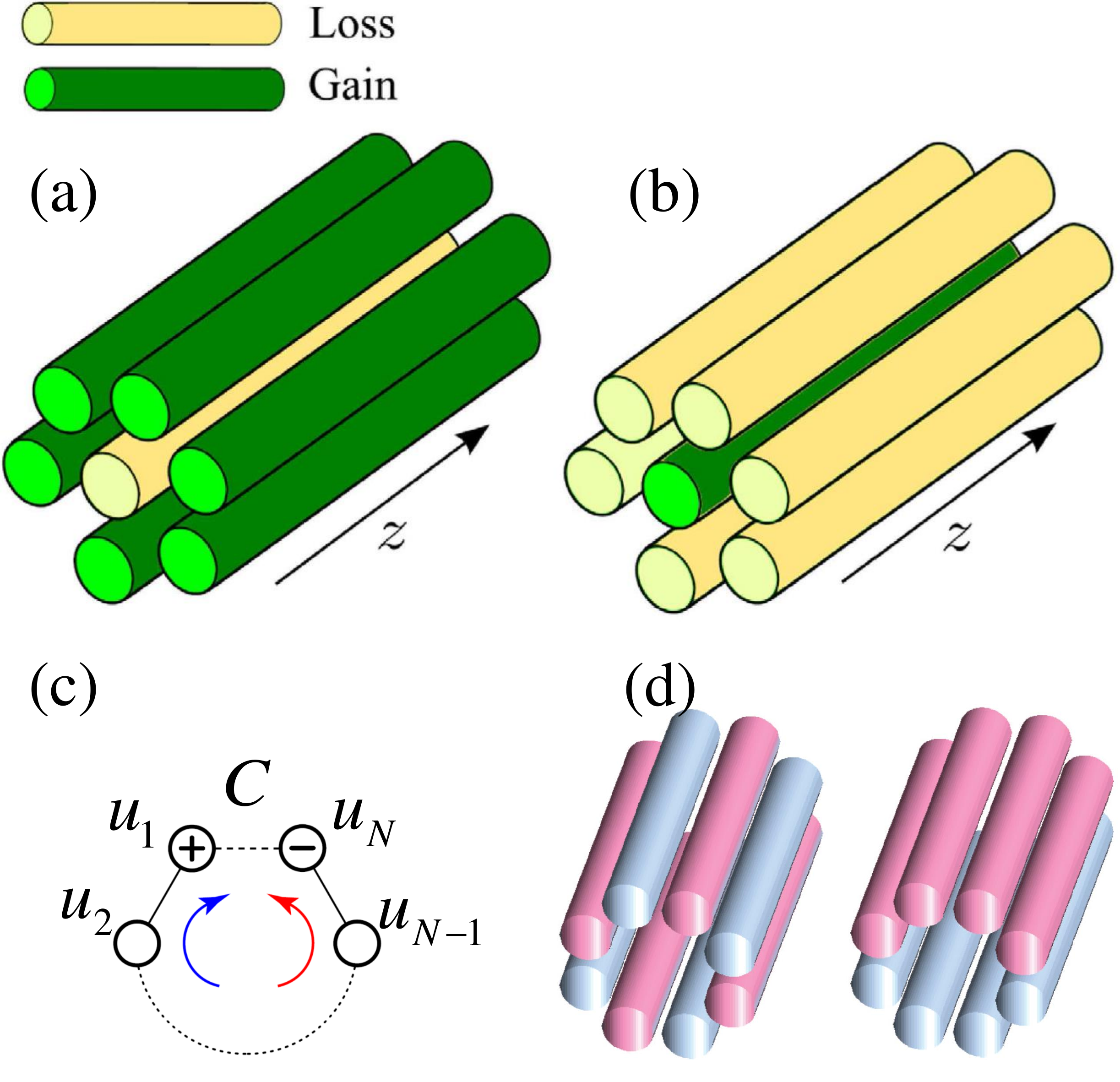}{Fig-sec6-F2}{
Schematic (a),(b) radially symmetric multicore fibers with balanced gain and loss. From Ref.~\cite{Martinez:2015-23822:PRA}, and periodic PT-symmetric configurations: (c) $N$ - site ring with embedded PT-symmetric coupler. Phase circulation direction of vortices with topological charge $m>0$ and $m<0$ is shown by anticlockwise and clockwise arrows, respectively; (d) PT-symmetric necklaces of alternating and clustered gain and loss waveguides. Adopted from~\cite{Leykam:2013-371:OL,Barashenkov:2013-33819:PRA}, respectively.
}

It has been demonstrated that the system can generally be reduced to an effective PT-symmetric dimer with asymmetric coupling. In the linear regime, $G=0$, the bounded dynamics can be observed in the limit of the effective dimer and for the active core waveguide. Here two modes with real propagation constants exist before a critical value of gain and loss for an onset of PT-symmetry breaking, while other modes always have imaginary propagation constants. This directly relates to a stable (unstable) propagation of modes when gain is localized in the core (ring) of the waveguide structure.

In the case of self-focusing nonlinearity ($G>0$), an interplay between nonlinearity, gain and loss results in a high degree of instability, and quasi-stable propagation may appear under some parameter conditions. By reducing to the effective dimer, several stationary modes were found, however these stationary modes turned out to be unstable in the presence of nonlinearity and nonzero gain and loss. More interestingly, when a spatially periodic modulation is applied, stabilization of nonlinear modes can be achieved.

%The considerable interest is also presented by the periodic structures including gain and loss waveguides~\cite{Barashenkov:2013-33819:PRA,Leykam:2013-371:OL}.
%-------------------------------------------------------------------------------------------------------------------------------------------------
%\pict{fig13}{F2}{
%Schematic periodic PT-symmetric configurations. (a) $N$ - site ring with embedded PT-symmetric coupler. Phase circulation direction of vortices with topological charge $m>0$ and $m<0$ is shown by anticlockwise and clockwise arrows, respectively. (b) PT-symmetric necklaces of alternating and clustered gain and loss waveguides.}

Nonzero energy flow between gain and loss waveguides may affect such entities as discrete vortices in the ring geometry including PT-symmetric elements [see Fig.~\rpict{Fig-sec6-F2}(c)]~\cite{Leykam:2013-371:OL}. One of the important characteristics of the vortex modes is a topological charge $m$ which is equal to the phase winding number in units of $2\pi$ around the vortex origin and the sign of $m$ determines the direction of power flow. It has been shown that degenerate linear vortex modes spontaneously break PT symmetry, however nonlinear propagation modes still can be found. By breaking T symmetry, the existence, stability and the dynamics of nonlinear vortex modes become sensitive to the sign of their charge $m$, offering an additional degree of freedom for all-optical control of discrete vortices.

Generalization of the two-channel dispersive coupler~\cite{Barashenkov:2012-53809:PRA} to a PT-symmetric arrangement of $2N$ dispersive waveguides has been done in~\cite{Barashenkov:2013-33819:PRA}. The arrays of waveguides in the form of periodic necklaces with alternating and clustered gain and loss distribution [see Fig.~\rpict{Fig-sec6-F2}(d)] were discussed and asymptotic behaviour of the PT breaking threshold with number of sites $N$ was derived. Thus, for the alternating array $\gamma_{\rm crit}\sim1/N$, while for the clustered geometry $\gamma_{\rm crit}\sim1/N^2$ as $N\rightarrow\infty$.

Note that, the questions of existence, stability and dynamical behaviour of light in a finite chain of alternating waveguides with gain and loss have been addressed in~\cite{Kevrekidis:2013-365201:JPA,Pelinovsky:2014-85204:JPA}. 

{\protect\renewcommand\sectpath{Solitons_couplers_periodic}
%-------------------------------------------
\section{Solitons in couplers and periodic potentials} \lsect{}
%-------------------------------------------
Since the discovery of solitary waves in shallow water by Scott Russell in 1834, solitons have attracted a lot of attention due to their importance in many nonlinear physical systems, ranging from solids to Bose-Einstein condensates~\cite{Filippov:2010:VersatileSoliton}. In optics, solitons were investigated and described in many research papers and books (see, e.g., Ref.~\cite{Kivshar:2003:OpticalSolitons}).
Conservative solitons requiring balance of nonlinear response and medium dispersion usually form families with different amplitudes. Nonlinear dissipative systems, however, demand an additional balance between gain and loss to support soliton solutions~\cite{Akhmediev:2005:DissipativeSolitons,Rosanov:2002:SpatialHysteresis}. This latter requirement is usually satisfied only for particular soliton amplitudes and shapes, thus no continuous families can generally be found. In their turn PT-symmetric systems belong to a subclass of dissipative systems, however due to the symmetry property they can commonly support families of solitons with different amplitudes~\cite{Yang:2014-367:PLA}. Accordingly, PT symmetric systems are unique, combining certain features of conservative and dissipative systems.

In this section,  we discuss how PT-symmetry affects the soliton properties compared
to conservative or dissipative systems.

\subsection{Solitons in PT-symmetric couplers}

First, we consider a planar PT-symmetric coupler where light experiences diffraction in the $x$ direction and propagates along the $z$ coordinate~\cite{Driben:2011-4323:OL,Alexeeva:2012-63837:PRA,Driben:2012-54001:EPL,Barashenkov:2012-53809:PRA}.  Figure~\rpict{Fig4_1CouplerScheme}~(a) show a schematic pretention of the coupler where the red plane stands for a waveguide with gain, while the blue plane shows a lossy waveguide.
%------------------------------------------------------------------------------------------------------------
\pict{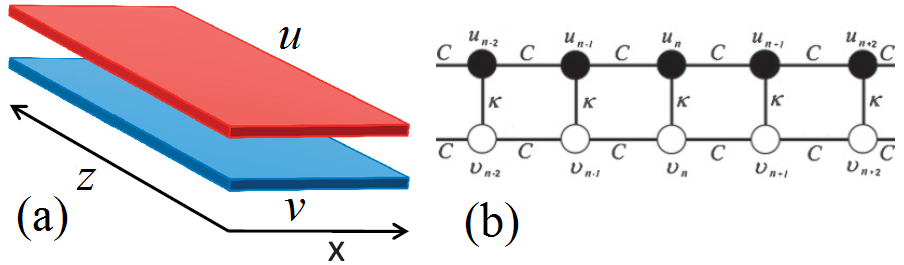}{Fig4_1CouplerScheme}{
  (a) Schematic of a planar PT-symmetric coupler and (b) a cross-section of its discrete analog. Red color indicates gain (black circles in (b)), while blue color indicates loss (white circles, respectively). In (b) couplings between waveguides are marked as $C$ and $\kappa$. Adopted from~\cite{Barashenkov:2012-53809:PRA,Suchkov:2011-46609:PRE}.
}
%------------------------------------------------------------------------------------------------------------

This structure can be described by the normalized system of coupled-mode equations:
\begin{eqnarray}\label{PlanarCoupler}
  i u_z+u_{xx}+|u|^2u-i\gamma u+v=0,\\
  i v_z+v_{xx}+|v|^2v+i\gamma v+u=0,\nonumber
\end{eqnarray}
where $u(z,x)$ and $v(z,x)$ are the mode amplitudes in the waveguides with gain and loss, respectively, $\gamma\ge0$ is the gain/loss coefficient. System~(\ref{PlanarCoupler}) supports two types of soliton solutions: symmetric and antisymmetric~\cite{Driben:2011-4323:OL,Alexeeva:2012-63837:PRA}, termed by analogy with conservative systems (when $\gamma=0$)~\cite{Kivshar:2003:OpticalSolitons}. These two solitons can be presented in the following form:
\begin{IEEEeqnarray}{l}\label{SolitonSolution}
    \left(
    \begin{array}{c}
    u\\
    v
    \end{array}\right)=A {\sech} \left[\frac{A}{\sqrt{2}}\left(x-Vz\right) \right]\e{i(\frac{Vx}{2}-\beta_{\pm} z)}\left(
    \begin{array}{c}
    \e{i\delta_{\pm}}\\
    1
    \end{array}\right).\IEEEeqnarraynumspace
\end{IEEEeqnarray}
Here $A$, $V$, $\beta_{\pm}$, are the soliton amplitude, velocity and propagation constant, respectively, satisfying the relation $\beta_{\pm}=-A^2-\cos{\delta_{\pm}}+V^2/4+A^2/2$ with $\displaystyle\cos\delta_{\pm}=\pm\sqrt{1-\gamma^2}$ and $\sin\delta_{\pm}=-\gamma$. The signs $\pm$ correspond to the symmetric and antisymmetric solitons, respectively. Stability analysis reveals that symmetric solitons are stable if $A^2\le(4/3)\sqrt{1-\gamma^2}$, while antisymmetric solitons are always unstable. However, their lifetimes are exponentially long if amplitudes are small enough, so they could be considered to be stable in practice. Generally speaking, it is easy to show that under assumption $u(z)=v(z) \exp(i\delta_{\pm})$ Eqs.~(\ref{SolitonSolution}) can be reduced to a single conservative nonlinear Schr\"{o}dinger equation. It means that, if the solitons~(\ref{SolitonSolution}) are stable, then their properties do not differ from ones of their conservative counterparts, and full energy of the system is exactly conserved. However, unstable behaviour of such solutions may demonstrate unique features. If the soliton amplitude exceeds the critical value,  two types of unstable dynamics for each type of solitons are possible: the emergence of breathers with oscillating energy or unbounded growth of the mode intensity in the waveguide with gain. In the conservative case, the latter type of dynamics leads to formation of an asymmetric soliton, which concentrates mostly in one waveguide. Since the asymmetric solitons break a balance of gain and loss in PT systems, it leads to the PT-symmetry breaking. Examples of these unstable behaviors are presented in Fig.~\rpict{Fig5_Sol} for a symmetric soliton.
\pict{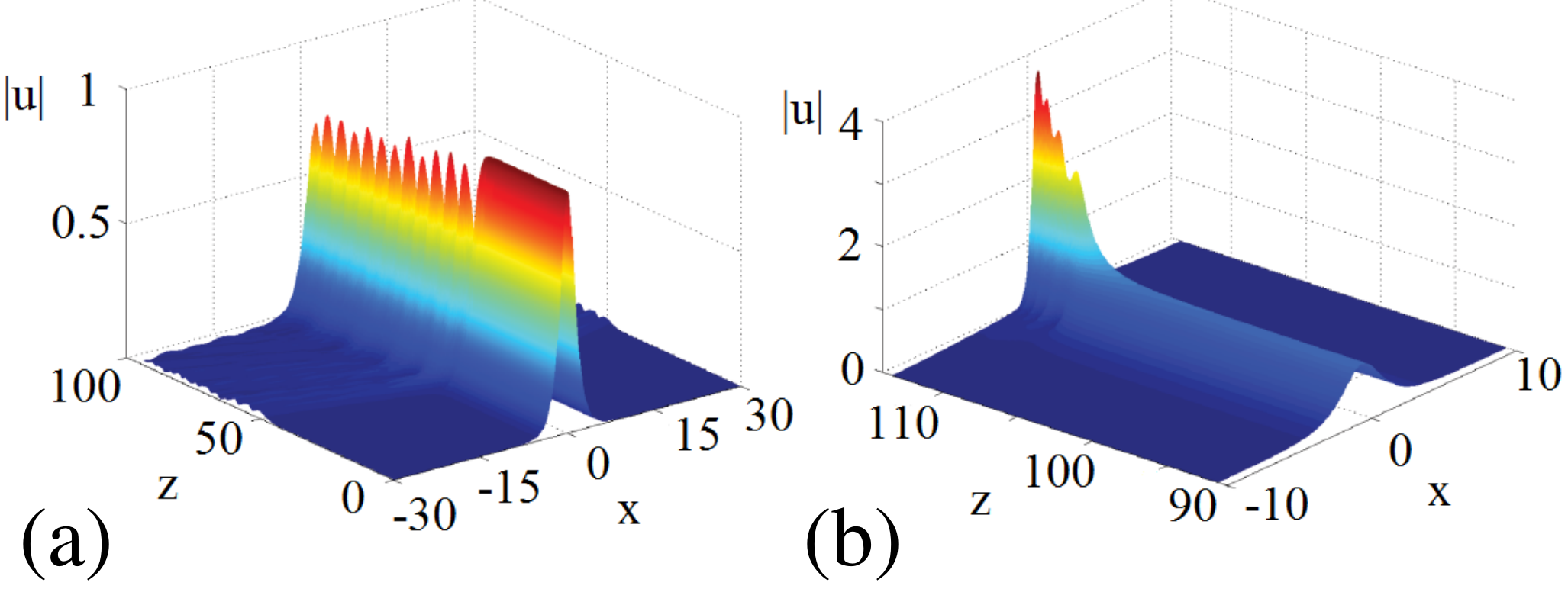}{Fig5_Sol}{
  Unstable dynamics of a symmetric soliton with (a) transformation to a breather, and (b) PT-symmetry breaking (blow up). The ${\displaystyle u}$ component is shown. Adopted from~\cite{Alexeeva:2012-63837:PRA}.
}

The considered system supports not only fundamental solitons, but also stable $N$-soliton complexes as well~\cite{Driben:2012-54001:EPL}. It turns out that unlike the fundamental solitons considered above, the stability region for the antisymmetric two-soliton is considerably larger than for its symmetric counterpart.

As mentioned above, unstable solitons can transform into breather-like objects~\cite{Barashenkov:2012-53809:PRA}. Such nonlinear modes are stable on the distances of $z\leq\varepsilon^{-2}$, where $\varepsilon^{1/2}$ gives the scale of the amplitude of the small-amplitude breathers. The breathers are more common objects in the planar PT-symmetric coupler in the following sense. Propagation of low intensity light (determined by $\varepsilon$) is governed by a conservative system of amplitude equations obtained in~\cite{Barashenkov:2012-53809:PRA}. This system admits soliton solutions, which correspond to breather solutions in original model~(\ref{PlanarCoupler}) as well as degenerate soliton solutions, which are symmetric and antisymmetric solitons of the model~(\ref{PlanarCoupler}). Breathers can be produced resulting from collision of symmetric and antisymmetric solitons [see Fig.~\rpict{Fig5_2new}(a)] as well as at fission of unstable solitons[see~Fig.\rpict{Fig5_Sol}~(a)]. Remarkably, although the breather energy is not conserved during propagation, it is conserved on average. An interesting example of breather dynamics with merging of two breathers into one is shown in Fig.~\rpict{Fig5_2new}(b). Breathers dynamics is very sensitive to a relative phase of interacting  breathers~\cite{Rysaeva:2014-577:JETPL}. Thus, for fixed initial conditions varying only a relative phase of the colliding breathers the different types of  dynamics such as merging two breather into one (temporal or permanent), elastic collision, collision with energy exchange and PT-symmetry breaking are observed.
\pict{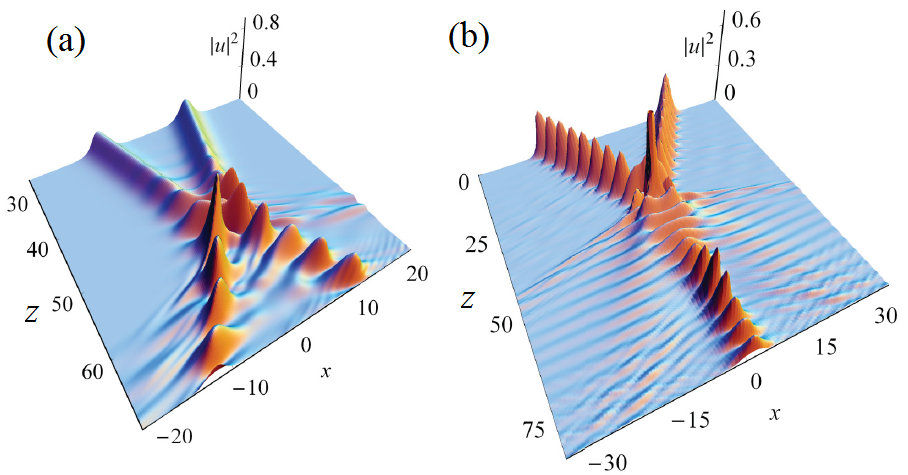}{Fig5_2new}{
  (a) Breather emergence in collision of symmetric and antisymmetric solitons. (b) Fusion of two breathers into one. Only $u$ component is shown. Adopted from~\cite{Barashenkov:2012-53809:PRA}.
}

A nonlinear dissipative coupler can demonstrate unique features due to PT-symmetric behaviour. "Zeno"-like effect was observed in numerical simulation~\cite{Abdullaev:2011-4566:OL} for a system obtained from Eq.~\eqref{PlanarCoupler} by replacing the waveguide with gain by conservative one. The Zeno phenomenon, introduced in quantum mechanics~\cite{Misra:1977-756:JMP} and consisting in strong suppression of the decay of an unstable particle by means of frequent measurements. In terms of considered model more frequent measures correspond to stronger loss in the system.

Generally speaking, the new system is no longer PT-symmetric, however, a linear analogue of such a system can be brought to PT-symmetric view by simple change of variables~\cite{Guo:2009-93902:PRL}. As discussed above, if the gain/loss parameter is below the PT-symmetry breaking threshold there appears beating of light between the waveguides, while in the opposite case this beating is suppressed, and light gets localized in the waveguide with gain. At the same time, the systems with loss only also exhibit similar dynamics. Thus, stronger loss in the system suppresses beating of light, locking it in the conservative waveguide and making the whole system more transparent (see Fig.~\rpict{Fig4Zeno}).

%------------------------------------------------------------------------------------------------------
\pict{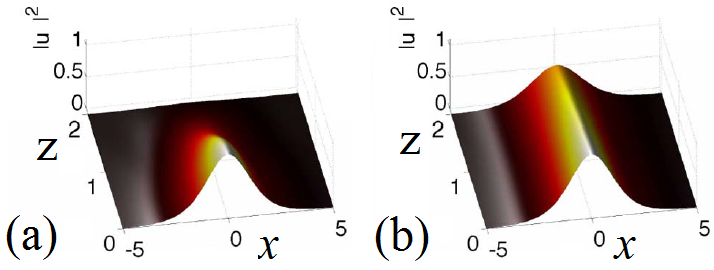}{Fig4Zeno}{
  Propagation of "Zeno" soliton in a coupler composed of lossy and conservative waveguides. Input signal was injected into the conservative waveguide only ($u$ component). (a) Loss coefficient $\gamma=1$, in (b) $\gamma=10$. Adopted from Ref.~\cite{Abdullaev:2011-4566:OL}.
}
%------------------------------------------------------------------------------------------------------

A discrete analogue of the model~(\ref{PlanarCoupler}) has been considered in Ref.~\cite{Suchkov:2011-46609:PRE}. The structure is a chain of linearly coupled PT-symmetric dimers shown schematically in Fig.~\rpict{Fig4_1CouplerScheme}(b). An important distinction of the discrete model from its continuous counterpart is that the former  model possesses discrete linear spectrum providing additional possibility for solitons to be stable. Two types of solitons, symmetric and antisymmetric,  were constructed numerically, and their properties were investigated. It turns out that the discrete PT-symmetric lattice supports freely moving solitons. Solitons behave quite similar to solitons in conservative systems provided they are stable and their amplitude does not exceed a critical value determined as $\displaystyle |A_{\rm crit}|^2=\sqrt{1-\gamma^2}$. If soliton amplitude exceeds $A_{\rm crit}$, the $u$ component of the soliton, which experiences gain, grows exponentially, while the component with loss, $v$, quickly decays.

The discussed model~(\ref{PlanarCoupler}) as well its discrete counterpart do not admit dark soliton solutions [see for example~\cite{Suchkov:2011-46609:PRE}]. However nonlinear coupling between waveguides (cross-phase modulation) can facilitate the formation of dark and bright solitons. Also stable rogue waves of the Peregrine type were identified~\cite{Bludov:2013-13816:PRA,Bludov:2013-64010:JOPT}. It was shown that gain/loss parameter modifies the region of stability of both background of dark solitons as well as solitons themselves.

\subsection{Solitons in localized potentials}

As was mentioned above, the necessary condition for one-dimensional optical system to be PT symmetric is $U(x)=U^*(-x)$~\cite{Musslimani:2008-30402:PRL}. The presence of nonlinear response of a medium, generally speaking, changes the real part of system potential making eigenvalues of the Hamiltonian complex-valued. However, judicious distribution of an intensity in such systems may retain eigenspectrum to be real.

Light propagation through the waveguide with a modulated refractive index in transverse direction ($x$) and experiencing gain and loss is described by the normalized equation
%-------------------------------------------------------------------------------------------------------
\begin{IEEEeqnarray}{l} \label{LocalizedPotential}
  i\psi_z+\psi_{xx}+U(x)\psi+\sigma|\psi|^2\psi=0,
\end{IEEEeqnarray}
%-------------------------------------------------------------------------------------------------------
where $U(x)=V(x)+iW(x)$ satisfies the condition above for PT-symmetric systems. In this section we consider the case, when both $V(x)$ and $W(x)$ are localized in transverse direction $x$.

Stable solitons in PT-symmetric systems with distributed gain and loss were first reported in Ref.~\cite{Musslimani:2008-30402:PRL}. Particularly, the potential of Scarff form  $V=V_0\sech^2(x)$, $W(x)=W_0\sech(x)\tanh(x)$ has been considered for focusing type of nonlinearity [see Fig.~\rpict{Fig4_4Soliton}~(a,b)]. It has been revealed that the system possesses real linear spectrum if $W_0\leq V_0+0.25$. However, even if this condition is broken (linear spectrum is complex), nonlinear modes can alter the amplitude of refractive index due to nonlinearity, thus bringing the system into PT-symmetric phase. Some analytical solutions have been obtained in~\cite{Musslimani:2008-244019:JPA} for 1D and 2D PT-symmetric wells and lattices.
\pict{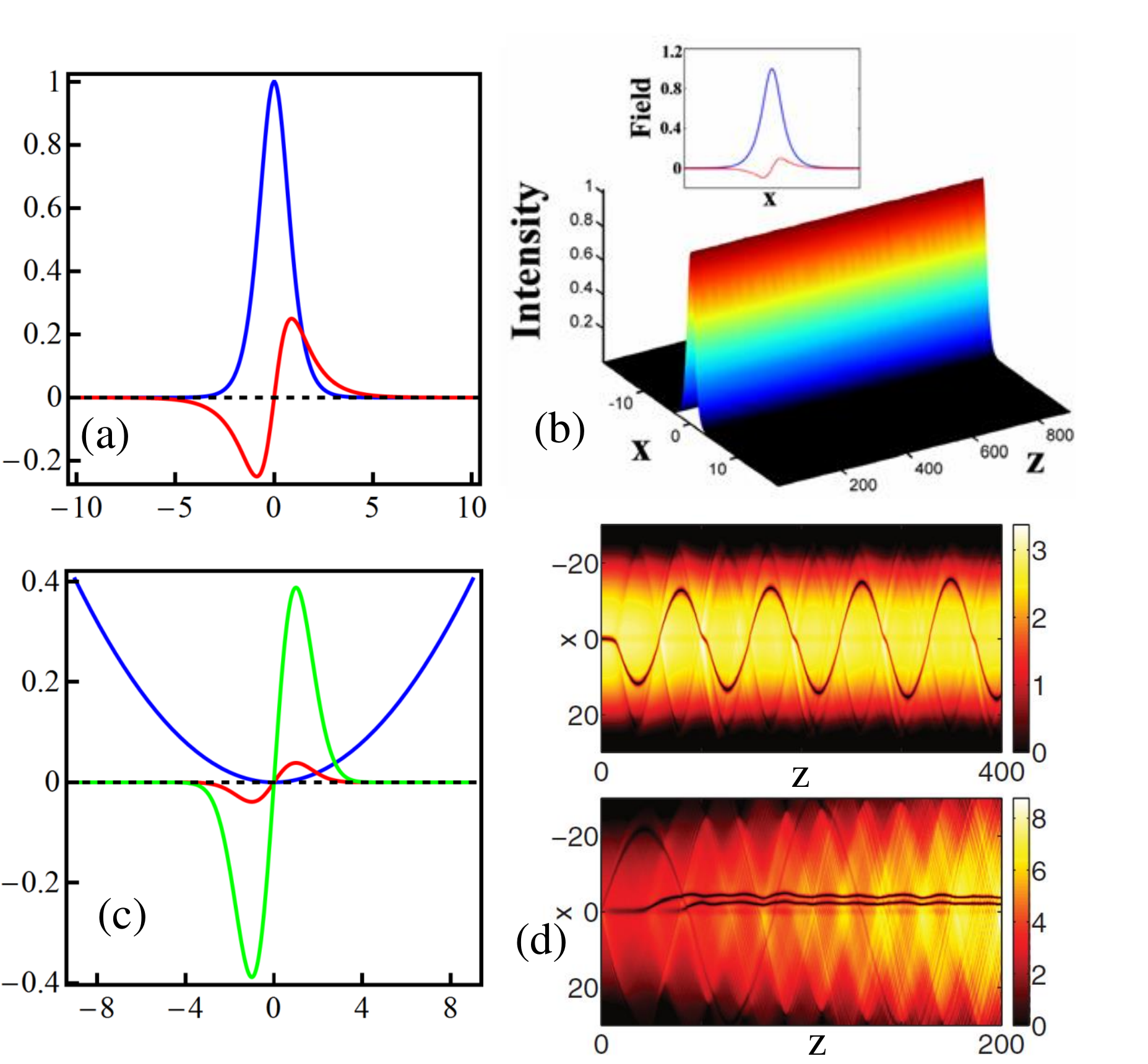}{Fig4_4Soliton}{
  Solitons in localized PT-symmetric potentials. (a) The Scarff type potential: blue indicates a real part of the potential, red  - its imaginary part. (b) Soliton profile and its intensity distribution (adopted from~\cite{Musslimani:2008-30402:PRL}). (c) Parabolic trap potential considered in Ref.~\cite{Achilleos:2012-13808:PRA} with $\Omega=0.1$ (blue) for two cases of gain/loss strength: $\varepsilon=\varepsilon_{\rm crit}^{(1)}=0.064$ (red) and $\varepsilon=0.64$ (green). (d) Dark-soliton dynamics corresponding to $\varepsilon=0.064$ (top) and $\varepsilon=0.64$ (bottom). Adopted from~\cite{Achilleos:2012-13808:PRA}.
}
Although the soliton in Fig.~\rpict{Fig4_4Soliton}(b) looks similar to ones determined by Eq.~\eqref{SolitonSolution} and depicted in Fig.~\rpict{Fig5_Sol}, in fact they are completely different. Each of the components, $u$ and $v$, of the soliton~\eqref{SolitonSolution} experiences pure gain or loss respectively, and the energy flow is directed from the gain waveguide to the lossy one. The soliton presented in Fig.~\rpict{Fig4_4Soliton} has only one component, which is partially situated in both gain and loss regions. This soliton has the transverse energy flow, which strongly affects its properties, e.g. stability and mobility.

In the case of defocusing nonlinearity~\cite{Li:2011-3290:OL,Shi:2011-53855:PRA,Shi:2012-64006:EPL}, gray solitons, nonlinear guided modes as well as bright solitons in nonlocal nonlinear media, have been found to be stable in PT-symmetric localized potential of the Scarff type.

In Ref.~\cite{Hu:2011-43818:PRA} stable fundamental soliton, as well as two- and three-soliton solutions, have been reported for the Gaussian PT-symmetric potential of the form $U(x)=(1+iW_0x)\exp (-x^2)$.

An interesting study of the behaviour of dark solitons near the phase transition points was reported in Ref.~\cite{Achilleos:2012-13808:PRA}. A parabolic trap potential of the form $V(x)=1/2\Omega^2x^2$ and $W(x)=\varepsilon x \exp(-x^2/2)$ was considered [see Fig.~\rpict{Fig4_4Soliton}(c)] and it has been found that dark solitons are stable if the amplitude of the imaginary potential, $\varepsilon$, is less than $\varepsilon_{\rm crit}^{(1)}=\sqrt{5/12}\Omega$. An unstable soliton of an oscillating type is shown in Fig.~\rpict{Fig4_4Soliton}(d)(top panel) for $\varepsilon=\varepsilon_{\rm crit}^{(1)}$ and $\Omega=0.1$. Above this critical value, the dark-soliton background demonstrates exponential growth corresponding to the blow-up due to PT-symmetry breaking. However, if $\varepsilon$ exceeds some critical value, namely $\varepsilon_{\rm crit}^{(2)}=0.62$ in the considered case, then instead of the exponential growth one observes the soliton generation from a background state [see Fig.~\rpict{Fig4_4Soliton}(d), bottom panel]. This critical point $\varepsilon_{\rm crit}^{(2)}$ corresponds to blue-sky bifurcation of the background state and the fundamental soliton branch, after which they cease to exist. The same type of bifurcation is also observed for soliton branches of higher orders, which are found for larger values of $\varepsilon$, i.e. 2-soliton branch bifurcates with 3-soliton branch at some $\varepsilon_{\rm crit}^{(3)}>\varepsilon_{\rm crit}^{(2)}$ etc. Thus, for larger values of $\varepsilon$ (strong gain/loss) we may observe spontaneous generation of multiple dark solitons from an input beam.

\subsection{Solitons in periodic potentials}

Next, we consider soliton solutions in structures having periodic modulation of both a real part of the refractive index as well as gain/loss regions. The governing equation for light propagation has the form of  Eq.~\eqref{LocalizedPotential}, but potential $U(x)$ is periodic.
\pict{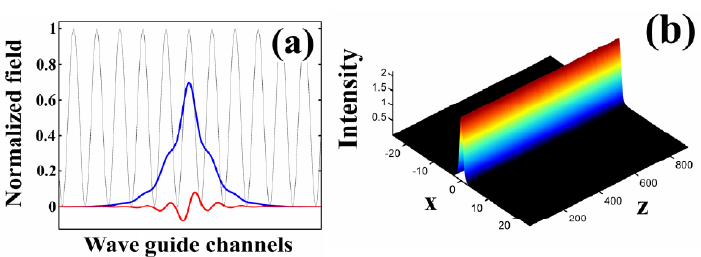}{Fig15_3}{
    (a) Soliton profile (blue curve - real part, red curve - imaginary part) and potential (refractive index) distribution. (b) Stable soliton propagation in a PT-symmetric periodic potential. Adopted from~\cite{Musslimani:2008-30402:PRL}.
}
Linear properties of such a system have been investigated in Ref.~\cite{Makris:2008-103904:PRL}. It has been shown that nonorthogonality of thed Floquet-Bloch modes for PT-symmetric potentials leads to effects such as double refraction, power oscillation (due to unfolding of nonorthogonalized Floquet-Bloch modes), nonreciprocity, and secondary emission in 2D case. These linear attributes of PT-symmetric system may also be observed in nonlinear models and are fundamental phenomenons that distinguish PT-symmetric systems from conservative ones.

In the case of self-focusing nonlinearity, $\sigma=1$, soliton solutions have been investigated for the potential of the form $\displaystyle V(x)=\cos^2(x)$, $W(x)=W_0\sin{(2x)}$~\cite{Musslimani:2008-30402:PRL}. It is interesting that nonlinear response in such systems can shift PT-symmetry breaking threshold enabling one to find nonlinear mode with real propagation constant even though the associated linear spectrum is complex. It turns out that the solitons can be stable over a wide range of model parameters. Furthermore, narrower self-trapped waves are more stable and it is possible to find stationary modes even above PT-breaking threshold, but this class of solitons unfortunately is unstable. Potential distribution as well as stable soliton propagation are shown in Fig.~\rpict{Fig15_3}.

A system of superlattices with $V(x)=\varepsilon\sin^2[x+\pi/2]+(1-\varepsilon)\sin^2[2(x+\pi/2)]$ and $W(x)=W_0\sin(2x)$  has been studied in~\cite{Zhu:2011-2680:OL}. Parameter $\varepsilon$ here controls the relative strength of the superlattices. Stable solitons propagating in the low power region were found in semi-infinite gap. The relative strength of the superlattices may significantly vary PT breaking threshold as well as region of soliton stability.

For the defocusing nonlinearity, $\sigma=-1$, multi-peak gap solitons exist and are stable over a wide range of parameters in a first gap of the potential of the form $V_0\sin^2(x)+W_0\sin(2x)$ ~\cite{Li:2012-4543:OL}. It is remarkable that stable solitons with larger number of peaks exist for a wider range of propagation constant.
\pict{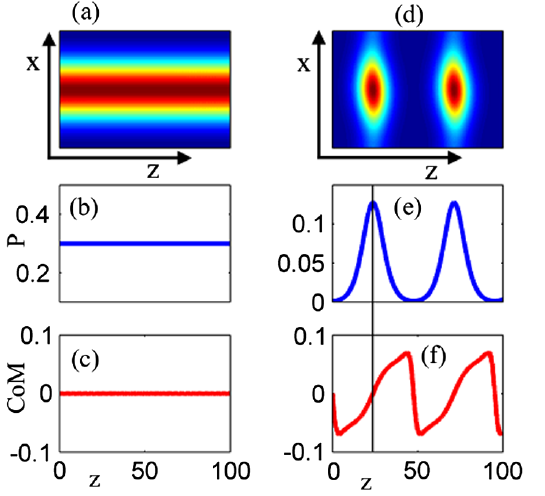}{Fig4_1}{
  Nonlinearity induced transition from PT broken to full PT phase.  (a-c) Invariant propagation of signal, the initial intensity of which is high enough to transit the system to PT symmetry phase. (a) Packet intensity, (b) power, (c) center of mass. (d-f) The same as in (a-c) but for the packet with insufficient initial intensity for transition. Adopted from~\cite{Lumer:2013-263901:PRL}.
}

Interesting finding has been obtained in Ref.~\cite{Lumer:2013-263901:PRL}. It turns out that nonlinearity may cause a transition from broken to PT symmetric phase for focusing nonlinearity and vice versa for defocusing one. Nonlinear band structure has been obtained for different strength of nonlinearity and it has been revealed, that the greater the nonlinearity strength, the smaller region in the Brilliouin zone where the eigenenergies are complex. When power reaches a certain threshold, the whole spectrum becomes real. Thus, two different dynamical regimes for nonlinear wave packet near the phase transition point were observed. First, if an initial power of a propagating packet is large enough for a transition of the system to PT-symmetric phase, then intensity and power are invariant throughout propagation. In Figs.~\rpict{Fig4_1}~(a-c) an intensity, power and center of mass of the propagating packet are presented for this case. Second, if the initial power of the packet is too low to induce transition, then the intensity and power start to grow [see Fig.~\rpict{Fig4_1}~(d-f)]. After power passes the level at which spectrum becomes real it continues to grow until it reaches some maximum value (indicated by black line in Figs.~\rpict{Fig4_1}~(e,f)) and after that it starts to decay bringing the system to the original state. This dynamics can be explained as follows: the evolution starts an the center of unit cell (we consider here only one period of the potential. Since the initial power is low (linear regime) we can decompose the packet into two eigenmodes, one of which is growing and the second one is decaying. The growing eigenmode resides mostly in the gain half of the unit cell, while decaying - in the loss half. After the decaying mode dies out, the center of mass of the packet is shifted along the potential minima to the gain region. Then it starts to move in opposite direction along with the power growth and ends up in the loss region. The power of the packet starts to decay, bringing the system to PT broken phase and the process repeats again. Both of these dynamical regimes are stable and numerical simulations indicate that such propagation should continue indefinitely. The second type of dynamics is associated with any continuous physical systems and is not observed in systems modeled by a standard tight-binding model, since for nonlinearity induced transition the space overlap of modes corresponding to gain and loss should be presented. This effect allows to design systems that can switch from being power conservative to amplification regime.

Behaviour of wavepackets near the phase transition point has been studied in~\cite{Nixon:2012-4874:OL}. In this case the envelope dynamics of the propagating packets can be described by nonlinear Klein-Gordon equation (Eq.~(13) in~\cite{Nixon:2012-4874:OL}). Based on this envelope equation a variety of phenomena such as wave blow up, periodic bound states and solitary waves are predicted and corroborated by comparison with numeric simulation of full model~(\ref{LocalizedPotential}).

A discrete counterpart of the PT-symmetric periodic potential can be realized by appropriate alteration of coupled waveguides with gain and loss. Such discrete structures are easier to fabrication then the waveguides with continuously varying refractive index as well as gain/loss strength. Solitons in discrete PT-symmetric lattices have been investigated in a series of works~\cite{Dmitriev:2010-2976:OL, Konotop:2012-56006:EPL, Kevrekidis:2013-365201:JPA}. A chain of alternating PT-symmetric couplers was considered in~\cite{Dmitriev:2010-2976:OL}. The field evolution along the waveguides is described by the system
\begin{eqnarray}\label{CouplerChain}
  i\frac{du_j}{dz}+i\gamma u_j+C_1v_j+C_2v_{j-1}+|u_j|^2u_j=0,\nonumber\\
  i\frac{dv_j}{dz}-i\gamma v_j+C_1u_j+C_2v_{j+1}+|v_j|^2v_j=0.
\end{eqnarray}
Here $u_j$ and $v_j$ are the mode amplitudes in waveguides with loss and gain, respectively, $C_1$ and $C_2$ - coupling coefficients. In the linear regime the system has real eigenspectrum if $||C_1|-|C_2||\ge|\gamma|$. It has been shown numerically that this lattice supports stable soliton solutions. These solitons might be actively reshaped or switched to PT-broken phase by a small-amplitude modulation of the gain/loss parameter $\gamma$ along the propagation direction $z$. Later, in the work~\cite{Konotop:2012-56006:EPL} an analytical proof of existence of such solitons was given and classification of solution families and their stability was provided. Moreover, by analytical continuation from the anticontinuum limit it is possible to show that localized discrete solitons exist in a network of PT-symmetric clusters (oligomers).

\subsection{Generalized nonlinear potentials}

In this section we consider PT-symmetric systems with generalized nonlinear response such as nonlocal nonlinearity, cubic-quintic competing nonlinearity, and mixed linear-nonlinear PT-symmetric potentials etc. The field evolution in one-dimensional systems with a general type of nonlinearity can be described by the following normalized equation
%-----------------------------------------------------------------------------------------------------------------
\begin{eqnarray}\label{NonlocNonlin}
    i\frac{\partial \psi}{\partial z}+\frac{1}{2}\frac{\partial \psi}{\partial x^2}+\left[V(x)+iW(x) \right]\psi+G(|\psi|^2,x)\psi=0,
\end{eqnarray}
%-----------------------------------------------------------------------------------------------------------------
where $G(|\psi|^2,x)$ is a functional determining the nature of nonlinear response.

Nonlocal nonlinear response  appears in such systems as photorefractive crystals, nematic liquid crystals, lead glasses, etc. This nonlinear response can significantly  modify the soliton properties such as existence, stability, and mobility. The function $G$ satisfies the following equation ${\displaystyle G-d(\partial^2 G)/(\partial x^2)=|\psi|^2}$. Here $d$ is a degree of nonlocality of the nonlinear response. Several papers were devoted to the study of gap solitons and defect modes in self-focusing nonlocal nonlinear media.  In Ref.~\cite{Hu:2012-43826:PRA}, the existence and stability of defect modes were studied for different values of the nonlocality parameter $d$, as well as for different types of defects. In Ref.~\cite{Li:2012-23840:PRA}, gap solitons were studied in dual periodic lattices, while Ref.~\cite{Hu:2012-14006:EPL} was devoted to the study of defect modes in PT-symmetric superlattices. In Ref.~\cite{Yin:2012-19355:OE} the effect of spatially modulated nonlocal nonlinearity has been addressed. It was shown that there exist regions where gap solitons and defect modes are stable, and that the degree of nonlocality affects this region drastically.

Defocusing type of nonlinearity has been considered in ~\cite{Jisha:2014-13812:PRA}. The existence and stability of gap solitons in media with diffusive-like nonlinear response were determined. The gap solitons are always oscillatory unstable for any degree of nonlocality in the presence of an imaginary potential.

It has also been shown that PT-symmetric periodic lattices can support a variety of stable vector solitons consisting of multiple bright beams in both focusing and defocusing media~\cite{Kartashov:2013-2600:OL}. Dynamics of such solitons is described by Eq.~(\ref{LocalizedPotential}) where $\mathbf\psi=(\psi_1;\psi_2)$ is a two component vector. In the component form Eq.~(\ref{LocalizedPotential}) becomes a system of two incoherently coupled-mode equations. Stable solitons can be found even if both components have different propagation constants, but to satisfy the PT-symmetry condition these solitons must have equal number of humps in each component. It also may happen, that one of the component is unstable itself, while the other is stable. In this case the soliton can be stable due to the coupling of these two components. Analysis of an effect of gain/loss strength on soliton stability revealed that stronger gain/loss coefficient can destabilize soliton in focusing media and stabilize it in defocusing one.

The case of spatial periodic modulation with Kerr-type nonlinearity also including gain/loss modulation has been considered in Ref.~\cite{Abdullaev:2011-41805:PRA}. In this case  $G(|\psi|^2, x)=V[(x)+iW(x)]|\psi|^2$ with zero linear potential and the functions $V(x)$ and $W(x)$ have the same modulation period. Analysis of the soliton existence and stability reveals that in the case of zero real part of the nonlinear potential there exists a non-vanishing region of parameters where stable solitons exist. The modes with the width less then a half of the lattice spacing appear to be stable. The system supports stable multihump solitons. Stable solitons were found in the lattice with mixed linear-nonlinear PT-symmetric optical potentials~\cite{He:2012-13831:PRA}. Some exact solutions to the system with mixed linear-nonlinear PT-symmetric potentials were obtained in Ref.~\cite{Abdullaev:2010-56606:PRE}.

An effect of multistability has been observed in a model with competing cubic-quintic nonlinear response with $\displaystyle G=(|\psi|^2-|\psi|^4)$~\cite{Liu:2012-1934:OC,Li:2012-16823:OE}. At this type of nonlinear response for low intensities of light the focusing nonlinearity prevails on defocusing one, while for high intensities defocusing dominates. It has been shown~\cite{Li:2012-16823:OE} that solitons with different symmetries and different number of humps can propagate with the same propagation constant. Another interesting point is that both even and odd hump solitons can be stable over some range of propagational constants, which is in contrast to conservative case, where only lowest branches of solitons of both types can be stable. Thus imaginary part of the potential can stabilize solitons in the first gap.

%-----------------------------------------------------------------------------------------------------------------
\pict{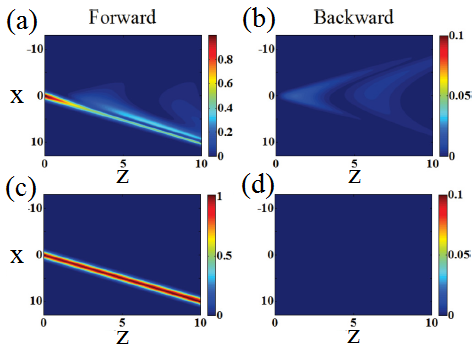}{Fig4_Bragg}{
  Bragg solitons in PT-symmetric potential. (a,b) Forward and backward waves below PT-symmetry breaking threshold, (c,d) the same but at the PT-symmetry transition point. Initially only forward wave was excited. Adopted from~\cite{Miri:2012-33801:PRA}. 
}
%------------------------------------------------------------------------------------------------------------------

In the work~\cite{Miri:2012-33801:PRA} the effect of PT-symmetric potential on Bragg solitons has been studied. These solitons are composed of interlocked both forward and backward propagation modes. Such interlocking open up forbidden band gap allowing energy transport. It is interesting that at the exact PT phase transition point the exchange of energy between forward and backward modes ceases, and full energy is conserved. Below PT breaking threshold, there is oscillatory power exchange between forward and backward modes [see Fig.~\rpict{Fig4_Bragg}].

Interesting observation has been made in~\cite{Yang:2014-367:PLA}. It has been shown analytically and corroborated numerically that for 1D nonlinear Schr\"{o}dinger equation PT-symmetry is a necessary condition for existence of a continuous family of solitons bifurcating out from real linear spectrum. In contrast for other dissipative systems based on non-PT symmetric potentials with all-real spectrum only isolated solitons can exist. From the other hand, the nonlinear Schr\"{o}dinger equation with a complex PT potential cannot admit a continuous family of non-PT-symmetric solitary waves~\cite{Yang:2014-332:STAM}.

\subsection{Two-dimensional potentials}

Here we describe the effect of PT-symmetric potentials on nonlinear modes in 2D case. Due to an additional degree of freedom there appear new types of nonlinear modes such as spatiotemporal solitons and vortex solitons.

Propagation of light in 2D nonlinear PT-symmetric systems with Kerr-type nonlinear response can be described by the equation similar to Eq.~(\ref{LocalizedPotential})
\begin{eqnarray}
  i\psi_z+\nabla^2\psi+U(x,y)\psi+\sigma|\psi|^2\psi=0,
  \label{2DPotential}
\end{eqnarray}
with two dimensional PT-symmetric potential $U(x,y)=V(x,y)+iW(x,y)$.

One of the first examples of 2D solitons in a PT-symmetric periodic potential has been provided in ~\cite{Musslimani:2008-30402:PRL}. Later, in Ref.~\cite{Zhu:2013-2723:OL} the multipeak gap solitons in a PT-symmetric periodic potential were considered. It was shown that the system~(\ref{2DPotential}) with defocusing Kerr-type nonlinearity and PT-symmetric potential of the form $V(x,y)=V_0\{\cos^2(x)+\cos^2(y)+iW_0[\sin(2x)+\sin(2y)]\}$ admits stable multipeak gap solitons. Remarkably, that not only even peak soliton might be stable, but odd-peak solitons as well.

The system~(\ref{2DPotential}) possesses distinct dynamics near the PT phase transition point. For the potential $U(x,y)=V_0^2[\cos(2x)+\cos(2y)+iW_0(\sin(2x)+\sin(2y))]$ it has been shown, that wave packets experience pyramid like diffraction near the transition point $W_0=1$~\cite{Nixon:2013-1933:OL}. Near this point diffraction surface intersects itself four-fold and wavepacket dynamics can be described by fourth-order envelope equation [see Eq.~(11) in Ref.~\cite{Nixon:2013-1933:OL}]. It was found that below certain threshold (which is equal to 3.2 in the considered case) a wave packet exhibits discrete (pyramid type) diffraction as in the linear case. Above this threshold an exponential growth is observed. It is quite surprising that the wave packet dynamics is insensitive to the type of nonlinearity (compared to one-dimensional case considered in Ref.~\cite{Nixon:2012-4874:OL}) as follows from the middle and right columns of Fig.~\rpict{Fig4_2}. In Fig.~\rpict{Fig4_2}, nonlinear dynamics for different initial amplitudes $A$ and two types of nonlinearity is presented. Upper row is obtained from simulations of the envelope equation, lower row from full model simulations.
\pict{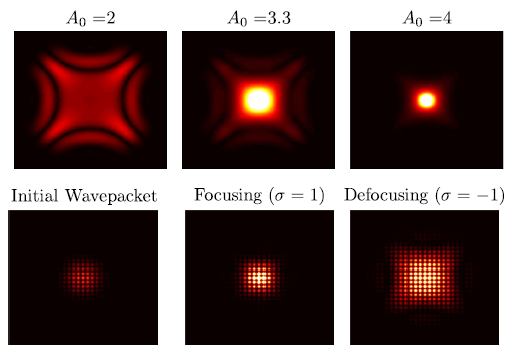}{Fig4_2}{
    Pyramid like diffraction in 2D PT-symmetric potential near the transition point $W_0=1$. Left column - below transition point, middle and right above. Upper row corresponds simulation of the envelop equation, lower row  - full model. Adopted from~\cite{Nixon:2013-1933:OL}.
}

The PT-symmetry potentials can be taken as a base for systems, which, generally speaking, are not PT invariant. In the work~\cite{Jovic:2012-4455:OL} the Anderson localization of light has been studied in the system~\eqref{2DPotential} with PT symmetric potential, and for introduced disordered potential depth $V_0$ along the waveguide. Although there are exponentially growing (in $z$ direction) modes in the system, for low level of disorder as well as for short propagation distance it is possible to observe an effect of PT-potential on transverse localization of light. It turns out that PT-symmetric potential enhances the transverse localization of light in comparison with conservative systems as well as systems with just gain or loss.

One promising observation that could find some applications was suggested in Ref.~\cite{Yang:2014-1133:OL}. A partially PT-symmetric potential, satisfying the following property $V^*(x,y)=V(-x,y)$, can possesses all-real spectra and support continuous families of stable solitons as in case of full PT-symmetry. Remarkably, this soliton family may exhibit multiple power branches with upper branches more stable then the lower ones. These results expand the concept of PT symmetry in multidimensions.

By analogy to the 1D discrete potential considered in~\cite{Suchkov:2011-46609:PRE}, a $2$D discrete nonlinear PT-symmetric lattice has been considered in~\cite{Chen:2014-29679:OE}. An array of coupled PT-symmetric dimers can be realized in form of a set of dual-core waveguides embedded into a photonic crystal. This lattice supports stable symmetric and antisymmetric fundamental and vortical discrete solitons. A symmetric fundamental soliton represents the ground state of the system, and this soliton can be stable at lower values of the total power, while antisymmetric fundamental solitons as well as on-site-centered vortices tend to be stable at higher powers.

{\protect\renewcommand\sectpath{Scattering}
%-------------------------------------------
\section{Scattering on PT-symmetric potentials} \lsect{}
%-------------------------------------------

Various optical systems can be probed in terms of light scattering, which delivers important information about their
spectral properties and reveals the consequences of their symmetries. PT potentials enable new regimes of linear scattering, which cannot be achieved in conservative structures, and furthermore facilitate efficient light switching including nonreciprocal response in the nonlinear regime.

%~\cite{Schomerus:2013-20120194:PTRSA, Lin:2011-213901:PRL, Mostafazadeh:2013-12103:PRA}.

\subsection{Linear scattering and invisibility}\lsect{linear}

Combination of active and lossy optical elements can be designed
in a special way to achieve counterintuitive effects including absence of reflections and invisibility~\cite{Lin:2011-213901:PRL, Mostafazadeh:2013-12103:PRA, Feng:2013-108:NMAT}.
The scattering properties of a one-dimensional system can be characterized by the matrix
\begin{equation}\label{Scattering01}
\mathbf{S} = \begin{bmatrix}
       T   & R^r           \\[0.3em]
       R^l & T
     \end{bmatrix},
\end{equation}
where $T$ and $R^{r,l}$ respectively are the (complex)
transmission and right/left reflection amplitudes, so that the
transmission and right/left reflection coefficients are given by
$|T|^2$ and $|R^{r,l}|^2$. The scattering matrix is formulated taking into account equality $T^l=T^r = T$, which reflects the {\em reciprocity of transmission} for any linear, stationary, and non-magnetic medium.

The scatterer is called {\em reflectionless from the left
(right)}, if $R^l=0$ and $R^r \neq 0$ ($R^r=0$ and $R^l \neq 0$).
The scatterer is called {\em invisible from the left (right)}, if
it is reflectionless from the left (right) and in addition $T=1$.
A thorough assessment of the role of {PT}-symmetry in
the phenomenon of invisibility of scattering potentials in one
dimension has been done by Mostafazadeh
\cite{Mostafazadeh:2013-12103:PRA}. PT-symmetric
nature of invisibility is established in this work by the
following {\em invisibility theorem.} Consider a general real or complex
one-dimensional scattering potential $v$ that is defined on $\mathbb{R}$. Let
$v^{\rm PT}$ be the {PT} transform of $v$ that
is given by $v^{\rm PT}(z):=v(-z)^*$, and $k_{\star}$ be a
positive real number. Then for an incident wave with wavenumber $k$, the following equivalent statements
hold:

(i) $v$ is invisible from the left (or right) for $k=k_{\star}$ if
and only if so is $v^{\rm PT}$.

(ii) $v$ is invisible from the left (right) for $k=k_{\star}$ if
and only if $v^*$ is invisible from the right (left) for
$k=k_{\star}$.

Invisibility does not require {PT}-symmetry of $v$ and
for non-{PT}-symmetric potentials this theorem implies a
pairing of the left/right-invisible configurations that are
related by the {PT} transformation. Interestingly, the
statement of the invisibility theorem also holds, if the term
"invisible" is replaced with "reflectionless"
\cite{Mostafazadeh:2013-12103:PRA}.

\pict{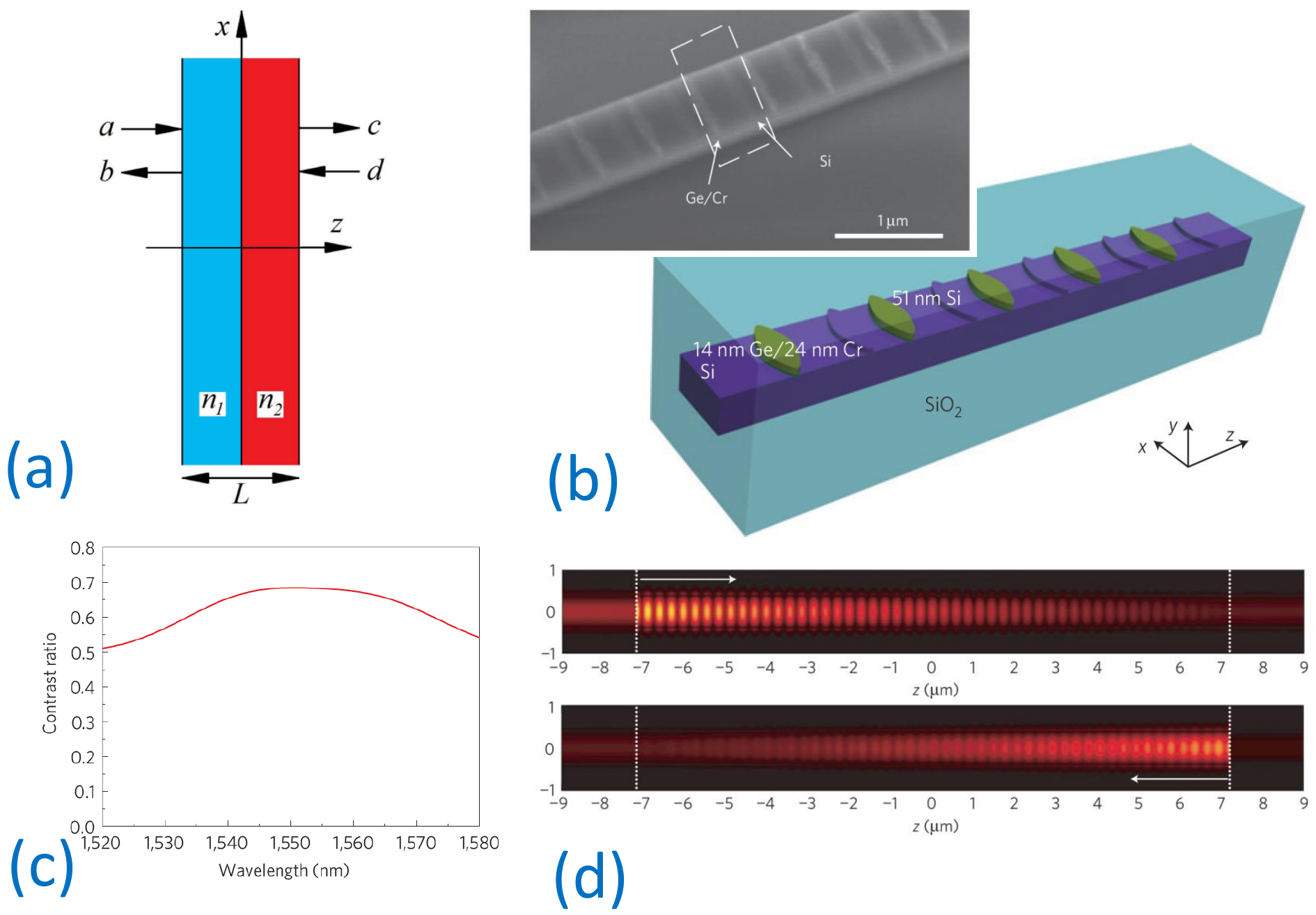}{Fig5_1}{
Linear scattering regimes.
(a)~Schematic of scattering by active two-layer slab, as formulated in Eqs.~(\ref{Scattering04}), (\ref{Scattering05}).
%view of the cross section of a two-layer infinite planar slab of gain material of thickness $L$
%that is aligned in the $x-y$ plane.
%The arrows marked by $a$, $b$,
%$c$, and $d$ represent the amplitudes of plane waves related to
%each other by Eq. (\ref{Scattering05}).
(b-d)~Experimental demonstration of unidirectional reflectionless scattering by active Bragg gratings. (b)~Design of passive structure and SEM picture of the fabricated device. (c)~Measured spectrum of contrast ratio of reflectivities for left and right incidence. (d)~Simulated electric field distribution for incidence from the left (top) and right (bottom). Adopted from Ref.~\cite{Feng:2013-108:NMAT}.
}

A simple two-layer structure shown in Fig.~\rpict{Fig5_1} represents a basic, exactly solvable, and experimentally realizable system which can realize invisibility~\cite{Mostafazadeh:2013-12103:PRA}. Infinite planar slab of optically active material consists of two layers having complex refractive indices $n_1$ and $n_2$ and equal thickness of $L/2$. The wave  equation for a linearly polarized time-harmonic electromagnetic wave that propagates along the direction normal to the slab reads
\begin{equation}\label{Scattering03}
\Psi ''(z)+k^2n(z)^2\Psi(z)=0,
\end{equation}
where
\begin{equation}\label{Scattering04}
  n(z) = \left\{
  \begin{array}{l l}
    n_1 & \quad \text{for} \quad -L/2 \le  z<0,\\
    n_2 & \quad \text{for} \quad \quad \quad \, 0 \le  z \le L/2, \\
    1 & \quad \text{for} \quad \quad \quad \quad \,\, |z|> L/2.
  \end{array} \right.
\end{equation}
The amplitudes of plane waves depicted in Fig.~\rpict{Fig5_1}~(a) are
related through a $2\times 2$ matrix $\mathbf{M}$
\begin{equation}\label{Scattering05}
  \begin{bmatrix}
  c \\  d
  \end{bmatrix}
  =\mathbf{M}
  \begin{bmatrix}
  a\\  b
  \end{bmatrix}.
\end{equation}
Particular cases of left-incident and right-incident scattering
solutions can be written as \cite{Mostafazadeh:2009-220402:PRL}
\begin{equation}\label{Scattering06}
 \Psi^l(z) = \left\{
  \begin{array}{l l}
    c_l(e^{ikz}+r_l e^{-ikz}) & \quad \text{for} \quad z< -L/2,\\
    c_l t_le^{ikz} & \quad \text{for} \quad z> L/2,
  \end{array} \right.
\end{equation}
and
\begin{equation}\label{Scattering07}
 \Psi^r(z) = \left\{
  \begin{array}{l l}
    c_r t_r e^{-ikz} & \quad \text{for} \quad z< -L/2,\\
    c_r(e^{-ikz}+ r_re^{ikz}) & \quad \text{for} \quad z> L/2,
  \end{array} \right.
\end{equation}
where the incident, transmitted and reflected light intensities
are $I_{l,r}=|c_{l,r}|^2$, $T_{l,r}=|t_{l,r}|^2$, and
$R_{l,r}=|r_{l,r}|^2$, respectively. The (complex) transmission
and left/right reflection amplitudes can be determined by the
elements of matrix $\mathbf{M}$ (note that $\det(\mathbf{M})=1$)
as follows
\begin{equation}\label{Scattering08}
  t_l=t_r=\frac{1}{M_{22}}, \quad r_l=-\frac{M_{21}}{M_{22}}, \quad
  r_r=\frac{M_{12}}{M_{22}},
\end{equation}
with $M_{ij}$ defined by Eq.~(16) in Ref.~\cite{Mostafazadeh:2013-12103:PRA}.
%The equality $t_l=t_r$ in
%(\ref{Scattering08}) reflects the {\em reciprocity of
%transmission}, which holds for any linear medium.
A {\em spectral singularity} appears whenever one can satisfy $M_{22}=0$ for a
real $k$, then the transmission and reflection amplitudes diverge.
%
%From the definition of invisibility given above and $M_{ij}$ given
%by Eq. (16) in Ref. \cite{Mostafazadeh:2013-12103:PRA} it is
%possible to derive
Then, the necessary and sufficient conditions for the
unidirectional invisibility are found as
\begin{IEEEeqnarray}{l}\label{Scattering09}
  n_+^2\cos a_+ - n_-^2 \cos a_- =(n_+^2-n_-^2) \cos(Lk), \nonumber \\
  \tilde{n}_+n_+\sin a_+ +\tilde{n}_-n_-\sin a_- =(n_+^2-n_-^2) \sin(Lk), \nonumber \\
  \cos a_+ \neq \cos a_-,
\end{IEEEeqnarray}
where
\begin{equation}\label{Scattering10}
  n_{\pm}=n_1 \pm n_2, \quad a_{\pm}=n_{\pm}kL/2, \quad
  \tilde{n}_{\pm}=n_1n_2\pm 1.
\end{equation}
The two-layer slab is invisible from both directions provided that
$n_1$ and $n_2$ are rational numbers, and the wavelength $\lambda$
at which the device is invisible is such that $L$ is the half
integer multiple of $\lambda$ \cite{Mostafazadeh:2013-12103:PRA}.
Moreover, existence of invisible configurations that remain
practically reflectionless within a very wide spectral range was
demonstrated \cite{Mostafazadeh:2013-12103:PRA}.

%It has been demonstrated that
Unidirectional invisibility can be also achieved for {PT}-symmetric locally periodic
potentials describing Bragg gratings with active layers~\cite{Lin:2011-213901:PRL}
\begin{equation}\label{Scattering02}
 v(z) = \left\{
  \begin{array}{l l}
    \alpha_0+\alpha e^{2i\beta z} & \quad \text{for} \quad |z|\le L/2,\\
    0 & \quad \text{for} \quad |z|> L/2,
  \end{array} \right.
\end{equation}
where $\alpha_0$, $\alpha$, $\beta>0$, and $L>0$ are particular real
parameters. It was then shown that invisibility can be achieved only for structure length below a certain threshold~\cite{Longhi:2011-485302:JPA}, and analytical solutions were derived in Ref.~\cite{Jones:2012-135306:JPA}.

The invisibility phenomenon was demonstrated experimentally in Ref.~\cite{Feng:2013-108:NMAT}. The optical non-Hermitian parity-time system was fabricated with only absorptive media on the Si-on-insulator (SOI) platform, see Fig.~\rpict{Fig5_1}(b).
This structure effectively realized a parity-time-symmetric periodic potential according to Eq.~(\ref{Scattering02}), but in a lossy background. Accordingly, the transmission is attenuated, but unidirectional reflection can still be observed. Indeed, experimental measurements show large contrast for reflections for incidence from the left and right, see Fig.~\rpict{Fig5_1}(c). The difference occurs due to distinct distributions of the electric field, which are visualized based on numerical modelling in Fig.~\rpict{Fig5_1}(d).
%, although  be expected at this exceptional point of the
%proposed passive parity-time m
%{\bf describe experiment!}

%Scattering formalism for the optical systems symmetrically combining lossy and active elements has been developed in the recent work by Schomerus \cite{Schomerus:2013-20120194:PTRSA} (see also references therein). In this work the

Suchkov {\em et al.} have studied wave scattering on a domain wall
introduced into a linear array composed of PT-symmetric dimers by
switching the gain and loss in a half of the array \cite{Suchkov:2012-33825:PRA}.
The linear spectrum of the defect-free array includes high-frequency
and low-frequency branches. The two major effects have been revealed:
amplification of both reflected and transmitted waves and excitation of the
reflected and transmitted low-frequency and high-frequency waves by the
incident high-frequency and low-frequency waves, respectively.

General approach to scattering theory for optical PT-symmetric systems with a wide range of geometries beyond the one-dimensional setting has been discussed in a recent work by Schomerus \cite{Schomerus:2013-20120194:PTRSA}. In particular, it is demonstrated that in the presence of the magneto-optical effects the symmetry of optical device becomes important. In addition to the PT-symmetry, the $\mathrm{PTT}'$-symmetry is  introduced to account for the variants of the geometrical and time-reversal operations. The latter symmetry imposes the same spectral constraints as the former one but, generally speaking, a different condition on the magnetic field. Scattering theory is applied to study spectral features  of the open optical systems possessing different symmetries. This leads to effective models that can be formulated in the energy domain (via  Hamiltonians) and in the time domain (via time evolution operators) giving the information about the related energy and time scales.

%-----------------------------------------------------------------------
%\subsection{Nonlinear PT defects}
\subsection{Nonlinear nonreciprocity}
%-----------------------------------------------------------------------

%\cite{Liu:2014-13824:PRA}.
%In this section we overview the works where the effects of nonlinearity of the {PT}-symmetric scatterers on the transmission properties of linear waveguides are discussed.
%~\cite{Lin:2012-50101:PRA, Liu:2014-13824:PRA, Bender:2013-234101:PRL, Miroshnichenko:2011-12123:PRA}.

Whereas transmission is known to be reciprocal in linear systems as discussed in the previous section, nonlinearity can lead to nonreciprocal transmission which depends on the direction of incidence, i.e. $T_l \neq T_r$.
%Another important result of the work \cite{Liu:2014-13824:PRA} is
%that the saturable nonlinearity of the defect makes $t_l \neq
%t_r$, i.e., results in nonreciprocity,
%~\cite{Peng:2014-394:NPHYS}.
% and, as has been shown
%recently, even in some nonlinear systems with Kerr-type
%nonlinearity \cite{Agarwal:2002-1205:OL}.
%In the case of saturable nonlinearity the nonreciprocity can be amplified to the extent
%that
Nonlinear systems can realize optical isolation by predominantly allowing only one-way transmission, i.e.
%(optical isolation).
it allows light to pass through, say, for left incidence, whereas it is blocked for light coming from the other side. Such nonlinear optical diode have been extensively investigated in conservative optical systems, however active PT structures can be used to dramatically reduce the optical power threshold for nonlinear nonreciprocity~\cite{Peng:2014-394:NPHYS}.
% then
%leading to the {\em optical diode} action.
%

\REMOVE{
\pict{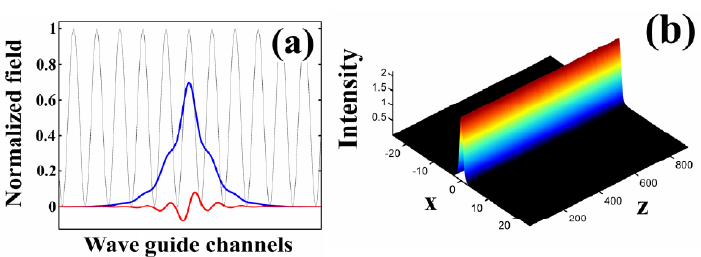}{Fig5_3}{
Experimental data of transmittances $T_l$ and $T_r$ for left
(solid blue line) and right (solid red line) incident waves. The
results of the numerical simulations $T_l$ ($T_r$) transmittances
are shown as dashed blue (red) lines. The arrow at 39.5 kHz shows
the position of maximal asymmetry. Adapted from Ref.~\cite{Bender:2013-234101:PRL}.
}
}

The effect of nonreciprocity was first observed experimentally for electric current transmission lines attached to PT-symmetric electronic circuit resonators playing the role of nonlinear scatterers~\cite{Lin:2012-50101:PRA,Bender:2013-234101:PRL}. As a reference model, Bender {\em et al.} used coupled nonlinear electronic Van der Pol oscillators with anharmonic parts consisting of a complementary amplifier (gain) and a dissipative conductor (loss) combined to preserve PT-symmetry.
% [see Fig. 3(a)]~\cite{Bender:2013-234101:PRL}.
%The authors suggest that similar setting can be implemented in optics by employing concatenated semiconductor optical amplifiers and semiconductor-doped two-photon absorber microcavities. In Fig.~\rpict{Fig5_3} the experimental and theoretical left and right transmittances for the {PT}-symmetric Van der Pol dimer are reported \cite{Bender:2013-234101:PRL}. The overall shape of the measured transmittances reasonably matches the numerical simulations.
A striking feature observed in measurements was that
%the results of Fig.~\rpict{Fig5_3} is the fact that
the transmittance from left to right
%$T_l(\nu)$
differs from the transmittance from right to left.
%$T_r(\nu)$, i.e. $T_l \neq T_r$.
The phenomenon is most pronounced in the regions of the resonances distorted by the nonlinearity, around $39.5$~kHz.
%, indicated by the arrow at $\nu=39.5$ kHz. A conservative nonlinear medium by itself cannot generate such transport asymmetries.
Furthermore, it was found that increasing the gain or loss parameter,
% which is responsible for the asymmetric transport,
maintains or even enhances the transmitted intensities while it leaves unaffected the resonance position. This has to be contrasted with other proposals of asymmetric transport which are based on conservative nonlinear schemes (see for example Ref.~\cite{Lepri:2011-164101:PRL}), where increase of asymmetry leads to reduced transmittances.

Nonreciprocal transmission in optics was observed in coupled-resonator structures, where nonlinear effects are enhanced due to strong field confinement~\cite{Peng:2014-394:NPHYS}.
%. It was shown that light self-action in PT microring resonator structures can lead to nonreciprocal transmission~\cite{Peng:2014-394:NPHYS}.
The system was composed of two coupled microtoroidal whispering-gallery-mode resonators [Fig.~\rpict{PTcouplerRings}(a)], where one was passive, and gain in the other one was provided in the 1,550nm wavelength band by optically pumping Er$^{3+}$ ions with a pump laser in the 1,460nm band. The coupled-resonator system is PT symmetric because under parity reflection $\mathbf{P}$ the resonators become interchanged and under time reversal $\mathbf{T}$ loss becomes gain and gain becomes loss. The microtoroids were fabricated at the edges of two separate chips placed on nanopositioning systems to control precisely the distance and accordingly the coupling, see Fig.~\rpict{PTcouplerRings}(b). For strong coupling, the modes are approximately PT-symmetric, whereas breaking of the mode symmetry is observed when the coupling is decreased beyond a critical value.

Transmission between ports 1 and 4 was monitored at different input power levels. At low powers
the input-output relation was linear and the system was reciprocal in both the broken- and unbroken-symmetry phases. Indeed, a linear static dielectric system, even with gain and loss, can not have non-reciprocal response.
At higher input powers, the system remained in the linear regime for the unbroken-symmetry phase.
However the nonlinear threshold was lower in the broken-symmetry phase, due to the stronger field localization in
the resonator with gain. As a result, nonlinear non-reciprocal effect was observed: the input signal at port 4 was transmitted to port 1 at resonance [Fig.~\rpict{PTcouplerRings}(d)], but the input signal at port 1 could not be transmitted to port 4 [Fig.~\rpict{PTcouplerRings}(c)].
The advantages of the PT microresonator systems for the realisation of nonreciprocal transmission include a significant reduction in the input power to $\sim$1 $\mu$W and higher contrast with a complete absence of the signal in one direction but resonantly enhanced transmission in the other direction.

%---------------------------------------------------------------------------------------------------
\pict{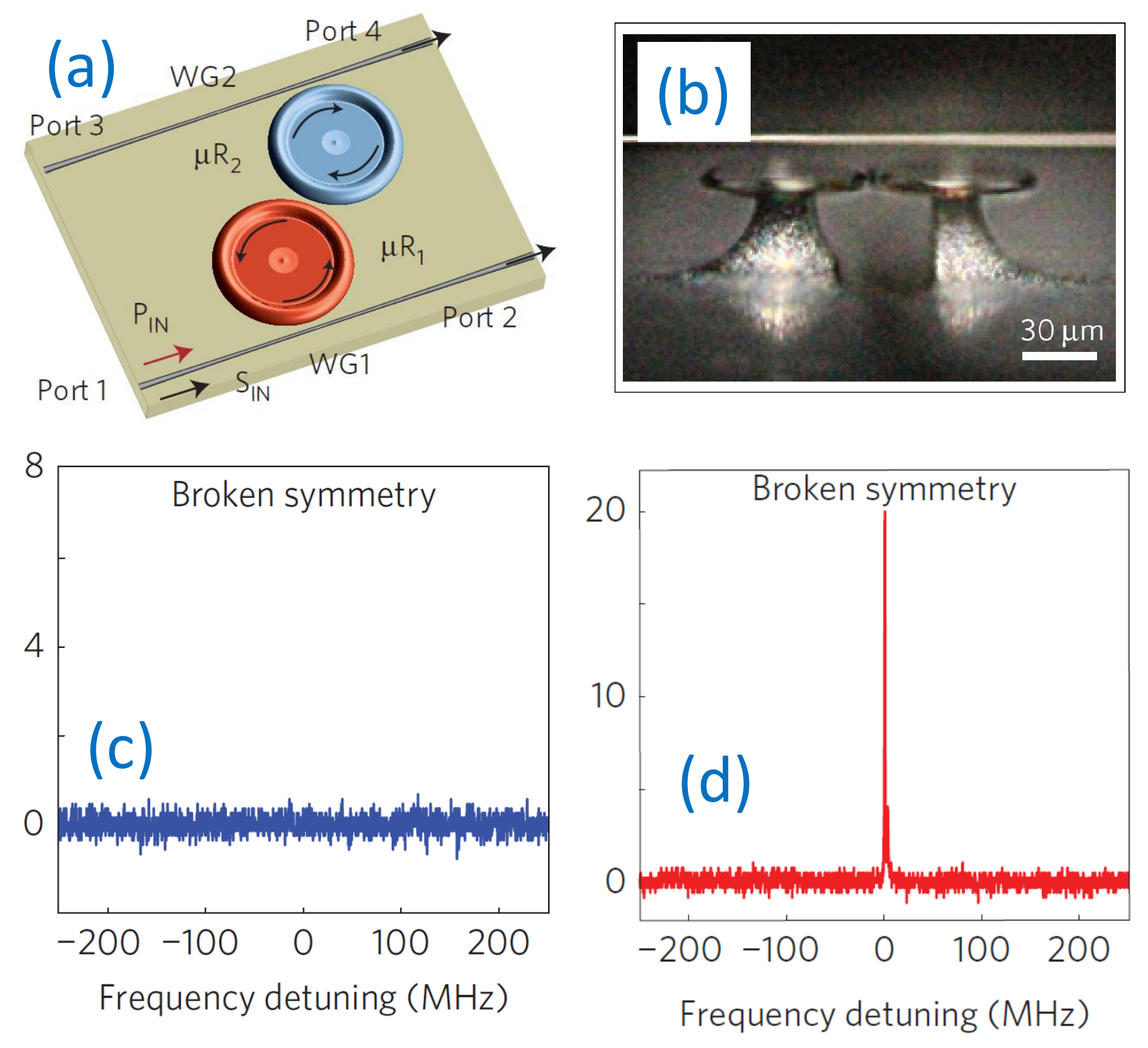}{PTcouplerRings}{
Parity-time-symmetric whispering-gallery microcavity resonators.
(a)~Schematic of two directly coupled resonators and fibre-taper waveguides. Gain is introduced in resonator $\mu$R$_1$ through Er$^{3+}$ doping. (b)~Side view of resonators.
(c,d)~Unidirectional transmission in the nonlinear regime: (c)~from port 1 to 4 and (d)~from port 4 to 1.
Figures after Ref.~\cite{Peng:2014-394:NPHYS}.
}
%---------------------------------------------------------------------------------------------------

PT-symmetric structures based on microresonators can also be used
to realize nonlinear Fano resonances that may give rise to
ultralow-power and high-contrast switching and non-reciprocity
due to their sharp asymmetric line shapes~\cite{Miroshnichenko:2011-12123:PRA, Nazari:2014-9574:OE}.
%Specifically, Fano resonances are created when two nonlinear micro-resonators with gain and loss are coupled to a single waveguide. In the nonlinear regime, the line-shape and Fano resonance position depends on the direction of the incident light, which can provide nonreciprocal transmission with 47~dBs contrast in the optical C-window around 1.55$\mu$m optical wavelength.
%
Miroshnichenko {\em et al.} have considered an optical
{PT}-symmetric gain/loss nonlinear dimer coupled to a
passive chain transmitting linear discrete waves $\phi_n(t)$, as
shown in Fig.~\rpict{Fig5_4}(a)
\cite{Miroshnichenko:2011-12123:PRA}. Propagation of light in this
system can be described by the following set of discrete nonlinear
Schr\"odinger equations
\begin{eqnarray}\label{Scattering13}
  i\dot{\psi}_A&=&E\psi_A+i(\gamma_0-\gamma_2|\psi_A|^2)\psi_A+V\psi_0,\\
  \label{Scattering14}
  i\dot{\psi}_n&=&C(\psi_{n-1}+\psi_{n+1})+V\delta_{n,0}(\psi_A+\psi_B),\\
  \label{Scattering15}
  i\dot{\psi}_B&=&E\psi_B-i(\gamma_0-\gamma_2|\psi_B|^2)\psi_B+V\psi_0,
\end{eqnarray}
where the overdot stands for the time derivative, $\gamma_0 > 0$
accounts for the linear gain and loss acting on complex variables
$\psi_A$ and $\psi_B$, respectively, $E$ is a frequency shift,
$\gamma_2$ accounts for the {PT}-symmetric nonlinear
loss and gain (as shown in \cite{Miroshnichenko:2011-12123:PRA},
stable eigenstates are obtained with $\gamma_2 > 0$, i.e., if the
nonlinear loss competes with the linear gain and vice versa), $C$
is the coupling constant in the linear chain, $V$ is a coupling
coefficient between the dimer and the chain, and
$\delta_{n,0}=1(0)$ for $n=0(\neq 0)$. Due to the symmetry, wave
scattering in this system does not depend on the side of
incidence.

\pict{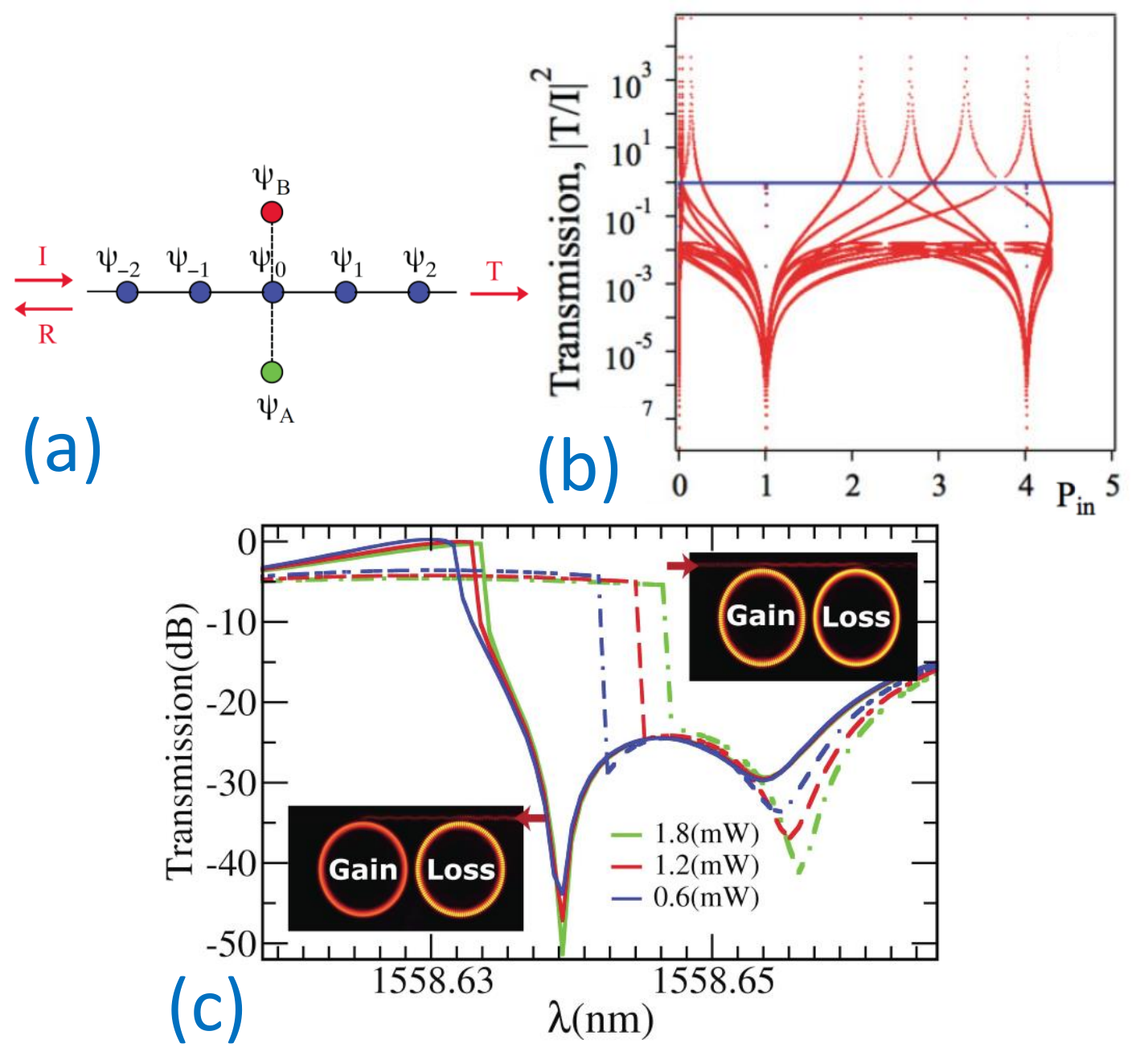}{Fig5_4}{
(a,b)~Fano resonances in a linear chain with the side-coupled elements featuring the nonlinear {PT}-symmetry described by
Eqs.~(\ref{Scattering13}-\ref{Scattering15}), adapted from Ref.~\cite{Miroshnichenko:2011-12123:PRA}.
(a)~Structure geometry, the arrows indicate incident, reflected, and transmitted waves.
(b) Normalized transmission coefficient
%and (c) excitation intensity of the side-coupled {PT} elements
as the functions of input power ($P_I=|I|^2$) for $\gamma_0 = 0.01$, $\gamma_2 = 0.0001$, $V = 0.2$,
$C = 1$, and $\omega = E = 0.1$. The nonlinear Fano resonances correspond to $T = 0$.
%The red and blue curves depict, respectively, the full multitude of asymmetric scattering regimes, produced by Eq. (\ref{Scattering19}), and the symmetric one corresponding to Eq. (\ref{Scattering20}). The inset in (c) shows the tristability region in the latter case.
(c)~Transmittance and Fano resonances for nonlinear microdisks with gain and loss coupled to an access waveguide, adapted from Ref.~\cite{Nazari:2014-9574:OE}.
%for different input powers.
Dashed lines correspond to transmission from the gain side, while solid lines to transmission from a lossy side, shown at different input power levels.
%The asymmetric transport is maintained for a broad range of input power levels.
The transport asymmetry is as high as 46.5 dB without compromising the outgoing optical intensity which is as high as -5 dBs.
}

The solution corresponding to the scattering of incident waves
with amplitude $I$ on the {PT} complex is looked for as
\begin{equation}\label{Scattering16}
  \psi_n = \left\{
  \begin{array}{l l}
    Ie^{i(kn-\omega t)}+Re^{-i(kn+\omega t)} & \quad \text{for} \quad n \le 0,\\
    Te^{i(kn-\omega t)} & \quad \text{for} \quad  n \ge 0,
  \end{array} \right.
\end{equation}
where wave number $k > 0$ is determined by the dispersion equation
for the linear chain, $k = \cos^{-1}(\omega/2C)$, while $R$ and
$T$ are the amplitudes of the reflected and transmitted waves.

\REMOVE{
A
straightforward analysis of Eqs. (\ref{Scattering14}) and
(\ref{Scattering16}) at $n = 0$ yields
\begin{equation}\label{Scattering17}
  R = \psi_0 - I, \quad T = \psi_0,
\end{equation}
and the expression for $\psi_0$ in terms of $I$ and
$\psi^{(0)}_{A,B}$:
\begin{equation}\label{Scattering18}
  \psi_0 = I+\frac{iV}{2C\sin k}(\psi^{(0)}_{A}+\psi^{(0)}_{B}).
\end{equation}
Substitution of (\ref{Scattering18}) into the stationary version
of Eqs. (\ref{Scattering13}) and (\ref{Scattering15}) leads to the
following set of complex cubic equations:
\begin{eqnarray}\label{Scattering19}
  (E-\omega)\psi_{A,B}^{(0)}+\frac{iV^2}{2C\sin k}(\psi_{A}^{(0)}+\psi_{B}^{(0)}) \nonumber \\
  \pm i(\gamma_0-\gamma_2|\psi_{A,B}^{(0)}|^2)\psi_{A,B}^{(0)}=-VI,
\end{eqnarray}
which should be solved for $\psi_{A,B}^{(0)}$ at given $I$ and
$\omega$. Then, $\psi_0$ can be found from Eq.
(\ref{Scattering18}), and, eventually, the reflection and
transmission coefficients can be found from Eq.
(\ref{Scattering17}).
}

\REMOVE{
For the linear system ($\gamma_2 = 0$),
%Eq. (\ref{Scattering19}) yields
there exist only symmetric solutions with $|\psi_{A}| = |\psi_{B}|$. In
the absence of gain/loss ($\gamma_0 =\gamma_2 = 0$) suppression of
the transmission by the degenerate side-coupled elements is
observed at $\omega=E$, which can be explained in terms of the
Fano resonance \cite{Miroshnichenko:2010-2257:RMP}.
}

In the nonlinear regime with $\gamma_2 > 0$, one can find symmetric solutions
with $\psi_{A} = -\psi_{B}=-i \phi$, where $\phi$ is real. For
$\phi \neq \sqrt{\gamma_0/\gamma_2}$, the symmetric mode exists at
$\omega=E$, with $\phi$ determined by the equation
\begin{equation}\label{Scattering20}
  \gamma_2\phi^3-\gamma_0 \phi +VI=0,
\end{equation}
which yields a single real root for $P_I = |I|^2 >
(4/27)\gamma^3_0 /(V^2\gamma_2)$, and three real solutions
(tristability) in the opposite case.
%According to Eq. (\ref{Scattering18}),
All these solutions realize the perfect
transmissivity with $T=1$ [the horizontal blue line in Fig.
\rpict{Fig5_4}(b)].
%The family of the symmetric states is displayed by the blue curve in Fig.~\rpict{Fig5_4}(c), where the tristability occurs at $P_I < 1/27$.}

In contrast to its linear counterpart, the nonlinear system may
support complete suppression of the transmission ($T = 0$), that
is, nonlinear Fano resonances \cite{Miroshnichenko:2010-2257:RMP}. The
nonlinear {PT}-symmetric scatterer attached to the
linear chain as shown in Fig.~\rpict{Fig5_4}(a) can support a {\em
continuous family} of symmetric nonlinear Fano resonances with
$\omega = E$ (for the details see
\cite{Miroshnichenko:2011-12123:PRA}). The family of the nonlinear
Fano resonances with the symmetrically excited side-coupled
{PT} elements exists in an {\em interval} of the
intensity of the incident wave $P_I$, even though Fano resonance is usually obtained as an isolated
solution.

It was further suggested that nonlinear nonreciprocity and optical isolation through Fano resonances can be realized with two nonlinear micro-resonators with gain and loss are coupled to a single waveguide~\cite{Nazari:2014-9574:OE}.
%Specifically, Fano resonances are created when two nonlinear micro-resonators with gain and loss are coupled to a single waveguide.
In the nonlinear regime, the line-shape and Fano resonance position depends on the direction of the incident light, which can provide nonreciprocal transmission with 47~dBs contrast in the optical C-window around 1.55$\mu$m optical wavelength.

Scattering of linear waves propagating in a waveguide array with an embedded
{\em nonlinear} {PT}-symmetric dimer was found to demonstrated distinct features~\cite{Zhang:2014-13927:OE}. It was found that, under certain conditions, nonlinearity can effectively suppress the nonconservative effects, which are generated by the linear gain and loss terms. In particular, at high intensities, the incident wave can almost entirely transmit across the defect.

%enabled the authors of
%the work \cite{Zhang:2014-13927:OE} to derive the discrete fundamental
%and dipole solitons in an exact analytical form.

Linear {PT}-symmetric systems can demonstrate spectral
singularities leading to infinite transmission and reflection
coefficients \cite{Mostafazadeh:2009-220402:PRL}. Divergence of
the fields in the medium breaks the assumptions of the linear
analysis and requires taking into account the dependence of the
refractive index of medium on the light intensity.
It was observed that simplest cubic Kerr-type nonlinearity does
not lead to regularization of spectral singularities
\cite{Mostafazadeh:2013-260402:PRL} but the saturable nonlinearity
does \cite{Liu:2014-13824:PRA}.

\subsection{Soliton scattering}

Collision of moving solitons with localized obstacles (defects) is the problem of fundamental significance in various nonlinear wave guiding systems. Recently in nonlinear optics a great deal of attention has been focused on the effect of soliton interaction with the complex {PT}-symmetric localized potentials embedded into the self-focusing or self-defocusing medium. In these settings the {PT}-symmetric gain and loss regions are spatially localized, e.g, in the form of nonlinear {PT}-symmetric dimers, trimers, or, more generally, oligomers embedded into nonlinear conservative lattice \cite{Dmitriev:2011-13833:PRA, Suchkov:2012-54003:EPL, DAmbroise:2012-444012:JPA, Zhang:2014-13927:OE}. Continuum versions of such models have been analyzed in the works \cite{Hu:2012-266:EPD,Abdullaev:2013-43829:PRA,Karjanto:2015-23112:CHA}.
%Nonlinear optical lattices with periodic {PT}-symmetric
%potentials support {PT} solitons
%\cite{Musslimani:2008-30402:PRL,Zhu:2011-2680:OL,Miri:2012-33801:PRA,Li:2012-16823:OE,Zezyulin:2011-64003:EPL}.
In other cases soliton dynamics is investigated in the two-core systems such as an array of coupled {PT}-symmetric dimers \cite{Suchkov:2011-46609:PRE, Suchkov:2011-222:RAR} and its continuum version in the form of coupled planar waveguides with gain in one of them and balanced loss in another \cite{Barashenkov:2012-53809:PRA, Bludov:2014-3382:OL}. Two-dimensional generalization of such models was studied in~\cite{Burlak:2013-62904:PRE}.

%Below the soliton scattering will be first discussed in continuum and then in discrete {PT}-symmetric systems.

%\subsection{Continuum systems}

\pict{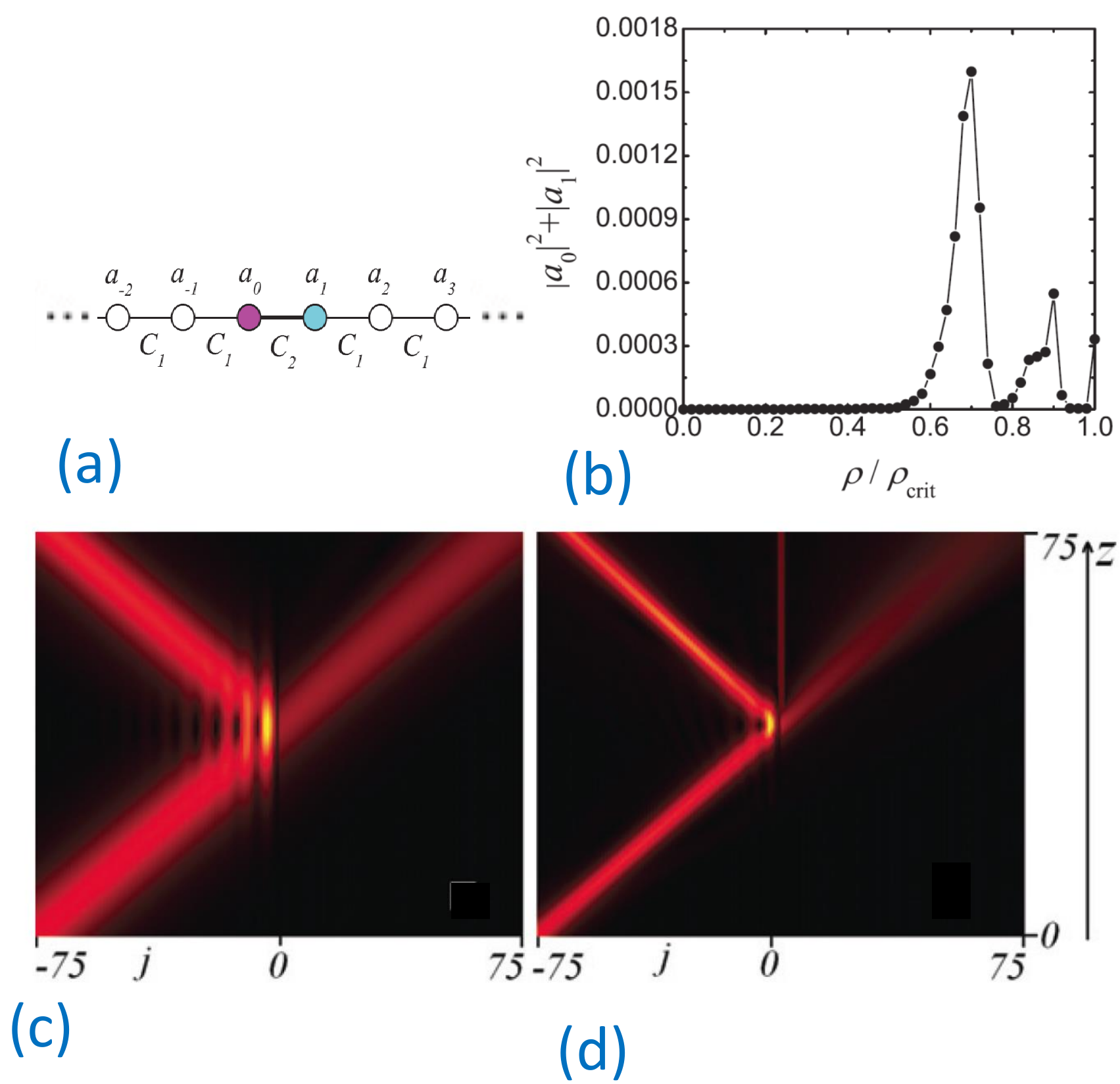}{Fig5_7}{
(a)~Schematic plot of the waveguide array structure.
The two {PT}-symmetric waveguides with balanced gain and
loss are at the sites $n=0,1$. The coupling constant between the modes of
waveguides is $C_1$ with the exception for the coupling constant $C_2$
between the active waveguides.
(b)~Total intensity of the localized mode excited by the incident soliton vs. $\rho/\rho_{\rm crit}$ for $A=0.2$.
(c,d)~Intensity distribution in the waveguide array vs. the propagation distance in the process of soliton
scattering for $\rho=1.5 = \rho_{\rm crit} / 3$ and the incident soliton amplitude (c)~$A=0.2$ and (d)~$A=0.5$. Other parameters are $C_1 =2$, $C_2 =4$, $v =0.5$.
Adapted from Refs.~\cite{Dmitriev:2011-13833:PRA, Suchkov:2012-54003:EPL}.
}
%(b) Example of the soliton scattering for the model parameters $C_1 = 1$, $C_2 = 4$, $\rho = -3.7108$, $\gamma = 1$ and the incident soliton parameters $A = 0.05$, $v = 0.5$, $n_0 = -150$. The 3D plot presents $|\psi_n (z)|^2$. Both reflected and transmitted solitons are considerably amplified. (c) Same as in (b) but for the opposite sign of $\rho$. }

Soliton scattering by PT defects was first investigated in a nonlinear waveguide array~\cite{Dmitriev:2011-13833:PRA, Suchkov:2012-54003:EPL}.
A local inhomogeneity was created by a pair of {PT}-symmetric
waveguides with balanced gain and loss [see Fig.~\rpict{Fig5_7}(a)].
% has been
%investigated in \cite{Dmitriev:2011-13833:PRA}. It is assumed that
Propagation of light in this system was modelled with the coupled discrete nonlinear
Schr\"odinger equations
\begin{eqnarray}\label{Scattering31}
  i\dot{\psi}_n &+& C_1(\psi_{n-1}+\psi_{n+1})+|\psi_n|^2\psi_n=0, \,\, n\neq 0,1,\\
  \label{Scattering32}  i\dot{\psi}_0&+&i\rho\psi_0+C_1\psi_{-1}+C_2\psi_{1}+|\psi_0|^2\psi_0=0,\\
  \label{Scattering33}
  i\dot{\psi}_1&-&i\rho\psi_1+C_2\psi_{0}+C_1\psi_{2}+|\psi_1|^2\psi_1=0,
\end{eqnarray}
where the overdot stands for the derivative with respect to the propagation distance $z$,
$n$ is the waveguide number, $\psi_n$ are the mode amplitudes at waveguides,
$\rho > 0(< 0)$ defines the rate of loss (gain) at zeroth and gain (loss) at first
waveguides, and $C_{1,2}$ are
the coupling coefficients between the modes of waveguides.

Far from the defect, i.e., for $|n|\gg 1$, Eq.~(\ref{Scattering31})
has an {\em approximate} soliton solution
\begin{eqnarray}\label{Scattering34}
 \psi_n =A\frac{\exp\{i[v(n-n_0)+(\delta^2-v^2)C_1z+\alpha]\}}{\cosh [\delta(n-n_0 - 2C_1 v z )]},\nonumber \\
\end{eqnarray}
which can be found by considering the continuum limit of
Eq.~(\ref{Scattering31}) in the form of the integrable nonlinear Schr\"odinger
equation. Here $A$, $\delta = A\sqrt{1/(2C_1)}$, $v$, $n_0$, and $\alpha$ are
parameters defining the soliton amplitude, inverse width, velocity, initial position, and initial phase, respectively.

It was found that solitons can be partially reflected and transmitted by the PT defect, and the total energy is generally not conserved in the scattering process.
%has been shown that the reflected
%and transmitted linear and nonlinear waves (solitons) can be substantially amplified by the {PT}-symmetric defect, and that the results of the linear theory give a good
%prediction for the scattering of relatively wide solitons in the regime of weak
%nonlinearity.
In addition, the soliton can excite the mode localized at the {PT}-symmetric defect, and this process resonantly depends on the structure parameters including the gain and loss, as illustrated in Fig.~\rpict{Fig5_7}(b). Example of soliton scattering without excitation of localized mode is shown in Fig.~\rpict{Fig5_7}(c), whereas under different parameters mode excitation is clearly visible, see Fig.~\rpict{Fig5_7}(d). The soliton interaction with the large-amplitude localized mode can result in {PT}-symmetry breaking \cite{Suchkov:2012-54003:EPL}.

%In Fig.~\rpict{Fig5_7}(b), show is the example of soliton scattering for model parameters $C_1 = 1$, $C_2 = 4$, $\rho = -3.7108$, $\gamma = 1$ and incident soliton parameters $A = 0.05$, $v = 0.5$, $n_0 =−150$. The 3D plot presents $|\psi_n (z)|^2$. It can be seen that both reflected and transmitted solitons are amplified. Figure \rpict{Fig5_7}(c) shows the same as in (b), but for $\rho$ having the opposite sign. Here again the reflected and transmitted solitons are amplified and, in addition, transmitted and reflected solitons bare the soliton internal mode.

Interestingly, a discrete model describing a waveguide array with energy
gain and loss that admits exact solitary wave solutions has been
found~\cite{Suchkov:2011-222:RAR}. Existence of the high-frequency
and the low-frequency solitons was demonstrated. The numerical study
of collisions between similar and dissimilar solitons has demonstrated
that solitons of the same type interact elastically, while for the dissimilar
ones the collision outcome significantly depends on the initial conditions.

\pict{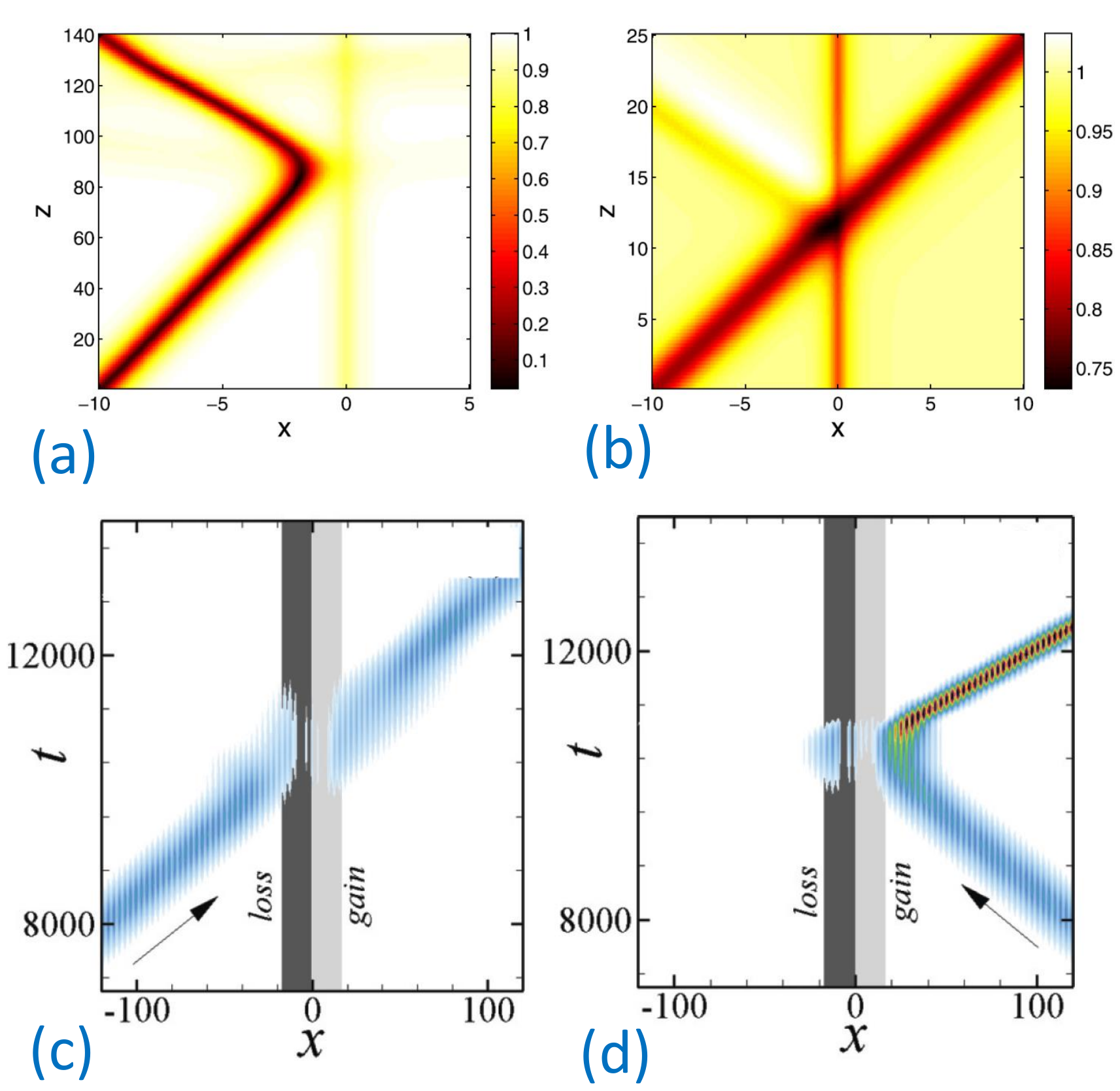}{Fig5_5}{
(a-b) Numerical results for the interaction of
%(a-d) bright and (e-h)
dark solitons with the {PT}-symmetric defect described by Eqs.~(\ref{Scattering21},\ref{Scattering22}).
%The cases of (a,b,e,f) self-focusing and (c,d,g,h) defocusing nonlinearity are presented.
Shown is the distribution of $|\Psi(x,z)|$.
%Insets in (a-d) show the relative reflected (blue) and transmitted (red) powers, $P_R/P_I$ and $P_T/P_I$, respectively. In (a) slow bright soliton ($v_0=0.8$) is trapped and blowups, while in (b) the fast soliton ($v_0=5$) passes through the defect. In (c) the slow bright soliton ($v_0=0.4$) bounces back from the dipole, while in (d) the fast one ($v_0=1$) passes through the defect.
(a)~Reflection of slow dark soliton ($v_0=0.1$) and (b) the passage of the fast dark solitons ($v_0=0.8$). In all cases $\epsilon=0$, $x_0 =-10$
%in (a,b,e,f)
$\gamma=0.3$. Adapted from Ref.~\cite{Karjanto:2015-23112:CHA}.
(c,d)~Scattering of gap solitons by PT defect according to Eqs.~(\ref{Scattering24}-\ref{Scattering26}).
Parameters are $\eta = −5.86$, $\xi/|\eta| = 0.02$, $E_s = −0.125$, $|v| = 0.05$.
Adapted from Ref.~\cite{Abdullaev:2013-43829:PRA}
}

Karjanto {\em et al.}~\cite{Karjanto:2015-23112:CHA} have studied the scattering of bright and
dark solitons on a strongly localized {PT}-symmetric
defect in frame of the nonlinear Schr\"odinger equation for the
local amplitude of the electromagnetic wave, $\Psi(x, z)$,
\begin{equation}\label{Scattering21}
   i \Psi_z +\frac{1}{2}\Psi_{xx}+g|\Psi|^2\Psi=[V(x)+iW(x)]\Psi,
\end{equation}
with a complex potential:
\begin{equation}\label{Scattering22}
   V(x)+iW(x)=\epsilon\delta(x)+i\gamma\delta^\prime(x),
\end{equation}
where $\delta$ and $\delta^\prime$ denote the Dirac delta function
and its derivative, $\epsilon$ and $\gamma$ being real constants,
positive or negative \cite{Karjanto:2015-23112:CHA}. The nonlinear
term in Eq.~(\ref{Scattering21}) represents
%the self-focusing ($g>0$) or
defocusing nonlinearity for $g<0$.

The collisions of incident dark soliton with the PT-symmetric defect were simulated, setting the initial conditions as
\begin{equation}\label{Scattering23}
   \Psi(x,0)=\Psi_{CW} \left[ \sqrt{1-v_0^2} {\rm tanh}[\sqrt{1-v_0^2} (x-x_0)] + i v_0 \right],
\end{equation}
where $x_0$ is the initial soliton position and $v_0 > 0$ is its velocity.
%, and $\eta= 1$ (once $\epsilon= 0$ was set, $\eta= 1$ may be always
%fixed by rescaling). For the dark solitons initial conditions can
%be found in \cite{Karjanto:2015-23112:CHA}.
%
Shown in Figs.~\rpict{Fig5_5}(a,b) are the examples of the interaction of the
%(a-d) bright and (e-h)
dark solitons with the defect for the cases of
%(a,b,e,f)
defocusing
%and (c,d,g,h) defocusing
nonlinearity. In case (a) the
%left column panels (a,c,e,g) the
soliton initial velocity is relatively small, and it is reflected by the defect.
In case (b) the soliton velocity
%the right column panels (b,d,f,h)
is relatively large, and it passes through the defect.
%Shown is the distribution of $|\Psi(x,z)|$.
%The insets in (a-d) show the relative reflected (blue) and transmitted (red) powers, $P_R/P_I$ and $P_T/P_I$, respectively.
%It can be seen that when the initial velocity is sufficiently large, the incoming
%bright and
%dark solitons pass through the defect, as seen in the right column panels. For the velocities smaller the threshold value, in (a) the soliton is trapped by the defect and subsequently blows up, and in (c,e,g) it is reflected by the defect. The following parameters were used: in all cases $\epsilon=0$ and in (a,b,e,f) $\gamma=0.3$, in (c,d) $\gamma=-0.5$, and (g,h) $\gamma=-0.3$. Soliton velocities were (a) $v_0=0.8$, (b) $v_0=5$, (c) $v_0=0.4$, (d) $v_0=1$, (e,g) $v_0=0.1$, and (f,h) $v_0=0.8$.
An essential difference from previously studied interactions of solitons with defects in conservative systems is that, in spite of the gain-loss balance in the {PT}-symmetric dipole, the collisions change the norm of the solitons making the interaction dynamics more complex.

Abdullaev {\em et al.}~\cite{Abdullaev:2013-43829:PRA} studied scattering of gap solitons by {PT} defects in the
framework of the following model,
\begin{eqnarray}\label{Scattering24}
   i \Psi_z + \Psi_{xx}&=&\sigma|\Psi|^2\Psi+[V_{\rm ol}(x)+V_{\rm
   d}(x)]\Psi, \\\label{Scattering25}
   V_{\rm ol}(x)&=&V_0\cos(2x), \\\label{Scattering26}
   V_{\rm  d}(x)&=&\frac{\eta+i\xi x}{\sqrt{2\pi}\Delta}\exp\Big(\frac{-(x-x_0)^2}{2\Delta^2}\Big),
\end{eqnarray}
that includes the periodic potential $V_{\rm ol}(x)$ (optical lattice) with the amplitude $V_0$ and the {PT}-symmetric localized defect $V_{\rm d}(x)$ that introduces gain and loss in the system. Here  $\eta$ is the strength of the conservative part of the defect, coefficient $\xi$ stands for the gain-dissipation parameter, while the width of the defect is fixed to $\Delta=5$ in all numerical calculations. In the absence of the defect potential, Eq.~\eqref{Scattering24} possesses families of exact gap soliton solutions.
%The authors present the transmission,  trapping, and reflection coefficients for the gap solitons of the semi-infinite gap and of the first band gap as the functions of $\eta$ for $\xi=\pm 0.02|\eta|$.
It is shown that by properly designing  the amplitudes of real and imaginary parts of the {PT} defect it is possible to achieve a resonant transmission of the gap solitons through the defect. This phenomenon occurs for potential parameters that support localized modes inside the {PT} defect potential with the same energy and norm of the incoming soliton. When the imaginary amplitude of the {PT} defect is increased, the possibility of  damping or amplification of transmitted or reflected gap soliton is demonstrated depending on the soliton incidence direction. Furthermore, unidirectional transmission or diode effect can be achieved  as illustrated in Figs.~\rpict{Fig5_5}(c,d). A pair of PT defects with oppositely positions gain and loss layers can then be used to trap a gap soliton.
%Unidirectional transmission (diode effect) of gap solitons and
%their amplification/depletion are discussed for the case of a pair
%of consecutive {PT} defects.

\pict{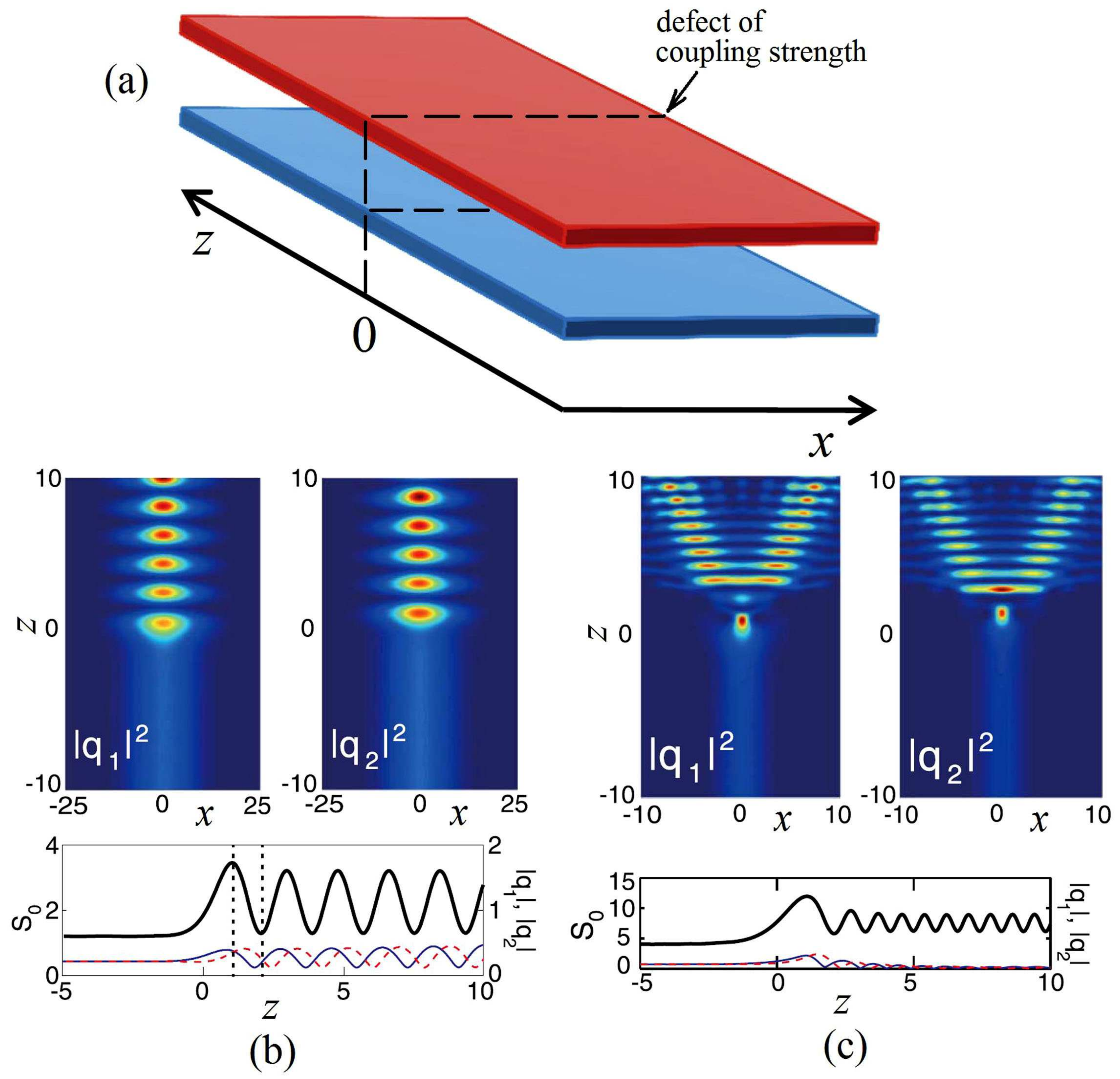}{Fig5_6}{
(a) Schematic plot of the planar coupler having
a defect of coupling exponentially localised around $z=0$ and defined by Eq.~\eqref{Scattering29}. (b)
Results for symmetric solitons ($\sigma=1$). Upper panels: field intensities with $\eta=0.15$ interacting with
defect characterised by $\kappa_{\min}=1$, $\kappa_0=2$ and located
at $z=0$. Lower panel: respective evolution of the total energy
flow $S_0$ for $l=1$ (thick solid lines) and soliton amplitudes
$|q_1|$ and $|q_2|$ (thin solid and dashed lines, respectively).
(c) Same as in (b), but for antisymmetric solitons ($\sigma=-1$)
and $\eta=0.5$, $l=2.7$, $\kappa_0=4$. Adapted from Ref.~\cite{Bludov:2014-3382:OL}.
}

Soliton scattering on a defect in the two-core system
schematically shown in Fig.~\rpict{Fig5_6}(a) was studied by Bludov
{\em et al.}~\cite{Bludov:2014-3382:OL}. The defect is
introduced in the planar {PT}-symmetric coupled
waveguides with gain (top waveguide) and loss (bottom waveguide)
by local modification of the coupling constant in the vicinity of
$z=0$. Light propagation in this system can be described by the
two coupled nonlinear Schr\"odinger equations
\begin{eqnarray}\label{Scattering27}
   i q_{1,z}&=&-q_{1,xx}+i\gamma q_1 -\kappa(z)q_2-|q_1|^2q_1, \\\label{Scattering28}
   i q_{2,z}&=&-q_{2,xx}-i\gamma q_2 -\kappa(z)q_1-|q_2|^2q_2, \\\label{Scattering29}
   \kappa(z)&=&\kappa_0-(\kappa_0-\kappa_{\min})\exp(-z^2/l^2).
\end{eqnarray}
The coupling coefficient $\kappa(z)$ attains the minimal value
$\kappa_{\min}$ at $z=0$ and tends to $\kappa_0$ at $z\rightarrow
\pm \infty$. The defect of coupling has the width $l$.
Far from the defect ($|z| \gg l$) one has $\kappa(z)=\kappa_0$, and the model supports symmetric and antisymmetric types of soliton solutions as described above.
\REMOVE{
. In this case
Eqs. (\ref{Scattering27}-\ref{Scattering29}) possess a soliton solution
\cite{Driben:2011-4323:OL}
\begin{eqnarray}\label{Scattering30}
   q_{1}^{(\sigma)}=\frac{\sqrt 2 \eta
   \exp[i(\eta^2+\sigma\kappa_0\cos\delta)z]}{\cosh(\kappa x)}
   =\sigma q_2^{(\sigma)}e^{-i\sigma\delta},
\end{eqnarray}
where $\delta=\arcsin(\gamma/\kappa_0)$ such that $0 \le \delta \le \pi/2$. The
soliton is parametrized by the positive parameter $\eta$,
and represents symmetric $(\sigma=1)$ and antisymmetric
$(\sigma=-1)$ solutions.
}
Examples the of the soliton-defect interactions are presented for the cases of symmetric and antisymmetric
solitons in Figs.~\rpict{Fig5_6}(b) and~(c), respectively. Upper panels give the field intensities, while
lower panels present the respective evolution of the total energy
flow $S_0$ (thick solid lines) and soliton amplitudes
$|q_1|$ and $|q_2|$ (thin solid and dashed lines, respectively). In Fig.~\rpict{Fig5_6}(b), the soliton passes through a relatively
short defect ($l=1$) transforming into a breather,
% \cite{Barashenkov:2012-53809:PRA},
which is characterized by the intensity oscillations between the two components:
minimum (maximum) in one component corresponds to maximum (minimum)
in the other. In Fig.~\rpict{Fig5_6}(c), for $l=2.7$ the splitting of the incident soliton in the two outward
propagating pulses is observed. There also exist other scenarios of the soliton-defect
interactions as discussed in Ref.~\cite{Bludov:2014-3382:OL}.

{\protect\renewcommand\sectpath{Modulated}
%-------------------------------------------
\section{Periodically Modulated Systems}\lsect{}

Periodic modulation is an effective tool for controlling tunneling dynamics~\cite{ Lin:1990-2927:PRL,Grossmann:1991-516:PRL, Grifoni:1998-229:PRP,Creffield:2007-110501:PRL,Luo:2011-53847:PRA}.
%Following the first theoretical studies~\cite{Driben:2011-51001:EPL, Moiseyev:2011-52125:PRA},
In recent years the PT symmetry of the periodically modulated systems have attracted increasing attention.

%-------------------------------------------
\subsection{Modulated waveguide couplers} \lsect{Couplers}
%-------------------------------------------

Modulation of structure parameters along the propagation direction of coupled optical waveguides with gain and loss can open new opportunities for optical signal control in both linear~\cite{Moiseyev:2011-52125:PRA, Joglekar:2014-40101:PRA, Gong:2015-42135:PRA} and nonlinear~\cite{Driben:2011-4323:OL, Driben:2011-51001:EPL, Horne:2013-485101:JPA, DAmbroise:2014-23136:CHA, Battelli:2015-353:NLD, Martinez:2015-23822:PRA} regimes.

%----------------------
\pict{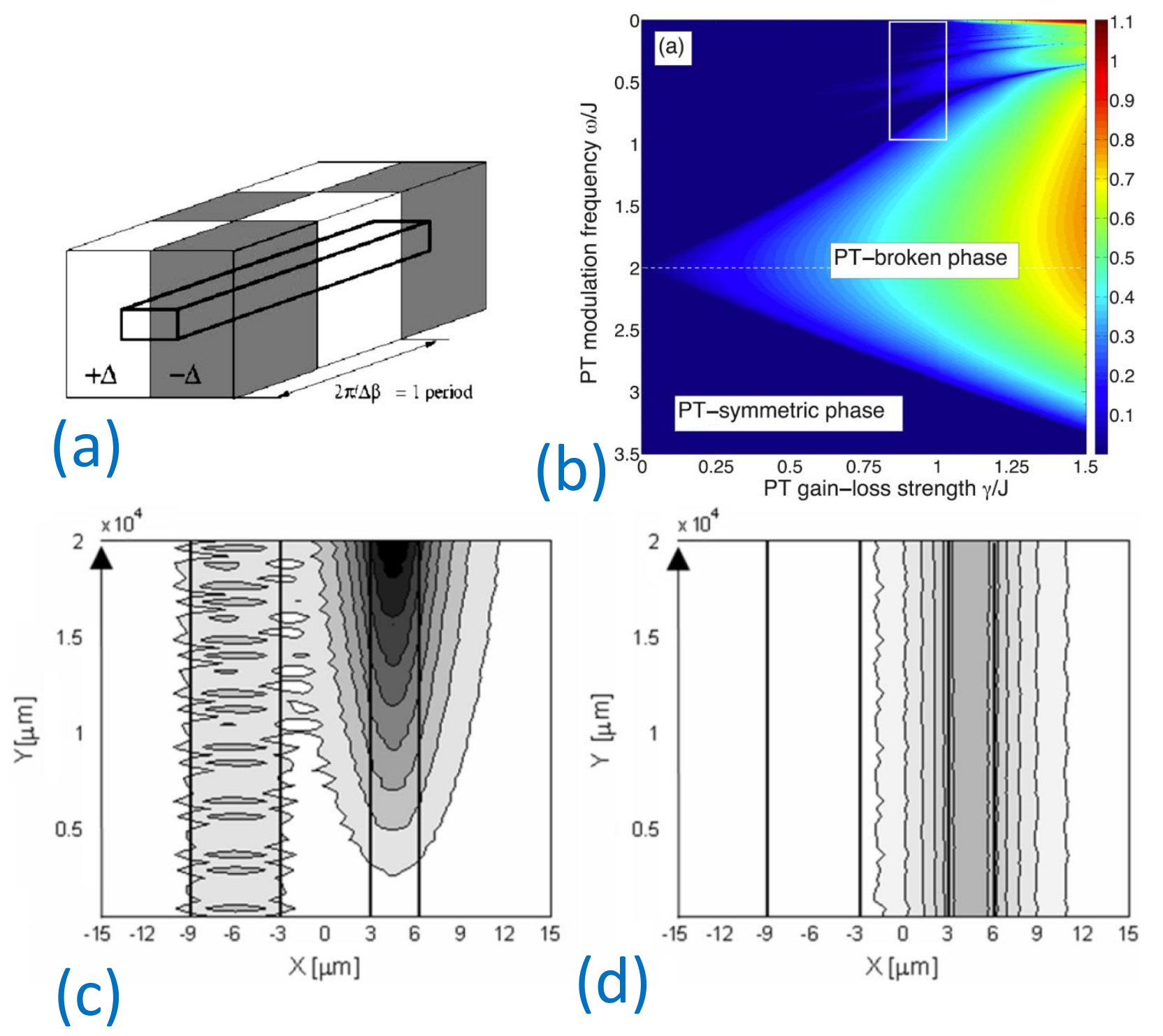}{couplerLinear}{
(a)~Modulation of gain and loss marked with shading along a waveguide, designed for irreversible coupling between two copropagating modes. Adopted from Ref.~\cite{Greenberg:2004-451:OL}.
(b)~PT-phase diagram showing the largest mode amplification vs. the normalised modulation frequency and gain/loss modulation strength. Modulation frequency "2" corresponds to the regime in plot~(a), where PT is in broken phase for arbitrary gain/loss.
Adopted from Ref.~\cite{Joglekar:2014-40101:PRA}.
(c,d)~Irreversible coupling between modes of two waveguides with different cross-sections, and gain/loss modulation in the left waveguide only. Light power evolution for input coupling to the (c)~left and (d)~right waveguide.
Adopted from Ref.~\cite{Greenberg:2005-1013:IQE}.
}
%----------------------

Interestingly, the first studies of couplers with modulated loss and gain were carried out without a consideration of PT symmetry~\cite{Greenberg:2004-451:OL, Greenberg:2004-4013:OE, Greenberg:2005-1013:IQE, West:2007-8052:AOP}. In these studies, it was shown that unidirectional energy transfer between co-propagating optical modes can be achieved in waveguides with longitudinally modulated gain and loss regions. An example of a two-mode active coupler is shown in Fig.~\rpict{couplerLinear}(a).
Examples of unidirectional behaviour are presented in Figs.~\rpict{couplerLinear}(c,d), where light gets localized and amplified in the right waveguide, irrespective of the input coupling port. Such irreversible coupling between the co-propagating modes cannot be achieved in conservative structures. This phenomenon relies on the breaking of the time-reversal symmetry through the modulation of the complex optical refractive index, being analogous to unidirectional scattering by PT-symmetric Bragg gratings discussed in Sec.~\rsect[Scattering]{linear}.

Light evolution in modulated two-mode couplers can be described within a framework of coupled-mode theory,
%
%
%Consider light propagation in a modulated optical coupler with balanced gain and loss.
%Making use of the coupled-mode theory, the electric field can be expressed via a two-mode ansatz with two localized waves $\{\psi_1(x),\psi_2(x)\}$ and the complex amplitudes $\{c_1(z),c_2(z)\}$.
%Thus one can obtain the coupled-mode equations,
%
\begin{equation} \leqt{Coupled-Eq}
    i\frac{d}{dz} \begin{pmatrix} a_1 \\ a_2 \end{pmatrix}
    =\begin{pmatrix} {i \gamma_1(z)}- {\beta_1(z)}
              & - C(z) \\ - C(z) & {i \gamma_2(z)}-{\beta_2(z)} \end{pmatrix} \begin{pmatrix} a_1 \\ a_2 \end{pmatrix}
\end{equation}
where $z$ is the propagation distance along the waveguide, $a_j$ are the mode amplitudes, $C$ is the coupling coefficient, $\gamma_j$ are the gain/loss coefficients, $\beta_j$ are the real propagation constants, and $j=1,2$ is the mode number.
%the biharmonic modulation $S(z)=-A F(z)=-A[\sin(\omega z)+f\sin(2\omega z +\phi)]$. Here, $\omega$ and $A$ denote the modulation frequency and amplitude, respectively. $\phi \in [0,2\pi)$ is the relative phase between the two harmonics, which determines whether the system is PT-symmetric or not. If $\phi=0$ or $\pi$, the original system is PT-symmetric. Otherwise, for $\phi \neq 0$ and $\pi$, it becomes non-PT-symmetric.
%More recently, the PT phase diagram of couplers with gain and loss
%the studies of the effect of longitudianl
Such type of models were found to possess PT symmetry, under certain restrictions on the modulation profiles~\cite{Moiseyev:2011-52125:PRA, Joglekar:2014-40101:PRA, Gong:2015-42135:PRA}.

The formulation of Eq.~\reqt{Coupled-Eq} is not unique, and it depends on the choice of the basis modes. For the structure presented in Fig.~\rpict{couplerLinear}(a), if the mode amplitudes are chosen as $a_1 = a_+ + a_-$  and  $a_2 = a_+ - a_-$, where $a_\pm$ correspond to symmetric and antisymmetric supermodes in the absence of gain or loss,
%when only the gain and loss is modulated in a balanced regime,
we have $\gamma_2(z) = -\gamma_1(z)$, $C(z) = C(0)$, $\beta_j(z) = \beta_0$.
%, if we choose $a_j$ to denote symmetric.
Such model is PT-symmetric in case of even modulation, $\gamma_j(z) = \gamma_j(-z)$. PT diagram for harmonic modulation, $\gamma_1 = - \gamma_2 \sim \cos(\omega z)$ was calculated in Ref.~\cite{Joglekar:2014-40101:PRA}, see Fig.~\rpict{couplerLinear}(b).
In PT-symmetric regime, average amplification is zero, however in PT-broken case, one mode experiences strong amplification. Boundaries between these regimes depend nontrivially both on the modulation amplitude and frequency.
%Whereas this general distinction is similar to unmodulated couplers, there is an important difference.
In case of resonant modulation amplitude matching the beating period between the modes, which corresponds to the normalized value ``2'' along the vertical axis in Fig.~\rpict{couplerLinear}(b), PT mode symmetry is broken for arbitrarily small values of gain/loss modulation. It is in this regime that unidirectional mode coupling is always realized, as illustrated in Figs.~\rpict{couplerLinear}(a,c,d). This is a key difference from unmodulated linear PT couplers, where PT breaking occurs only above a certain threshold of gain/loss.

%----------------------
\pict{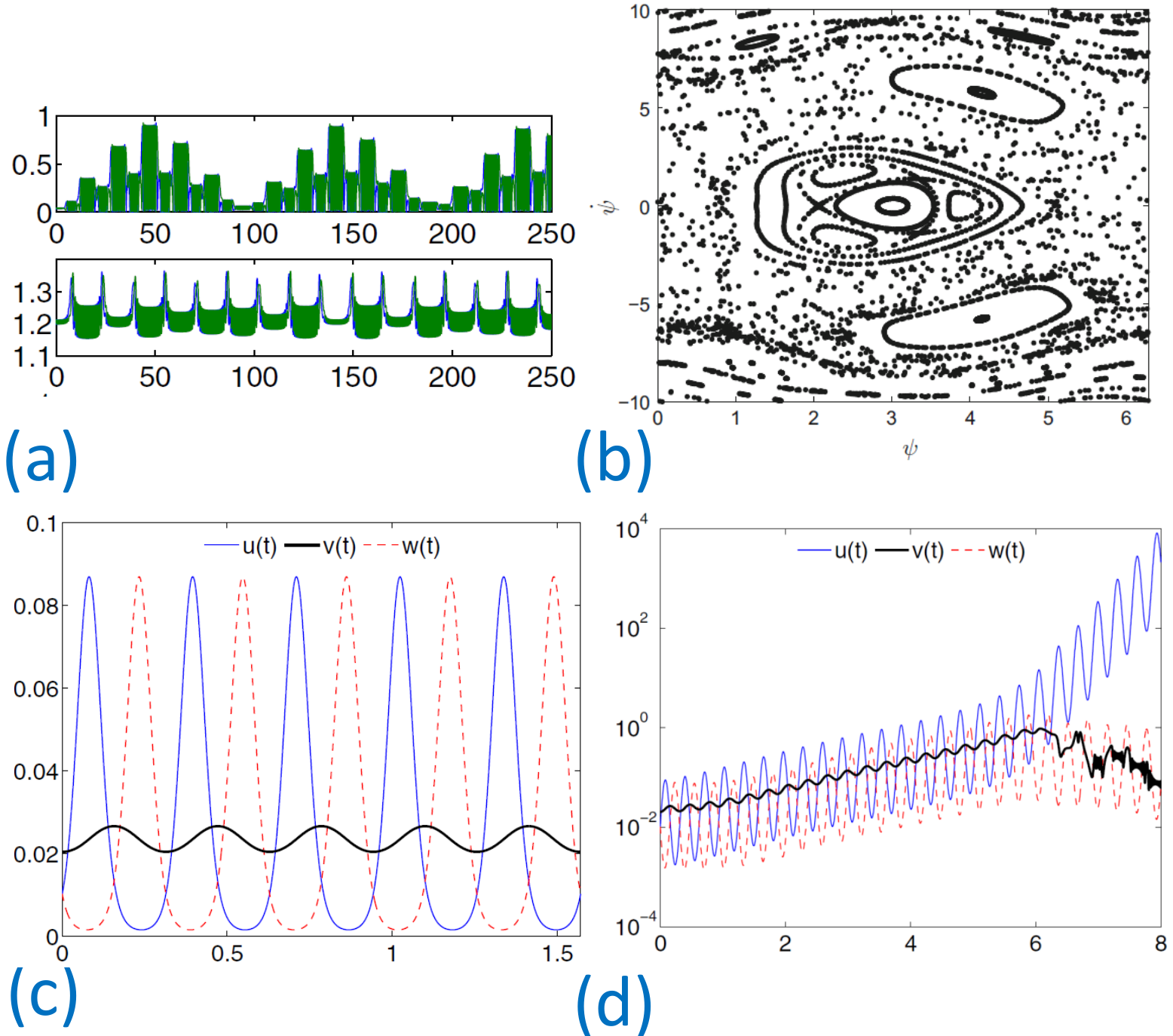}{couplerNonlin}{
Dynamics in nonlinear modulated PT structures.
(a)~Evolution of intensities in a PT dimer with periodically modulated coupling. Parameters correspond to nonlinear stabilisation in a linearly unstable system, adopted from Fig.~8 in Ref.~\cite{DAmbroise:2014-23136:CHA}.
(b)~Stroboscopic Poincare map for a dimer with periodically modulated gain and loss. Adopted from Ref.~\cite{Battelli:2015-353:NLD}.
(c,d)~Dynamics in a trimer demonstrating (c)~stable and (d)~unstable evolution. Adopted from Ref.~\cite{Horne:2013-485101:JPA}.
}
%----------------------

Nonlinear optical interactions can have a nontrivial effect on the mode evolution. It was shown that local Kerr-type nonlinearity can preserve or change the stability properties. Of particular interest is the regime when the system is unstable in the linear regime, but becomes stable due to conservative nonlinear self-action, as illustrated in Fig.~\rpict{couplerNonlin}(a). Such stabilising effect can enable design of laser sources, similar to the case of actively coupled PT waveguides with fixed parameters discussed in Sec.~\rsect[PT-couplers]{PT-couplers-active}, however periodic modulation can be realized in a simple way whily allowing fine control over the emerging nonlinear dynamics, which can be periodic or aperiodic. Indeed, chaotic evolution was predicted for a coupler (dimer) with periodically modulated gain and loss~\cite{Battelli:2015-353:NLD}. This is illustrated in Fig.~\rpict{couplerNonlin}(b), where the phase plane is presented for variables $(\psi, \dot{\psi})$, with $\dot{\psi})(z) = |a_2(z)|^2 - |a_1(z)|^2$.
Periodic modulation can be further applied to multiple coupled nonlinear waveguides, or oligomers~\cite{Horne:2013-485101:JPA, Martinez:2015-23822:PRA}. Examples of stable periodic and unstable dynamics in a nonlinear trimer with periodically modulated gain and loss are show in Figs.~\rpict{couplerNonlin}(c) and~(d), respectively.  Furthermore, periodic modulation can stabilize temporal solitons in coupled waveguides, in the supersymmetric regime when loss and gain are at the linear PT breaking threshold~\cite{Driben:2011-51001:EPL}.

\pict{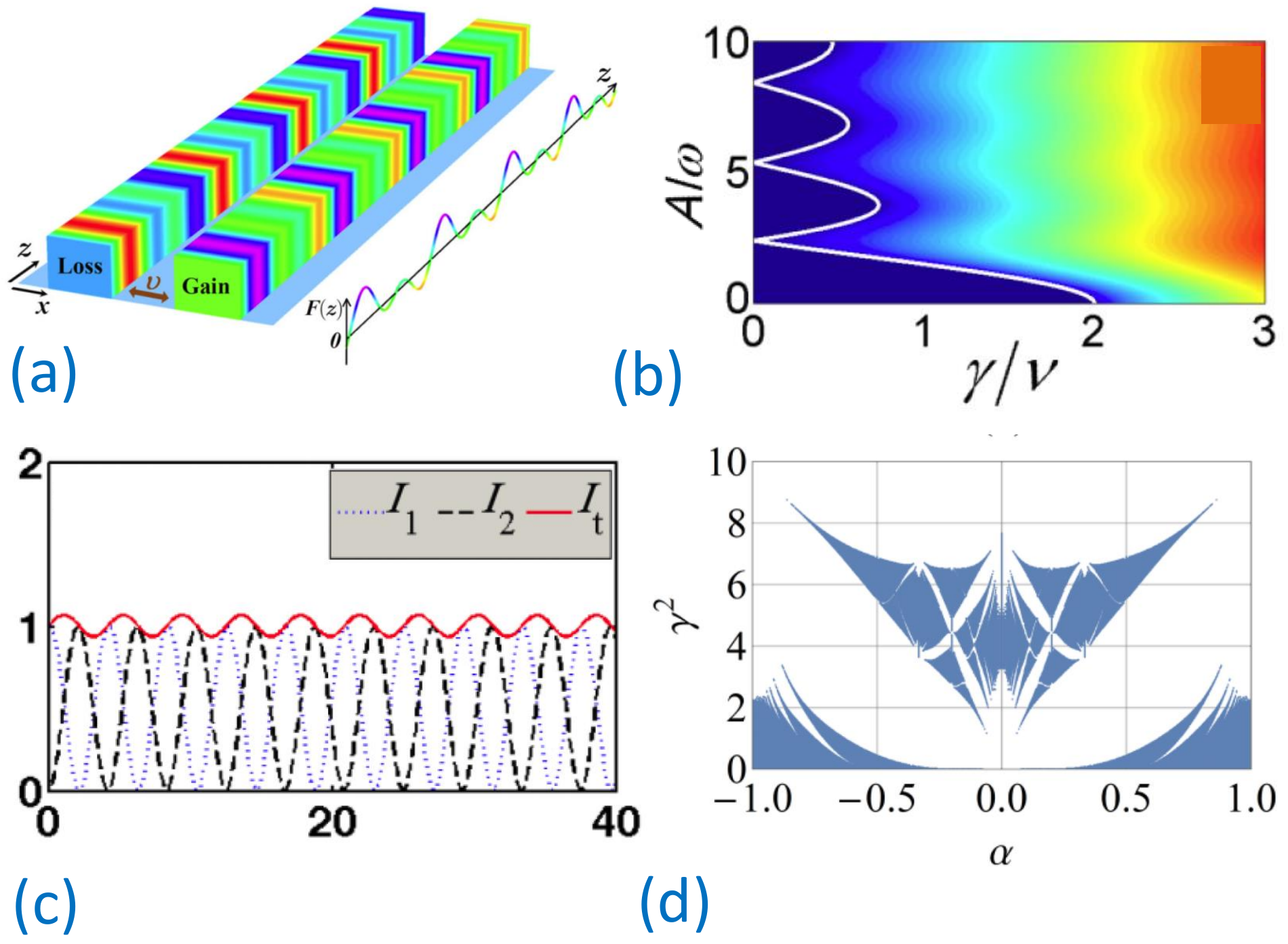}{couplerAperiodic}{
Quasi-PT symmetry in linear structures.
% with aperiodic modulation of gain and loss.
(a-c)~Adopted from Ref.~\cite{Luo:2013-243902:PRL}.
(a)~Schematic of a coupler with symmetric gain and loss, but with asymmetric  modulation of the propagation constants.
(b)~The mode growth rate vs. the normalized gain/loss amplitude (horizontal axis) and the modulation strength (vertical axis), for fast modulations. White line marks the boundary of quasi-PT transition.
(c)~Example of almost periodic evolution with very small growth rate in quasi-PT symmetric regime.
(d)~Phase diagram analogous to Hofstadter’s butterfly spectrum for a PT-symmetric Hamiltonian with aperiodic driving, adopted from Ref.~\cite{Gong:2015-42135:PRA}.
}
%----------------------

When system parameters are modulated such that underlying system is not PT-symmetric at any modulation cross-section, the overall dynamics can still demonstrate the features of
%, for example with biharmonic pattern, then such systems can demonstrate
pseudo-PT symmetry~\cite{Luo:2013-243902:PRL}.
The concept of pseudo-PT symmetry corresponds to the PT-symmetry in the effective system, which manifests as quasi-stationary propagation.
%, while the original system is non-PT-symmetric.
Such regime can be achieved
%
%In Ref.~\cite{Luo:2013-243902:PRL}, it has been demonstrated that whether or not the original system is PT-symmetric, by applying periodic modulations, the effective system obtained from the high-frequency Floquet method may be PT-symmetric.
under high-frequency periodic modulation, when the modulated system can be mapped into an effectively unmodulated one with rescaled parameters.
%.
%The effective system may show PT-symmetry if it could be
%described by a PT-symmetric Hamiltonian.
%Under the condition $\nu \ll \max[\omega,\sqrt{|A|\omega}]$, one can use the high-frequency Floquet analysis.
%By averaging the high-frequency terms, averaged equations corresponding to a stationary PT coupler are obtained.
Following Ref.~\cite{Luo:2013-243902:PRL}, consider a coupler described by Eq.~\reqt{Coupled-Eq} with balanced gain and loss, $\gamma_1(z) = \gamma/2$ and $\gamma_2(z) = - \gamma/2$, fixed coupling $C(z) = - \nu$, and modulated propagation constants $\beta_1(z) = - S(z) / 2$, $\beta_2(z) = S(z) / 2$. A biharmonic modulation was analyzed, $S(z) = -A [ \sin(\omega z) + f \sin(2 \omega z + \phi)]$. For $\phi \ne 0,\pi$, the corresponding Hamiltonian is not PT-symmetric. Nevertheless, the averaged system obtained after averaging the high-frequency modulations features PT symmetry, and the quasi-energies for such effective system are found as
\begin{equation}\label{Quasienergies}
    \epsilon=\pm|J|\sqrt{1-[\gamma/(2J)]^2},
\end{equation}
with the rescaled coupling strength
\begin{equation}\label{JJ}
    J=\nu \sum^{\infty}_{m=-\infty} (i)^{-m}J_{-2m}\left(\frac{A}{\omega}\right)J_m\left(\frac{Af}{2\omega}\right)\exp(im\phi).
\end{equation}
The two quasi-energies are real if $\gamma<2|J|$, and they are complex if $\gamma>2|J|$ with the critical point $\gamma_{c}=2|J|$.
There exists a spontaneous PT-symmetry-breaking transition in the effective system when the imaginary part $Im(\epsilon)$ changes from zero to nonzero, as illustrated in Fig.~\rpict{couplerAperiodic}(b). It appears surprising that the quasi-energies can be real even when the modulated system~\reqt{Coupled-Eq} is non-PT-symmetric, however the numerical analysis reveals that the eigenvalues of the original modulated structure have a very small but nonzero imaginary part. Accordingly, at intermediate propagation distances the dynamics can be almost exactly periodic in the quasi-PT symmetric regime, as show in Fig.~\rpict{couplerAperiodic}(c).

Further interesting possibilities can arise in case of aperiodic driving. It was found that a phase diagram analogous to Hofstadter’s butterfly spectrum can emerge in PT Hamiltonians~\cite{Gong:2015-42135:PRA}, see Fig.~\rpict{couplerAperiodic}(d). The horizontal axis in the plot corresponds to the ration of two driving frequencies, and the vertical axis characterizes detuning between the modes.

\subsection{Modulated lattices}

%----------------------
\pict{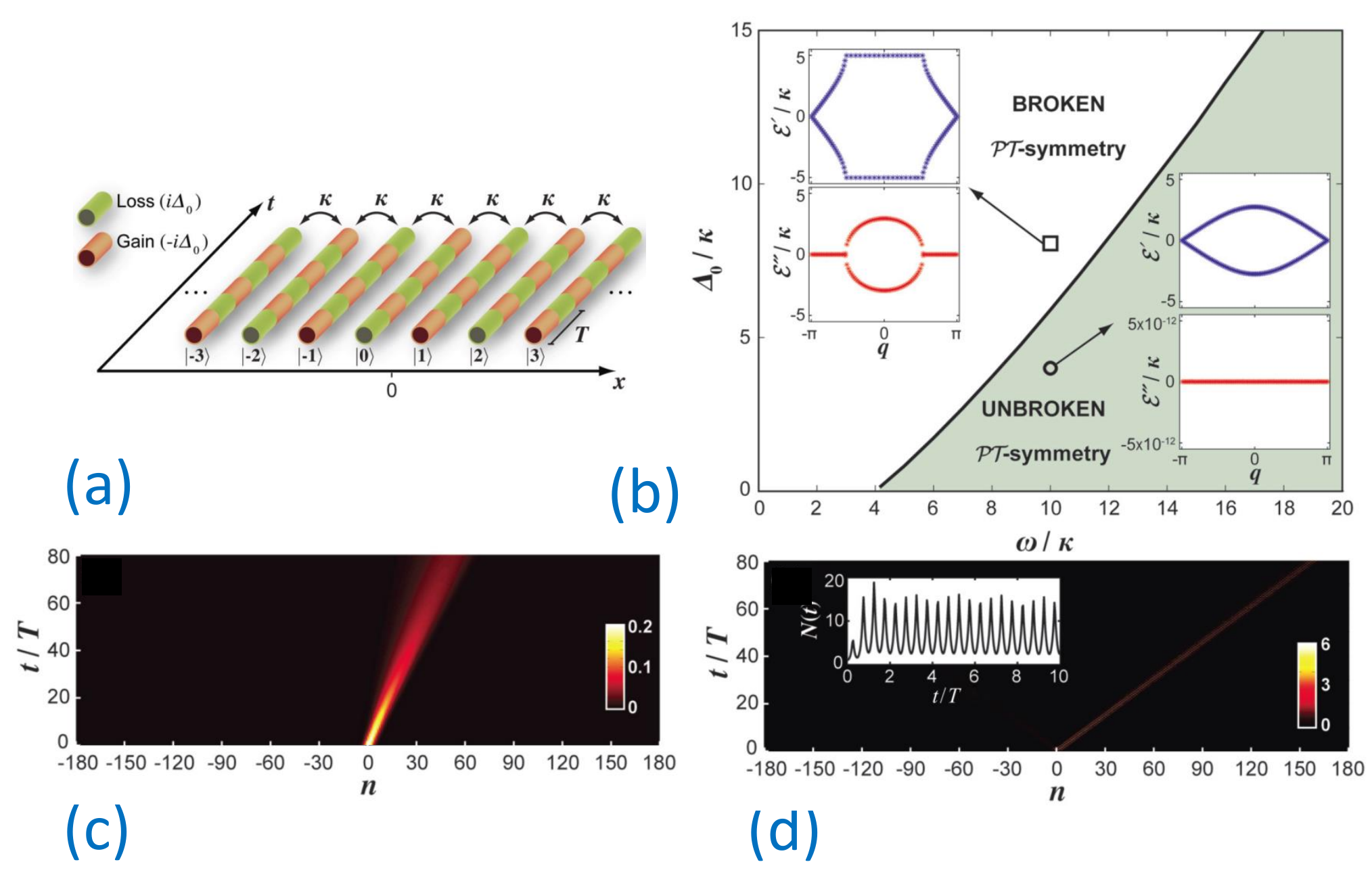}{array}{
(a)~Schematic of a waveguide array with modulated gain and loss according to Eq.~\reqt{array}.
(b)~Phase diagram of PT mode symmetry. The inset shows the real (upper panel) and imaginary part (lower panel) of the mode quasi-energies.
(c)~Beam intensity profiles for (c)~conventional propagation across the lattice in a conservative array and (b)~hyper-ballistic regime in a modulated lattice close to PT threshold.
Adopted from Ref.~\cite{DellaValle:2013-22119:PRA}.
}
%----------------------

Arrays of coupled modulated waveguides with gain and loss open new opportunities to control the rate of beam propagation and spreading through the lattice~\cite{DellaValle:2013-22119:PRA}. Detailed theoretical analysis was performed for a lattice with periodic modulation of gain and loss, which is out-of-phase between the neighbouring waveguides as illustrated in Fig.~\rpict{array}(a). Linear beam evolution in such arrays is described by the coupled-mode equations,
\begin{equation} \leqt{array}
   i \frac{d a_n}{d z} = - \kappa (a_{n-1} + a_{n+1}) + i (-1)^n \Delta(t) a_n,
\end{equation}
where $n$ is the waveguide number, $a_n$ are the mode amplitudes, $t$ is the propagation coordinate, $\kappa$ is the coupling between the neighbouring waveguides, and $\Delta(t) = \Delta_0 {\rm square}(\omega t + \theta)$ is a square modulation profile of gain and loss.
Solutions of this system can be found analytically in Fourier space, characterized by the transverse wavenumber $q$. For a range of modulation parameters indicated with shading in Fig.~\rpict{array}(b), the eigenmodes with all $q$ have unbroken PT symmetry. Outside of this parameter region, some modes have broken PT symmetry. Characteristic dispersion dependencies corresponding to these two regimes are shown as insets in Fig.~\rpict{array}(b). The imaginary part at the lower insets represents the mode gain or loss, and we see that it is zero for all modes in the unbroken PT regime. The real part shown in the upper insets defines the propagation angle (proportional to the first derivative) and strength of diffraction (proportional to the second derivative). It is seen that these propagation characteristics can be very strongly affected due to PT transition. A comparison of beam propagation in a conservative array and array with gain/loss modulation parameters chosen close to PT threshold are presented in Figs.~\rpict{array}(c) and~(d), respectively. The input beam is inclined such that the phase gradient between the neighbouring waveguides is $\pi/4$, which corresponds to the fastest propagation across a conventional conservative lattice, limited by the coupling rate $\kappa$ [Fig.~\rpict{array}(c)]. However in a PT lattice with modulated gain and loss, hyper-ballistic transport can be realised with much higher propagation velocity [Fig.~\rpict{array}(c)]. Additionally, the beam spreading due to diffraction is strongly reduced.

%----------------------
\pict{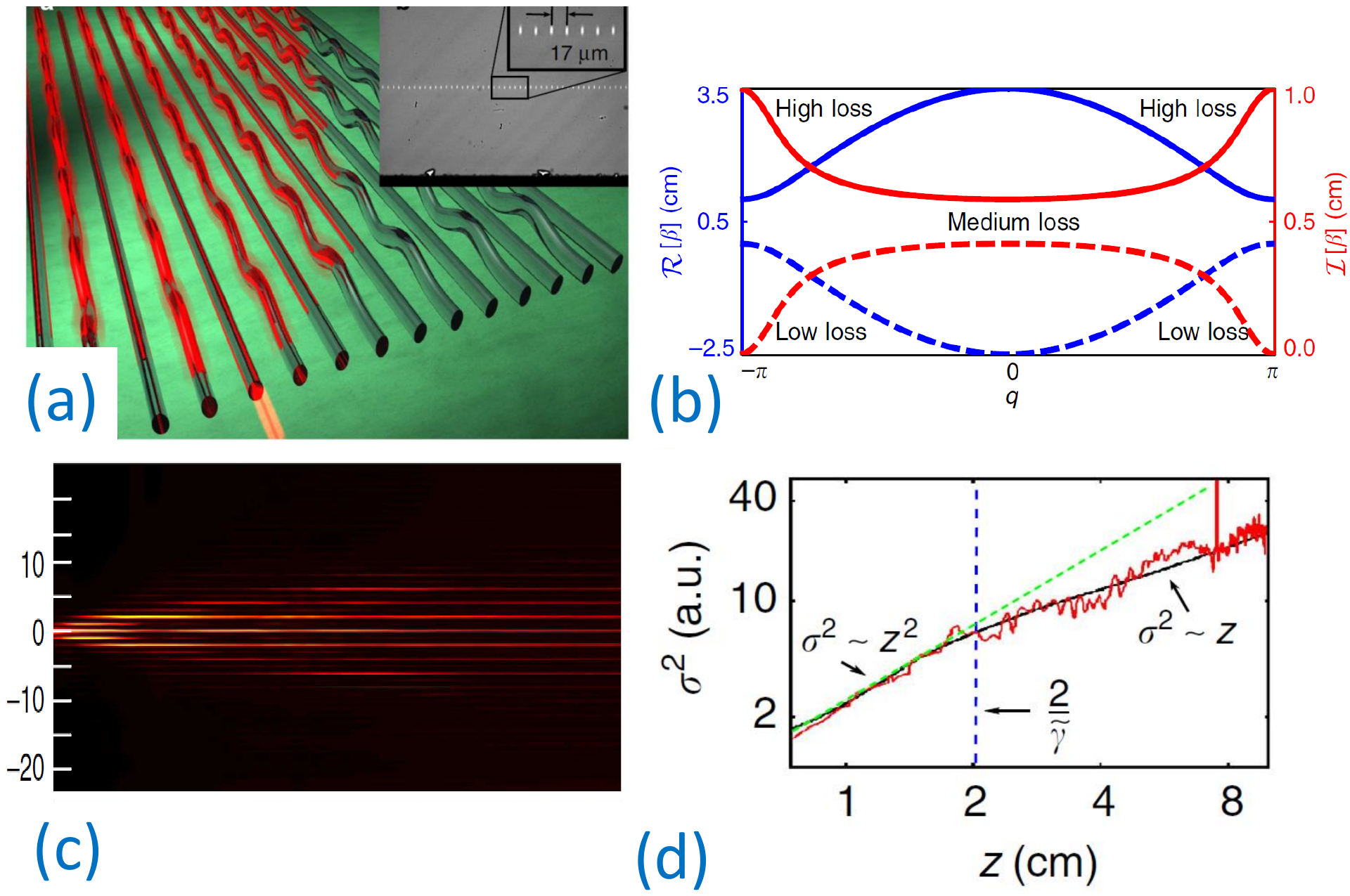}{arrayExperim}{
(a)~Schematic of a PT waveguide array with bending loss in every second waveguide.
(b)~The eigenvalue spectrum vs. the transverse lattice wavenumber $q$.
(c)~Experimental fluorescence microscopy image of the light beam as propagating through the lattice.
(d)~The variance of the beam vs. the propagation distance $z$ (red line, experimental data; black line, simulation).
Adopted from Ref.~\cite{Eichelkraut:2013-2533:NCOM}.
}
%----------------------

Experimental realisation of PT lattices based on fs laser-written modulated conservative waveguides was reported in Ref.~\cite{Eichelkraut:2013-2533:NCOM}. As shown in Fig.~\rpict{arrayExperim}, every second waveguide was periodically curved, which resulted in bending losses. In the regime of high-frequency modulation, the loss can be considered as effectively constant along the waveguides. The corresponding spectrum of eigenmodes is presented in Fig.~\rpict{arrayExperim}(b). Eigenmodes in different regions of the spectrum will experience different losses, depending on the value of the imaginary part of their propagation constant. As a consequence, in both bands modes in the centre of the spectrum (where $q \simeq 0$) experience an intermediate loss, which is close to the average loss in the system. At the edge of the spectrum (around $q \simeq \pi$), the modes in the upper band suffer from a loss that is much higher than the systems’s average loss, whereas in the lower band the modes experience much less loss. Hence, because of decay, the modes in the upper band at the edge of the spectrum will disappear after a
relatively short propagation distance z, whereas the modes in the centre of the spectrum will disappear somewhat later. Therefore, at long propagation distances. the spectrum will be getting narrower, and only a part of the spectrum will contribute to transport. Experimental image of the beam propagation is presented in Fig.~\rpict{arrayExperim}(c), and Fig.~\rpict{arrayExperim}(d) presents the beam variance characterising its width, $\sigma^2(z) = \sum_n n^2 |a_n|^2 / \sum_n |a_n|^2$. It is observed that initially, the wave exhibits ballistic spreading with $\sigma^2 \sim z^2$, as all eigenmodes still contribute to the transverse. transport. However after $z_{\rm crit} \simeq 2 cm$, transition from ballistic to diffusive transport is visible, with slower beam spreading as $\sigma^2 \sim z$.
Such coexistence of ballistic and diffusive transport in a static, ordered system, could only happen due to non-Hermitically. More recently, topological phase transition was reported in such waveguide arrays~\cite{Zeuner:2015-40402:PRL}.

The unconventional properties of modulated PT lattices based on waveguide arrays suggest that such systems would demonstrate novel features in the nonlinear regime, and we anticipate theoretical and experimental studies targeting these topics in the near future. In particulatr, due to the flexibility of waveguide array writing with fs lasers~\cite{Meany:2015-363:LPR}, this platform could enable the observation of linear and nonlinear effects in modulated PT structures.

%have been investigated theoretically. Spectral and dynamical properties of time-periodic linear PT-symmetric tight-binding lattices were discussed in Ref.~\cite{DellaValle:2013-22119:PRA}, while nonlinear effects were Theoretical studies included PT-symmetric models with supersymmetry by periodic management~\cite{Driben:2011-51001:EPL}, and the

\subsection{Mesh lattices}

\pict{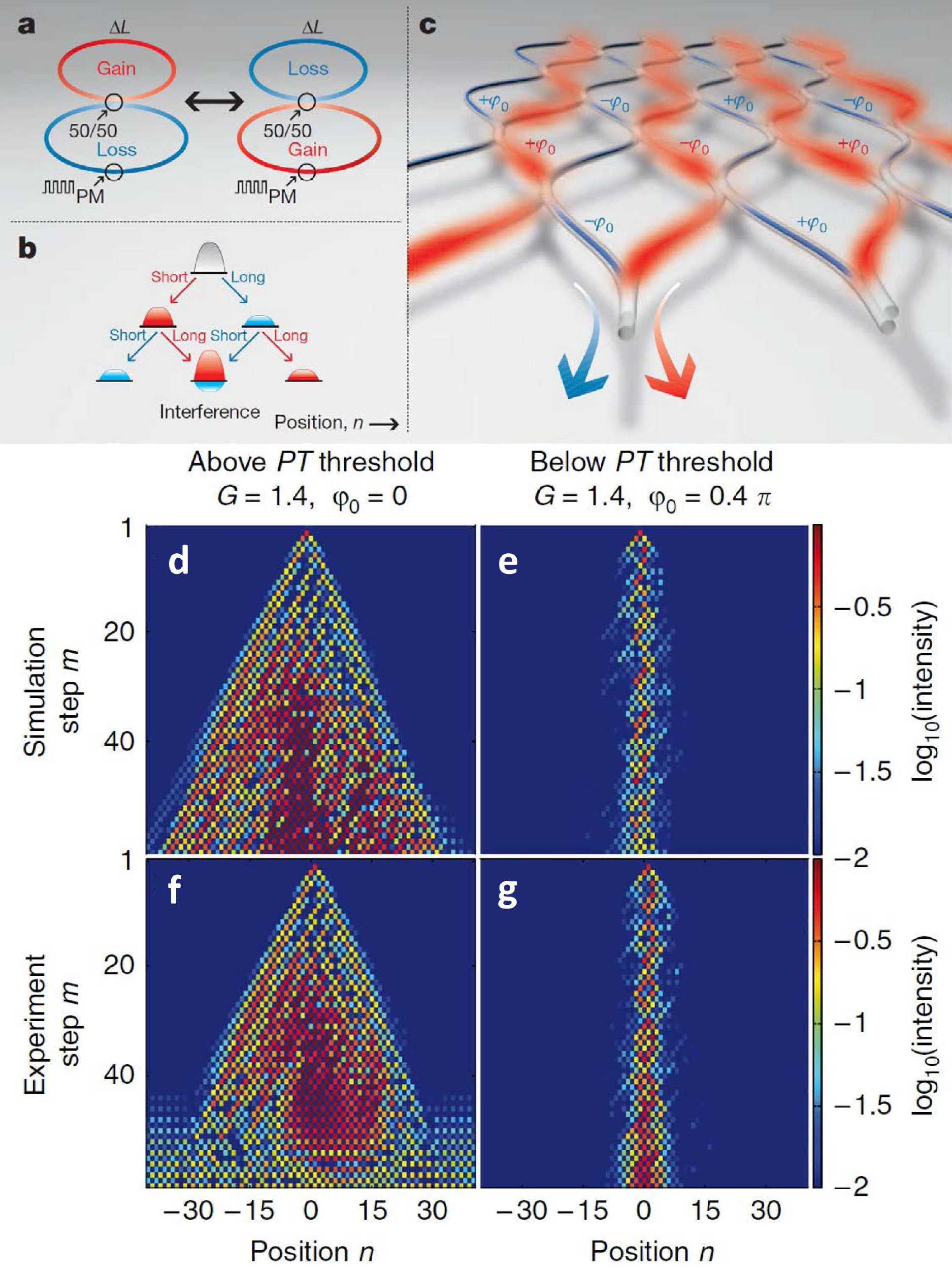}{Fig-sec6-F1}{
(a) Schematic diagram of the temporal fibre networks. Two coupled fibre loops periodically switching between gain and loss as used in the experiment. Pulses are delayed or advanced owing to a length difference $\Delta L$ between the short and long loops. (b) The evolution of the pulses in the temporal networks. Passages through the short and long loops are indicated. (c) The equivalent PT-symmetric spatial mesh lattices. Gain (red) and loss (blue) channels are arranged anti-symmetrically and are periodically coupled. The real part of the potential is introduced by using phase modulation with $\pm \varphi_0$. From Ref.~\cite{Regensburger:2012-167:NAT}.
(d) and (f): Simulation and experiment results in the nonlinear regime above PT threshold. (e) and (g): Simulation and experiment results in the nonlinear regime below PT threshold. Optical soliton and internal oscillations are observed in the experiment. From Ref.~\cite{Wimmer:2015-7782:NCOM}.
}

Regensburger {\em et al.}~\cite{Regensburger:2012-167:NAT} demonstrated the first experimental realization of a large-scale PT-symmetric lattice, which was a new kind of optical PT synthetic devices. They reported the experimental observation of light transport in large-scale temporal lattices which are parity-time symmetric. In addition, the periodic structures respecting PT symmetry can act as unidirectional invisible media when operated near their exceptional points.

To realize the specially artificial structures and study the light transportation, they have used an elegant experimental arrangement that operates in the temporal domain, see Fig.~\rpict{Fig-sec6-F1}~(a) and (b).
In most cases, the refractive index is a complex quantity. The real part of the refractive index affects the velocity of the propagating light, while the imaginary part can lead to the amplification or absorbtion of light within a material.
By temporally modulating the real and imaginary part of the refractive index, they were able to realize the PT-symmetric optical network in the temporal domain, whose system is analogous to a spatially periodic mesh network with gain and loss.
Each node of the spatial network corresponds to a specific 'time slot' of the temporal lattice, see Fig.~\rpict{Fig-sec6-F1}~(c).
A sequence of light pulses are injected into two connected optical fiber loops that are designed to exhibit PT symmetry. The anti-symmetric imaginary part of refractive index profile is attained by alternating gain and loss in the two loops via using optical amplifiers and amplitude modulators. While the even, real component of the index profile is introduced by using phase modulators.

The observation of the unusual unidirectional invisibility in their system is also interesting.
Since the left-right symmetry is broken in this structure and propagation is no longer invariant when gain and loss are exchanged in time.
More specifically, light propagating in such a system can experience reduced or enhanced reflections dependent on the direction of propagation while it has little effect on the transmitted light. The system can become totally invisible when light traverses it from one side, whereas it can still be seen when it is illuminated from the other.

In Ref.~\cite{Miri:2012-23807:PRA}, they investigated a class of optical mesh periodic structures that are discretized in both the transverse and longitudinal directions, which is equivalent to the above ones in the temporal domain.
The mesh arrangements are composed of an array of waveguides with each one being discretely and periodically coupled to its adjacent neighbors. The practical appealing of this type of lattice is the physical separation between the coupling and amplification within each building block. And the band structure of these systems have been systematically analyzed and their dispersion relation is analytically obtained. And the optical dynamics including the unidirectional invisibility in these PT-symmetric mesh lattices are also examined.

By introducing nonlinearity, the observation of stable optical discrete solitons in the PT-symmetric mesh lattices system has been experimentally demonstrated~\cite{Wimmer:2015-7782:NCOM}. The structure of lattices is the same as shown in Fig.~\rpict{Fig-sec6-F1}~(c), which is globally PT-symmetric. In the linear regime, when a single position is excited, the wavepacket spreading experiences a great amplification in the broken PT-symmetry regime and remains neutral below the PT threshold. Interestingly, when the power of single initial pulse is raised, the nonlinearity comes into play and the discrete solitons may appear in the system. Above the PT threshold, the nonlinearity cannot lead to stable solitary waves and the wavepacket still spreads exponentially, see Fig.~\rpict{Fig-sec6-F1}~(d) and (f). However, below the PT threshold, the nonlinearity can compensate the spreading and the optical soliton appear, see Fig.~\rpict{Fig-sec6-F1}~(e) and (g). It has been demonstrated that these solitons belong to a continuous family of solitons which is not typically the case for dissipative solitary waves and this class of discrete solitons can avoid instabilities. Based on such mesh lattices system, the behavior of nonlinear waves provides a possibility for realizing saturable absorbers which are widely used in Q-switched laser cavities and ultra-short optical pulse arrangements.

These results, obtained for both temporal and spatial modulated systems, represent an important step towards applications of the parity-time symmetry concept to a new generation of multi-functional optical devices and networks.

{\protect\renewcommand\sectpath{Conclusion}
%-------------------------------------------
\section{Conclusion and outlook} \lsect{}
%-------------------------------------------

Future technologies will demand a substantial increase in a density of photonic integration and energy efficiency far surpassing that of bulk optical components and modern silicon photonics. Such advances can be achieved only by embedding the photonic functionalities at material’s level taking advantages of subwavelength nanophotonics and metadevices. Many metadevices based on plasmonic technologies would require an engineered loss compensation that, as we expect, can be achieved in a clever way through the concepts of non-Hermitian dynamics and the PT symmetric photonic systems. It is believed that the PT symmetric photonic systems can offer promising solutions of a range of important problems providing an efficient and smart loss compensation. Photonic PT systems feature alternating and symmetrically distributed regions of gain and loss, and in such systems losses are as important as gain so that, depending on the structural parameters, gain can either compensate losses or amplify optical pulses. In addition, tunable photonic PT-symmetric systems can add many new exciting functionalities being potentially able to find even broader applications beyond the field of photonics, e.g. in the field of Bose-Einstein condensates~\cite{Li:2013-13604:PRA}. Also, the soliton scattering on PT-symmetric defects has been studied for otehr types of models, for example for the Klein-Gordon system~\cite{Kevrekidis:2014-10102:PRA, Saadatmand:2014-52902:PRE}.

As an example of the most recent developments we mention a very general study of constant-intensity waves in non-Hermitian potentials. It is well known that any finite-amplitude wave propagating in a nonlinear system undergoes modulational instability in any Hermitian potential with self-focusing nonlinearity. Markis {\em et al.}~\cite{Makris:2015-7257:NCOM} demonstrated that this fundamental restriction is conveniently lifted when working with non-Hermitian potentials. In particular, they presented a whole class of waves that have constant intensity in the presence of linear and nonlinear inhomogeneous media with gain and loss. These results suggest new directions for the experiments on nonlinear non-Hermitian scattering.

Many recent developments on non-Hermitian and PT symmetric systems occur beyond the field of photonics, and one of the examples is exciton-polariton condensate which is composed of hybrid light-matter quasiparticles formed by strongly interacting photons and excitons (electron-hole pairs) in semiconductor microcavities. The exciton-polaritons always exist in a balanced potential landscape of gain and loss and, as was shown recently~\cite{Gao:1504.00978:ARXIV}, the non-Hermitian nature of the system can modify dramatically the structure of modes and spectral degeneracies in exciton-polariton systems, and, therefore, will affect their quantum transport, localisation, and dynamical properties. These findings pave the way for studies of non-Hermitian quantum dynamics of exciton-polaritons, which can uncover novel operating principles for polariton-based devices.

As discussed above, PT symmetric systems demonstrate many nontrivial non-conservative wave interaction and phase transitions, which can be employed for signal filtering and switching, opening new prospects for active control of light. In this review, we have discussed only some problems involving nonlinear PT-symmetric photonic systems with an intensity-depend refractive index. Nonlinearity in such PT symmetric systems provides a basis for many effects such as formation of localized modes, nonlinearly induced PT-symmetry breaking, and all-optical switching of signals. Nonlinear PT-symmetric systems can serve as powerful building blocks for the development of novel photonic devices targeting an active light control.

As follows from the results summarized above, the study of photonic systems with the PT symmetry is an active research area which is expanding very rapidly leaving the boundaries outlined by the photonic systems and penetrating to other fields such as the physics of Bose-Einstein condensates.  However, the photonics is a cornerstone of those research due to rapidly expanding experimental demonstrations. Here, we have reviewed only a few of many intriguing properties and applications of PT-symmetric photonics systems being concentrated on the less studied nonlinear properties, but the field is quickly growing and many new ideas and demonstrations appear almost every week.

\begin{acknowledgement}
S.V.S., A.A.S., and Y.S.K. were supported by the Australian Research Council, including Discovery Project DP130100135 and Future Fellowship FT100100160.
S.V.S. and S.V.D. acknowledge financial support from the Russian Foundation
for Basic Research, grant No$15-31-20037$ mol\_a, and the D.I. Mendeleev fund of the Tomsk State University. C. Lee and J. Huang are supported by the National Basic Research Program of China (NBRPC) under Grant No. 2012CB821305 and the National Natural Science Foundation of China (NNSFC) under Grants No. 11374375.
\end{acknowledgement}

\begin{biographies}
  \authorbox{cvSuchkov}{Sergey V. Suchkov}{graduated from Ufa State Aviation Technical University, Ufa, Russia in 2009. After his graduation he worked as an engineer researcher at the Institute for Metals Superplasticity problems of RAS, Ufa. In 2014 he joined the Nonlinear Physics Centre at Australian National University where he is currently carrying out his research as a PhD student. The area of interests includes optical microresonators, nonlinear \PT-symmetric systems, discrete nonlinear models.}
  \authorbox{cvSukhorukov}{Andrey A. Sukhorukov}{graduated from the Physics Faculty of the Lomonosov Moscow State University, Russia and received Ph.D. degree in physics from the Australian National University (ANU) in 2002. He currently leads Nonlinear and Quantum Optics Group at the Nonlinear Physics Centre, ANU. His research interests include nonlinear optical switching, frequency conversion, generation of non-classical states of light, and opto-mechanical interactions in waveguides and nano-structures. He received Queen Elizabeth II and Future Felowships from the Australian Research Council and Humboldt Research Fellowship.
  %He has published 5 book chapters and 160 journal articles. His current stay in Germany is supported through Humboldt Research Fellowship.
  }
  \authorbox{cvHuang}{Jiahao Huang}{graduated from the School of Physics and Engineering, Sun Yat-Sen University (Guangzhou, China). Currently he is a PhD student at the same school. During his PhD, he performed several theoretical studies in quantum metrology and quantum simulation. His research interests include quantum-optical analog in optical waveguides and quantum metrology with ultracold atoms.}
  \authorbox{cvDmitriev}{Sergey V. Dmitriev}{was born in 1961 in Tomsk, Russia. S.V. Dmitriev graduated from Tomsk State University with a M.S. degree in 1984. Completed the PhD degree in structural mechanics at Tver State University, Russia, in 1988. Gained the second PhD degree in computer modeling of incommensurate phases in dielectrics from the University of Electro-Communications, Tokyo, Japan, in 1999. Received a Habilitation in 2008 with the thesis on solitary wave dynamics in discrete systems from the Altai State Technical University, Barnaul, Russia. Since 2008, S.V. Dmitriev has been serving as the Head of Laboratory of the Institute for Metals Superplasticity Problems, Ufa, Russia. His scientific interests are in solitary waves in discrete systems; discrete breathers; evolution of defect structures in solids during plastic deformation; mechanical properties of graphene; atomistic simulations}
  \authorbox{cvLee}{Chaohong Lee}{received a PhD degree in atomic and molecular physics in 2003 from the Wuhan Institute of Physics and Mathematics, Chinese Academy of Sciences (Wuhan, China). From 2003 to 2009 he worked in the Max Planck Institute for the Physics of Complex Systems (Dresden, Genmany) and the Australian National University (Canberra, Australia). In 2009, he joined the Sun Yat-Sen University (Guangzhou, China) as a full professor and established a research group on cold atomic physics and quantum technologies. His research activities span over several branches of quantum atomic gases and quantum photonics, including nonlinear matter-waves, macroscopic quantum phenomena, many-body-quantum physics, non-equilibrium quantum dynamics, quantum metrology, and quantum simulation.}
  \authorbox{cvKivshar}{Yuri S. Kivshar}{received a PhD degree in theoretical physics in 1984 from the Institute for Low Temperature Physics and Engineering (Kharkov, Ukraine). From 1988 to 1993 he worked at different research centers in USA, France, Spain, and Germany, and in 1993 he moved to the Australian National University (ANU) where he founded Nonlinear Physics Center being currently its Head and ANU Distinguished Professor. His research interests include nonlinear photonics, optical solitons, nanophotonics, and metamaterials. He is Fellow of the Australian Academy of Science, the Optical Society of America, the American Physical Society, the Institute of Physics (UK), Deputy Director of the Center of Excellence for Ultrahigh-bandwidth Devices for Optical Systems CUDOS (Australia) and Research Director of Metamaterial Laboratory (Russia). He received many prestigious international awards including the Lyle Medal of the Australian Academy of Science (Australia), the State Prize in Science and Technology (Ukraine), the Lebedev Medal of the International Rozhdestvensky Society (Russia), and the Harrie Massey Medal of the Institute of Physics (UK).}
\end{biographies}

\providecommand{\WileyBibTextsc}{}
\let\textsc\WileyBibTextsc
\providecommand{\othercit}{}
\providecommand{\jr}[1]{#1}
\providecommand{\etal}{~et~al.}


\begin{thebibliography}{[100]}

\bibitem{Bender:1998-5243:PRL}% article
 \textsc{C.\,M. Bender} and  \textsc{S.~Boettcher},
Real spectra in non-hermitian hamiltonians having {PT} symmetry,
\hrefBib[ ]{http://dx.doi.org/10.1103/PhysRevLett.80.5243}{ \jr{Phys. Rev.
  Lett.} \textbf{80}, 5243--5246 (1998)}\hrefBibPDF[
  ]{PRL_1998_80_05243.pdf}{PDF}.


\bibitem{Bender:2002-270401:PRL}% article
 \textsc{C.\,M. Bender},  \textsc{D.\,C. Brody},  and  \textsc{H.\,F. Jones},
Complex extension of quantum mechanics,
\hrefBib[ ]{http://dx.doi.org/10.1103/PhysRevLett.89.270401}{ \jr{Phys. Rev.
  Lett.} \textbf{89}, 270401--4 (2002)}\hrefBibPDF[
  ]{PRL_2002_89_270401.pdf}{PDF}.


\bibitem{Bender:2007-947:RPP}% article
 \textsc{C.\,M. Bender},
Making sense of non-hermitian hamiltonians,
\hrefBib[ ]{http://dx.doi.org/10.1088/0034-4885/70/6/R03}{ \jr{Rep. Prog.
  Phys.} \textbf{70}, 947--1018 (2007)}\hrefBibPDF[
  ]{RPP_2007_70_00947.pdf}{PDF}.


\bibitem{Ruschhaupt:2005-L171:JPA}% article
 \textsc{A.~Ruschhaupt},  \textsc{F.~Delgado},  and  \textsc{J.\,G. Muga},
Physical realization of {PT}-symmetric potential scattering in a planar slab
  waveguide,
\hrefBib[ ]{http://dx.doi.org/10.1088/0305-4470/38/9/L03}{ \jr{J. Phys. A}
  \textbf{38}, L171--L176 (2005)}\hrefBibPDF[ ]{JPA_2005_38_L00171.pdf}{PDF}.


\bibitem{El-Ganainy:2007-2632:OL}% article
 \textsc{R.~{E}l {G}anainy},  \textsc{K.\,G. Makris},  \textsc{D.\,N.
  Christodoulides},  and  \textsc{Z.\,H. Musslimani},
Theory of coupled optical {PT}-symmetric structures,
\hrefBib[ ]{http://dx.doi.org/10.1364/OL.32.002632}{ \jr{Opt. Lett.}
  \textbf{32}, 2632--2634 (2007)}\hrefBibPDF[ ]{OL_2007_32_02632.pdf}{PDF}.


\bibitem{Ruter:2010-192:NPHYS}% article
 \textsc{C.\,E. Ruter},  \textsc{K.\,G. Makris},  \textsc{R.~{E}l {G}anainy},
  \textsc{D.\,N. Christodoulides},  \textsc{M.~Segev},  and  \textsc{D.~Kip},
Observation of parity-time symmetry in optics,
\hrefBib[ ]{http://dx.doi.org/10.1038/NPHYS1515}{ \jr{Nature Physics}
  \textbf{6}, 192--195 (2010)}\hrefBibPDF[ ]{NPHYS_2010_06_00192.pdf}{PDF}.


\bibitem{Guo:2009-93902:PRL}% article
 \textsc{A.~Guo},  \textsc{G.\,J. Salamo},  \textsc{D.~Duchesne},
  \textsc{R.~Morandotti},  \textsc{M.~{Volatier-Ravat}},  \textsc{V.~Aimez},
  \textsc{G.\,A. Siviloglou},  and  \textsc{D.\,N. Christodoulides},
Observation of {PT}-symmetry breaking in complex optical potentials,
\hrefBib[ ]{http://dx.doi.org/10.1103/PhysRevLett.103.093902}{ \jr{Phys. Rev.
  Lett.} \textbf{103}, 093902--4 (2009)}\hrefBibPDF[
  ]{PRL_2009_103_93902.pdf}{PDF}.


\bibitem{Feng:2014-972:SCI}% article
 \textsc{L.~Feng},  \textsc{Z.\,J. Wong},  \textsc{R.\,M. Ma},
  \textsc{Y.~Wang},  and  \textsc{X.~Zhang},
Single-mode laser by parity-time symmetry breaking,
\hrefBib[ ]{http://dx.doi.org/10.1126/science.1258179}{ \jr{Science}
  \textbf{346}, 972--975 (2014)}\hrefBibPDF[ ]{SCI_2014_346_00972.pdf}{PDF}.


\bibitem{Hodaei:2014-975:SCI}% article
 \textsc{H.~Hodaei},  \textsc{M.\,A. Miri},  \textsc{M.~Heinrich},
  \textsc{D.\,N. Christodoulides},  and  \textsc{M.~Khajavikhan},
Parity-time-symmetric microring lasers,
\hrefBib[ ]{http://dx.doi.org/10.1126/science.1258480}{ \jr{Science}
  \textbf{346}, 975--978 (2014)}\hrefBibPDF[ ]{SCI_2014_346_00975.pdf}{PDF}.


\bibitem{Zyablovsky:2014-1063:PUS}% article
 \textsc{A.\,A. Zyablovsky},  \textsc{A.\,P. Vinogradov},  \textsc{A.\,A.
  Pukhov},  \textsc{A.\,V. Dorofeenko},  and  \textsc{A.\,A. Lisyansky},
{PT}-symmetry in optics,
\hrefBib[ ]{http://dx.doi.org/10.3367/UFNr.0184.201411b.1177}{ \jr{Phys. Usp.}
  \textbf{57}, 1063--1082 (2014)}\hrefBibPDF[ ]{PUS_2014_57_01063.pdf}{PDF}.


\bibitem{Chen:1992-239:IQE}% article
 \textsc{Y.\,J. Chen},  \textsc{A.\,W. Snyder},  and  \textsc{D.\,N. Payne},
Twin core nonlinear couplers with gain and loss,
\hrefBib[ ]{http://dx.doi.org/10.1109/3.119519}{ \jr{IEEE J. Quantum Electron.}
  \textbf{28}, 239--245 (1992)}\hrefBibPDF[ ]{IQE_1992_28_00239.pdf}{PDF}.


\bibitem{Malomed:1996-330:OL}% article
 \textsc{B.\,A. Malomed},  \textsc{G.\,D. Peng},  and  \textsc{P.\,L. Chu},
Nonlinear-optical amplifier based on a dual-core fiber,
\hrefBib[ ]{http://dx.doi.org/10.1364/OL.21.000330}{ \jr{Opt. Lett.}
  \textbf{21}, 330--332 (1996)}\hrefBibPDF[ ]{OL_1996_21_00330.pdf}{PDF}.


\othercit
\bibitem{Kivshar:2003:OpticalSolitons}% book
 \textsc{Y.\,S. Kivshar} and  \textsc{G.\,P. Agrawal},
\hrefBib[ ]{http://www.sciencedirect.com/science/book/9780124105904}{{Optical
  Solitons: From Fibers to Photonic Crystals}} (Academic Press, San Diego,
  2003).


\othercit
\bibitem{Rosanov:2002:SpatialHysteresis}% book
 \textsc{N.\,N. Rosanov},
\hrefBib[ ]{dx.doi.org/10.1007/978-3-662-04792-7}{Spatial Hysteresis and
  Optical Patterns}, Springer Series in Synergetics (Springer, New York, 2002).


\othercit
\bibitem{Akhmediev:2005:DissipativeSolitons}% book
 \textsc{N.~Akhmediev} and  \textsc{A.~Ankiewicz} (eds.),
\hrefBib[ ]{http://dx.doi.org/10.1007/b11728}{Dissipative Solitons}, Lecture
  Notes in Physics (Springer, New York, 2005).


\bibitem{Ramezani:2010-43803:PRA}% article
 \textsc{H.~Ramezani},  \textsc{T.~Kottos},  \textsc{R.~{E}l {G}anainy},  and
  \textsc{D.\,N. Christodoulides},
Unidirectional nonlinear {PT}-symmetric optical structures,
\hrefBib[ ]{http://dx.doi.org/10.1103/PhysRevA.82.043803}{ \jr{Phys. Rev. A}
  \textbf{82}, 043803--6 (2010)}\hrefBibPDF[ ]{PRA_2010_82_43803.pdf}{PDF}.


\bibitem{Sukhorukov:2010-43818:PRA}% article
 \textsc{A.\,A. Sukhorukov},  \textsc{Z.\,Y. Xu},  and  \textsc{Y.\,S.
  Kivshar},
Nonlinear suppression of time reversals in {PT}-symmetric optical couplers,
\hrefBib[ ]{http://dx.doi.org/10.1103/PhysRevA.82.043818}{ \jr{Phys. Rev. A}
  \textbf{82}, 043818--5 (2010)}\hrefBibPDF[ ]{PRA_2010_82_43818.pdf}{PDF}.


\bibitem{Kevrekidis:2013-365201:JPA}% article
 \textsc{P.\,G. Kevrekidis},  \textsc{D.\,E. Pelinovsky},  and  \textsc{D.\,Y.
  Tyugin},
Nonlinear dynamics in {PT}-symmetric lattices,
\hrefBib[ ]{http://dx.doi.org/10.1088/1751-8113/46/36/365201}{ \jr{J. Phys. A}
  \textbf{46}, 365201--17 (2013)}\hrefBibPDF[ ]{JPA_2013_46_365201.pdf}{PDF}.


\bibitem{Barashenkov:2014-45802:PRA}% article
 \textsc{I.\,V. Barashenkov},
Hamiltonian formulation of the standard {PT}-symmetric nonlinear
  {S}chr\"odinger dimer,
\hrefBib[ ]{http://dx.doi.org/10.1103/PhysRevA.90.045802}{ \jr{Phys. Rev. A}
  \textbf{90}, 045802--4 (2014)}\hrefBibPDF[ ]{PRA_2014_90_45802.pdf}{PDF}.


\bibitem{Barashenkov:2013-53817:PRA}% article
 \textsc{I.\,V. Barashenkov},  \textsc{G.\,S. Jackson},  and
  \textsc{S.~Flach},
Blow-up regimes in the {PT}-symmetric coupler and the actively coupled dimer,
\hrefBib[ ]{http://dx.doi.org/10.1103/PhysRevA.88.053817}{ \jr{Phys. Rev. A}
  \textbf{88}, 053817--8 (2013)}\hrefBibPDF[ ]{PRA_2013_88_53817.pdf}{PDF}.


\bibitem{Pickton:2013-63840:PRA}% article
 \textsc{J.~Pickton} and  \textsc{H.~Susanto},
Integrability of {PT}-symmetric dimers,
\hrefBib[ ]{http://dx.doi.org/10.1103/PhysRevA.88.063840}{ \jr{Phys. Rev. A}
  \textbf{88}, 063840--8 (2013)}\hrefBibPDF[ ]{PRA_2013_88_63840.pdf}{PDF}.


\bibitem{Barashenkov:2015-325201:JPA}% article
 \textsc{I.\,V. Barashenkov},  \textsc{D.\,E. Pelinovsky},  and
  \textsc{P.~Dubard},
Dimer with gain and loss: Integrability and {PT}-symmetry restoration,
\hrefBib[ ]{http://dx.doi.org/10.1088/1751-8113/48/32/325201}{ \jr{J. Phys. A}
  \textbf{48}, 325201--28 (2015)}\hrefBibPDF[ ]{JPA_2015_48_325201.pdf}{PDF}.


\bibitem{Lupu:2014-305:PNFA}% article
 \textsc{A.~Lupu},  \textsc{H.~Benisty},  and  \textsc{A.~Degiron},
Using optical {PT}-symmetry for switching applications,
\hrefBib[ ]{http://dx.doi.org/10.1016/j.photonics.2014.05.003}{ \jr{Photonics
  Nanostruct.: Fundam. Appl.} \textbf{12}, 305--311 (2014)}\hrefBibPDF[
  ]{PNFA_2014_12_00305.pdf}{PDF}.


\bibitem{Schindler:2011-40101:PRA}% article
 \textsc{J.~Schindler},  \textsc{A.~Li},  \textsc{M.\,C. Zheng},
  \textsc{F.\,M. Ellis},  and  \textsc{T.~Kottos},
Experimental study of active lrc circuits with {PT} symmetries,
\hrefBib[ ]{http://dx.doi.org/10.1103/PhysRevA.84.040101}{ \jr{Phys. Rev. A}
  \textbf{84}, 040101--5 (2011)}\hrefBibPDF[ ]{PRA_2011_84_40101.pdf}{PDF}.


\bibitem{Cuevas:2013-32108:PRA}% article
 \textsc{J.~Cuevas},  \textsc{P.\,G. Kevrekidis},  \textsc{A.~Saxena},  and
  \textsc{A.~Khare},
{PT}-symmetric dimer of coupled nonlinear oscillators,
\hrefBib[ ]{http://dx.doi.org/10.1103/PhysRevA.88.032108}{ \jr{Phys. Rev. A}
  \textbf{88}, 032108--11 (2013)}\hrefBibPDF[ ]{PRA_2013_88_32108.pdf}{PDF}.


\bibitem{Duanmu:2013-20120171:PTRSA}% article
 \textsc{M.~Duanmu},  \textsc{K.~Li},  \textsc{R.\,L. Horne},  \textsc{P.\,G.
  Kevrekidis},  and  \textsc{N.~Whitaker},
Linear and nonlinear parity-time-symmetric oligomers: a dynamical systems
  analysis,
\hrefBib[ ]{http://dx.doi.org/10.1098/rsta.2012.0171}{ \jr{Philos. Trans. R.
  Soc. A} \textbf{371}, 20120171--19 (2013)}\hrefBibPDF[
  ]{PTRSA_2013_371_20120171.pdf}{PDF}.


\bibitem{Miroshnichenko:2011-12123:PRA}% article
 \textsc{A.\,E. Miroshnichenko},  \textsc{B.\,A. Malomed},  and  \textsc{Y.\,S.
  Kivshar},
Nonlinearly {PT}-symmetric systems: Spontaneous symmetry breaking and
  transmission resonances,
\hrefBib[ ]{http://dx.doi.org/10.1103/PhysRevA.84.012123}{ \jr{Phys. Rev. A}
  \textbf{84}, 012123--4 (2011)}\hrefBibPDF[ ]{PRA_2011_84_12123.pdf}{PDF}.


\bibitem{Zezyulin:2011-64003:EPL}% article
 \textsc{D.\,A. Zezyulin},  \textsc{Y.\,V. Kartashov},  and  \textsc{V.\,V.
  Konotop},
Stability of solitons in {PT}-symmetric nonlinear potentials,
\hrefBib[ ]{http://dx.doi.org/10.1209/0295-5075/96/64003}{ \jr{Europhys. Lett.}
  \textbf{96}, 64003--6 (2011)}\hrefBibPDF[ ]{EPL_2011_96_64003.pdf}{PDF}.


\bibitem{Alexeeva:2014-13848:PRA}% article
 \textsc{N.\,V. Alexeeva},  \textsc{I.\,V. Barashenkov},  \textsc{K.~Rayanov},
  and  \textsc{S.~Flach},
Actively coupled optical waveguides,
\hrefBib[ ]{http://dx.doi.org/10.1103/PhysRevA.89.013848}{ \jr{Phys. Rev. A}
  \textbf{89}, 013848--5 (2014)}\hrefBibPDF[ ]{PRA_2014_89_13848.pdf}{PDF}.


\bibitem{Barashenkov:2014-282001:JPA}% article
 \textsc{I.\,V. Barashenkov} and  \textsc{M.~Gianfreda},
An exactly solvable {PT}-symmetric dimer from a hamiltonian system of nonlinear
  oscillators with gain and loss,
\hrefBib[ ]{http://dx.doi.org/10.1088/1751-8113/47/28/282001}{ \jr{J. Phys. A}
  \textbf{47}, 282001--18 (2014)}\hrefBibPDF[ ]{JPA_2014_47_282001.pdf}{PDF}.


\bibitem{Sarma:2014-52918:PRE}% article
 \textsc{A.\,K. Sarma},  \textsc{M.\,A. Miri},  \textsc{Z.\,H. Musslimani},
  and  \textsc{D.\,N. Christodoulides},
Continuous and discrete {S}chr\"odinger systems with parity-time-symmetric
  nonlinearities,
\hrefBib[ ]{http://dx.doi.org/10.1103/PhysRevE.89.052918}{ \jr{Phys. Rev. E}
  \textbf{89}, 052918--7 (2014)}\hrefBibPDF[ ]{PRE_2014_89_52918.pdf}{PDF}.


\bibitem{Moreira:2012-53815:PRA}% article
 \textsc{F.\,C. Moreira},  \textsc{F.\,K. Abdullaev},  \textsc{V.\,V. Konotop},
   and  \textsc{A.\,V. Yulin},
Localized modes in $\chi^{(2)}$ media with {PT}-symmetric localized potential,
\hrefBib[ ]{http://dx.doi.org/10.1103/PhysRevA.86.053815}{ \jr{Phys. Rev. A}
  \textbf{86}, 053815--7 (2012)}\hrefBibPDF[ ]{PRA_2012_86_53815.pdf}{PDF}.


\bibitem{Li:2013-53820:PRA}% article
 \textsc{K.~Li},  \textsc{D.\,A. Zezyulin},  \textsc{P.\,G. Kevrekidis},
  \textsc{V.\,V. Konotop},  and  \textsc{F.\,K. Abdullaev},
{PT}-symmetric coupler with $\chi^{(2)}$ nonlinearity,
\hrefBib[ ]{http://dx.doi.org/10.1103/PhysRevA.88.053820}{ \jr{Phys. Rev. A}
  \textbf{88}, 053820--11 (2013)}\hrefBibPDF[ ]{PRA_2013_88_53820.pdf}{PDF}.


\othercit
\bibitem{Boyd:2008:NonlinearOptics}% book
 \textsc{R.\,W. Boyd},
\hrefBib[ ]{http://www.sciencedirect.com/science/book/9780123694706}{{Nonlinear
  Optics}}, 3rd edition (Academic Press, San Diego, 2008).


\bibitem{Antonosyan:1506.02143:ARXIV}% article
 \textsc{D.\,A. Antonosyan},  \textsc{A.\,S. Solntsev},  and  \textsc{A.\,A.
  Sukhorukov},
Parity-time anti-symmetric parametric amplifier,
\hrefEmpty[ ]{Empty}{ \jr{arXiv} \textbf{\mdseries 1506.02143}
  (2015)}.


\bibitem{Tsoy:2012-3441:OC}% article
 \textsc{E.\,N. Tsoy},  \textsc{S.\,S. Tadjimuratov},  and  \textsc{F.\,K.
  Abdullaev},
Beam propagation in gain-loss balanced waveguides,
\hrefBib[ ]{http://dx.doi.org/10.1016/j.optcom.2012.03.027}{ \jr{Opt. Commun.}
  \textbf{285}, 3441--3444 (2012)}\hrefBibPDF[ ]{OC_2012_285_03441.pdf}{PDF}.


\bibitem{Cartarius:2012-444008:JPA}% article
 \textsc{H.~Cartarius},  \textsc{D.~Haag},  \textsc{D.~Dast},  and
  \textsc{G.~Wunner},
Nonlinear {S}chr\"odinger equation for a {PT}-symmetric delta-function double
  well,
\hrefBib[ ]{http://dx.doi.org/10.1088/1751-8113/45/44/444008}{ \jr{J. Phys. A}
  \textbf{45}, 444008--15 (2012)}\hrefBibPDF[ ]{JPA_2012_45_444008.pdf}{PDF}.


\bibitem{Rodrigues:2013-5:ROMRP}% article
 \textsc{A.\,S. Rodrigues},  \textsc{K.~Li},  \textsc{V.~Achilleos},
  \textsc{P.\,G. Kevrekidis},  \textsc{D.\,J. Frantzeskakis},  and
  \textsc{C.\,M. Bender},
{PT}-symmetric double-well potentials revisited: bifurcations, stability and
  dynamics,
\hrefBib[ ]{http://www.rrp.infim.ro/2013_65_1/art01Rodrigues.pdf}{ \jr{Rom.
  Rep. Phys.} \textbf{65}, 5--26 (2013)}\hrefBibPDF[
  ]{ROMRP_2013_65_00005.pdf}{PDF}.


\bibitem{Mayteevarunyoo:2013-22919:PRE}% article
 \textsc{T.~Mayteevarunyoo},  \textsc{B.\,A. Malomed},  and
  \textsc{A.~Reoksabutr},
Solvable model for solitons pinned to a parity-time-symmetric dipole,
\hrefBib[ ]{http://dx.doi.org/10.1103/PhysRevE.88.022919}{ \jr{Phys. Rev. E}
  \textbf{88}, 022919--11 (2013)}\hrefBibPDF[ ]{PRE_2013_88_22919.pdf}{PDF}.


\bibitem{Suchkov:2014-443:APA}% article
 \textsc{S.\,V. Suchkov},  \textsc{S.\,V. Dmitriev},  \textsc{A.\,A.
  Sukhorukov},  \textsc{I.\,V. Barashenkov},  \textsc{E.\,R. Andriyanova},
  \textsc{K.\,M. Badgetdinova},  and  \textsc{Y.\,S. Kivshar},
Phase sensitivity of light dynamics in {PT}-symmetric couplers,
\hrefBib[ ]{http://dx.doi.org/10.1007/s00339-013-8036-1}{ \jr{Appl. Phys. A}
  \textbf{115}, 443--447 (2014)}\hrefBibPDF[ ]{APA_2014_115_00443.pdf}{PDF}.


\bibitem{Zezyulin:2012-43840:PRA}% article
 \textsc{D.\,A. Zezyulin} and  \textsc{V.\,V. Konotop},
Nonlinear modes in the harmonic {PT}-symmetric potential,
\hrefBib[ ]{http://dx.doi.org/10.1103/PhysRevA.85.043840}{ \jr{Phys. Rev. A}
  \textbf{85}, 043840--6 (2012)}\hrefBibPDF[ ]{PRA_2012_85_43840.pdf}{PDF}.


\bibitem{Lin:2011-213901:PRL}% article
 \textsc{Z.~Lin},  \textsc{H.~Ramezani},  \textsc{T.~Eichelkraut},
  \textsc{T.~Kottos},  \textsc{H.~Cao},  and  \textsc{D.\,N. Christodoulides},
Unidirectional invisibility induced by {PT}-symmetric periodic structures,
\hrefBib[ ]{http://dx.doi.org/10.1103/PhysRevLett.106.213901}{ \jr{Phys. Rev.
  Lett.} \textbf{106}, 213901--4 (2011)}\hrefBibPDF[
  ]{PRL_2011_106_213901.pdf}{PDF}.


\bibitem{Li:2011-66608:PRE}% article
 \textsc{K.~Li} and  \textsc{P.\,G. Kevrekidis},
{PT}-symmetric oligomers: Analytical solutions, linear stability, and nonlinear
  dynamics,
\hrefBib[ ]{http://dx.doi.org/10.1103/PhysRevE.83.066608}{ \jr{Phys. Rev. E}
  \textbf{83}, 066608--7 (2011)}\hrefBibPDF[ ]{PRE_2011_83_66608.pdf}{PDF}.


\bibitem{DAmbroise:2012-444012:JPA}% article
 \textsc{J.~D'Ambroise},  \textsc{P.\,G. Kevrekidis},  and  \textsc{S.~Lepri},
Asymmetric wave propagation through nonlinear {PT}-symmetric oligomers,
\hrefBib[ ]{http://dx.doi.org/10.1088/1751-8113/45/44/444012}{ \jr{J. Phys. A}
  \textbf{45}, 444012--16 (2012)}\hrefBibPDF[ ]{JPA_2012_45_444012.pdf}{PDF}.


\bibitem{Zezyulin:2012-213906:PRL}% article
 \textsc{D.\,A. Zezyulin} and  \textsc{V.\,V. Konotop},
Nonlinear modes in finite-dimensional {PT}-symmetric systems,
\hrefBib[ ]{http://dx.doi.org/10.1103/PhysRevLett.108.213906}{ \jr{Phys. Rev.
  Lett.} \textbf{108}, 213906--5 (2012)}\hrefBibPDF[
  ]{PRL_2012_108_213906.pdf}{PDF}.


\bibitem{Li:2012-444021:JPA}% article
 \textsc{K.~Li},  \textsc{P.\,G. Kevrekidis},  \textsc{B.\,A. Malomed},  and
  \textsc{U.~Gunther},
Nonlinear {PT}-symmetric plaquettes,
\hrefBib[ ]{http://dx.doi.org/10.1088/1751-8113/45/44/444021}{ \jr{J. Phys. A}
  \textbf{45}, 444021--23 (2012)}\hrefBibPDF[ ]{JPA_2012_45_444021.pdf}{PDF}.


\bibitem{Bendix:2009-30402:PRL}% article
 \textsc{O.~Bendix},  \textsc{R.~Fleischmann},  \textsc{T.~Kottos},  and
  \textsc{B.~Shapiro},
Exponentially fragile {PT} symmetry in lattices with localized eigenmodes,
\hrefBib[ ]{http://dx.doi.org/10.1103/PhysRevLett.103.030402}{ \jr{Phys. Rev.
  Lett.} \textbf{103}, 030402--4 (2009)}\hrefBibPDF[
  ]{PRL_2009_103_30402.pdf}{PDF}.


\bibitem{Pelinovsky:2014-85204:JPA}% article
 \textsc{D.\,E. Pelinovsky},  \textsc{D.\,A. Zezyulin},  and  \textsc{V.\,V.
  Konotop},
Nonlinear modes in a generalized {PT}-symmetric discrete nonlinear
  {S}chr\"odinger equation,
\hrefBib[ ]{http://dx.doi.org/10.1088/1751-8113/47/8/085204}{ \jr{J. Phys. A}
  \textbf{47}, 085204--20 (2014)}\hrefBibPDF[ ]{JPA_2014_47_85204.pdf}{PDF}.


\bibitem{Barashenkov:2013-33819:PRA}% article
 \textsc{I.\,V. Barashenkov},  \textsc{L.~Baker},  and  \textsc{N.\,V.
  Alexeeva},
{PT}-symmetry breaking in a necklace of coupled optical waveguides,
\hrefBib[ ]{http://dx.doi.org/10.1103/PhysRevA.87.033819}{ \jr{Phys. Rev. A}
  \textbf{87}, 033819--5 (2013)}\hrefBibPDF[ ]{PRA_2013_87_33819.pdf}{PDF}.


\bibitem{Martinez:2015-23822:PRA}% article
 \textsc{A.\,J. Martinez},  \textsc{M.\,I. Molina},  \textsc{S.\,K. Turitsyn},
  and  \textsc{Y.\,S. Kivshar},
Nonlinear multicore waveguiding structures with balanced gain and loss,
\hrefBib[ ]{http://dx.doi.org/10.1103/PhysRevA.91.023822}{ \jr{Phys. Rev. A}
  \textbf{91}, 023822--8 (2015)}\hrefBibPDF[ ]{PRA_2015_91_23822.pdf}{PDF}.


\bibitem{Li:2013-375304:JPA}% article
 \textsc{K.~Li},  \textsc{P.\,G. Kevrekidis},  \textsc{D.\,J. Frantzeskakis},
  \textsc{C.\,E. Ruter},  and  \textsc{D.~Kip},
Revisiting the {PT}-symmetric trimer: bifurcations, ghost states and associated
  dynamics,
\hrefBib[ ]{http://dx.doi.org/10.1088/1751-8113/46/37/375304}{ \jr{J. Phys. A}
  \textbf{46}, 375304--12 (2013)}\hrefBibPDF[ ]{JPA_2013_46_375304.pdf}{PDF}.


\bibitem{Leykam:2013-371:OL}% article
 \textsc{D.~Leykam},  \textsc{V.\,V. Konotop},  and  \textsc{A.\,S.
  Desyatnikov},
Discrete vortex solitons and parity time symmetry,
\hrefBib[ ]{http://www.opticsinfobase.org/abstract.cfm?URI=ol-38-3-371}{
  \jr{Opt. Lett.} \textbf{38}, 371--373 (2013)}\hrefBibPDF[
  ]{OL_2013_38_00371.pdf}{PDF}.


\bibitem{Mostafazadeh:2002-205:JMP}% article
 \textsc{A.~Mostafazadeh},
Pseudo-hermiticity versus {PT} symmetry: The necessary condition for the
  reality of the spectrum of a non-hermitian hamiltonian,
\hrefBib[ ]{http://dx.doi.org/10.1063/1.1418246}{ \jr{J. Math. Phys.}
  \textbf{43}, 205--214 (2002)}.


\bibitem{Mostafazadeh:2003-7081:JPA}% article
 \textsc{A.~Mostafazadeh},
Exact {PT}-symmetry is equivalent to hermiticity,
\hrefBib[ ]{http://dx.doi.org/10.1088/0305-4470/36/25/312}{ \jr{J. Phys. A}
  \textbf{36}, 7081--7091 (2003)}\hrefBibPDF[ ]{JPA_2003_36_07081.pdf}{PDF}.


\bibitem{Li:2013-33812:PRA}% article
 \textsc{K.~Li},  \textsc{D.\,A. Zezyulin},  \textsc{V.\,V. Konotop},  and
  \textsc{P.\,G. Kevrekidis},
Parity-time-symmetric optical coupler with birefringent arms,
\hrefBib[ ]{http://dx.doi.org/10.1103/PhysRevA.87.033812}{ \jr{Phys. Rev. A}
  \textbf{87}, 033812--8 (2013)}\hrefBibPDF[ ]{PRA_2013_87_33812.pdf}{PDF}.


\bibitem{Benisty:2015-53825:PRA}% article
 \textsc{H.~Benisty},  \textsc{A.~Lupu},  and  \textsc{A.~Degiron},
Transverse periodic {PT} symmetry for modal demultiplexing in optical
  waveguides,
\hrefBib[ ]{http://dx.doi.org/10.1103/PhysRevA.91.053825}{ \jr{Phys. Rev. A}
  \textbf{91}, 053825--11 (2015)}\hrefBibPDF[ ]{PRA_2015_91_53825.pdf}{PDF}.


\bibitem{Turitsyn:2012-31804:PRA}% article
 \textsc{S.\,K. Turitsyn},  \textsc{A.\,M. Rubenchik},  \textsc{M.\,P.
  Fedoruk},  and  \textsc{E.~Tkachenko},
Coherent propagation and energy transfer in low-dimension nonlinear arrays,
\hrefBib[ ]{http://dx.doi.org/10.1103/PhysRevA.86.031804}{ \jr{Phys. Rev. A}
  \textbf{86}, 031804--4 (2012)}.


\bibitem{Rubenchik:2013-4232:OL}% article
 \textsc{A.\,M. Rubenchik},  \textsc{E.\,V. Tkachenko},  \textsc{M.\,P.
  Fedoruk},  and  \textsc{S.\,K. Turitsyn},
Power-controlled phase-matching and instability of cw propagation in multicore
  optical fibers with a central core,
\hrefBib[ ]{http://dx.doi.org/10.1364/OL.38.004232}{ \jr{Opt. Lett.}
  \textbf{38}, 4232--4235 (2013)}.


\bibitem{Barashenkov:2012-53809:PRA}% article
 \textsc{I.\,V. Barashenkov},  \textsc{S.\,V. Suchkov},  \textsc{A.\,A.
  Sukhorukov},  \textsc{S.\,V. Dmitriev},  and  \textsc{Y.\,S. Kivshar},
Breathers in {PT}-symmetric optical couplers,
\hrefBib[ ]{http://dx.doi.org/10.1103/PhysRevA.86.053809}{ \jr{Phys. Rev. A}
  \textbf{86}, 053809--12 (2012)}\hrefBibPDF[ ]{PRA_2012_86_53809.pdf}{PDF}.


\othercit
\bibitem{Filippov:2010:VersatileSoliton}% book
 \textsc{A.\,T. Filippov},
\hrefBib[ ]{http://dx.doi.org/10.1007/978-0-8176-4974-6}{{The Versatile
  Soliton}}, 2nd edition (Birkh\"{a}user, Boston, 2010).


\bibitem{Yang:2014-367:PLA}% article
 \textsc{J.\,K. Yang},
Necessity of {PT} symmetry for soliton families in one-dimensional complex
  potentials,
\hrefBib[ ]{http://dx.doi.org/10.1016/j.physleta.2013.11.033}{ \jr{Phys. Lett.
  A} \textbf{378}, 367--373 (2014)}\hrefBibPDF[ ]{PLA_2014_378_00367.pdf}{PDF}.


\bibitem{Driben:2011-4323:OL}% article
 \textsc{R.~Driben} and  \textsc{B.\,A. Malomed},
Stability of solitons in parity-time-symmetric couplers,
\hrefBib[ ]{http://www.opticsinfobase.org/abstract.cfm?URI=ol-36-22-4323}{
  \jr{Opt. Lett.} \textbf{36}, 4323--4325 (2011)}\hrefBibPDF[
  ]{OL_2011_36_04323.pdf}{PDF}.


\bibitem{Alexeeva:2012-63837:PRA}% article
 \textsc{N.\,V. Alexeeva},  \textsc{I.\,V. Barashenkov},  \textsc{A.\,A.
  Sukhorukov},  and  \textsc{Y.\,S. Kivshar},
Optical solitons in {PT}-symmetric nonlinear couplers with gain and loss,
\hrefBib[ ]{http://dx.doi.org/10.1103/PhysRevA.85.063837}{ \jr{Phys. Rev. A}
  \textbf{85}, 063837--13 (2012)}\hrefBibPDF[ ]{PRA_2012_85_63837.pdf}{PDF}.


\bibitem{Driben:2012-54001:EPL}% article
 \textsc{R.~Driben} and  \textsc{B.\,A. Malomed},
Dynamics of higher-order solitons in regular and {PT}-symmetric nonlinear
  couplers,
\hrefBib[ ]{http://dx.doi.org/10.1209/0295-5075/99/54001}{ \jr{Europhys. Lett.}
  \textbf{99}, 54001--6 (2012)}\hrefBibPDF[ ]{EPL_2012_99_54001.pdf}{PDF}.


\bibitem{Suchkov:2011-46609:PRE}% article
 \textsc{S.\,V. Suchkov},  \textsc{B.\,A. Malomed},  \textsc{S.\,V. Dmitriev},
  and  \textsc{Y.\,S. Kivshar},
Solitons in a chain of parity-time-invariant dimers,
\hrefBib[ ]{http://dx.doi.org/10.1103/PhysRevE.84.046609}{ \jr{Phys. Rev. E}
  \textbf{84}, 046609--8 (2011)}\hrefBibPDF[ ]{PRE_2011_84_46609.pdf}{PDF}.


\bibitem{Rysaeva:2014-577:JETPL}% article
 \textsc{L.\,K. Rysaeva},  \textsc{S.\,V. Suchkov},  and  \textsc{S.\,V.
  Dmitriev},
Probability of breaking of the pj symmetry at collision between breathers with
  random phases in the model of a pj symmetric planar coupler,
\hrefBib[ ]{http://dx.doi.org/10.1134/S0021364014100099}{ \jr{JETP Lett.}
  \textbf{99}, 577--580 (2014)}.


\bibitem{Abdullaev:2011-4566:OL}% article
 \textsc{F.\,K. Abdullaev},  \textsc{V.\,V. Konotop},  \textsc{M.~Ogren},  and
  \textsc{M.\,P. Sorensen},
Zeno effect and switching of solitons in nonlinear couplers,
\hrefBib[ ]{http://www.opticsinfobase.org/abstract.cfm?URI=ol-36-23-4566}{
  \jr{Opt. Lett.} \textbf{36}, 4566--4568 (2011)}\hrefBibPDF[
  ]{OL_2011_36_04566.pdf}{PDF}.


\bibitem{Misra:1977-756:JMP}% article
 \textsc{B.~Misra} and  \textsc{E.\,C.\,G. Sudarshan},
Zenos paradox in quantum-theory,
\hrefBib[ ]{http://dx.doi.org/10.1063/1.523304}{ \jr{J. Math. Phys.}
  \textbf{18}, 756--763 (1977)}.


\bibitem{Bludov:2013-13816:PRA}% article
 \textsc{Y.\,V. Bludov},  \textsc{V.\,V. Konotop},  and  \textsc{B.\,A.
  Malomed},
Stable dark solitons in {PT}-symmetric dual-core waveguides,
\hrefBib[ ]{http://dx.doi.org/10.1103/PhysRevA.87.013816}{ \jr{Phys. Rev. A}
  \textbf{87}, 013816--7 (2013)}\hrefBibPDF[ ]{PRA_2013_87_13816.pdf}{PDF}.


\bibitem{Bludov:2013-64010:JOPT}% article
 \textsc{Y.\,V. Bludov},  \textsc{R.~Driben},  \textsc{V.\,V. Konotop},  and
  \textsc{B.\,A. Malomed},
Instabilities, solitons and rogue waves in {PT}-coupled nonlinear waveguides,
\hrefBib[ ]{http://dx.doi.org/10.1088/2040-8978/15/6/064010}{ \jr{J. Opt.}
  \textbf{15}, 064010--7 (2013)}\hrefBibPDF[ ]{JOPT_2013_15_64010.pdf}{PDF}.


\bibitem{Musslimani:2008-30402:PRL}% article
 \textsc{Z.\,H. Musslimani},  \textsc{K.\,G. Makris},  \textsc{R.~{E}l
  {G}anainy},  and  \textsc{D.\,N. Christodoulides},
Optical solitons in {PT} periodic potentials,
\hrefBib[ ]{http://dx.doi.org/10.1103/PhysRevLett.100.030402}{ \jr{Phys. Rev.
  Lett.} \textbf{100}, 030402--4 (2008)}\hrefBibPDF[
  ]{PRL_2008_100_30402.pdf}{PDF}.


\bibitem{Musslimani:2008-244019:JPA}% article
 \textsc{Z.\,H. Musslimani},  \textsc{K.\,G. Makris},  \textsc{R.~{E}l
  {G}anainy},  and  \textsc{D.\,N. Christodoulides},
Analytical solutions to a class of nonlinear {S}chr\"odinger equations with
  {PT}-like potentials,
\hrefBib[ ]{http://dx.doi.org/10.1088/1751-8113/41/24/244019}{ \jr{J. Phys. A}
  \textbf{41}, 244019--12 (2008)}\hrefBibPDF[ ]{JPA_2008_41_244019.pdf}{PDF}.


\bibitem{Achilleos:2012-13808:PRA}% article
 \textsc{V.~Achilleos},  \textsc{P.\,G. Kevrekidis},  \textsc{D.\,J.
  Frantzeskakis},  and  \textsc{R.~Carretero-Gonzalez},
Dark solitons and vortices in {PT}-symmetric nonlinear media: From spontaneous
  symmetry breaking to nonlinear {PT} phase transitions,
\hrefBib[ ]{http://dx.doi.org/10.1103/PhysRevA.86.013808}{ \jr{Phys. Rev. A}
  \textbf{86}, 013808--7 (2012)}\hrefBibPDF[ ]{PRA_2012_86_13808.pdf}{PDF}.


\bibitem{Li:2011-3290:OL}% article
 \textsc{H.\,G. Li},  \textsc{Z.\,W. Shi},  \textsc{X.\,J. Jiang},  and
  \textsc{X.~Zhu},
Gray solitons in parity-time symmetric potentials,
\hrefBib[ ]{http://www.opticsinfobase.org/abstract.cfm?URI=ol-36-16-3290}{
  \jr{Opt. Lett.} \textbf{36}, 3290--3292 (2011)}\hrefBibPDF[
  ]{OL_2011_36_03290.pdf}{PDF}.


\bibitem{Shi:2011-53855:PRA}% article
 \textsc{Z.\,W. Shi},  \textsc{X.\,J. Jiang},  \textsc{X.~Zhu},  and
  \textsc{H.\,G. Li},
Bright spatial solitons in defocusing {K}err media with {PT}-symmetric
  potentials,
\hrefBib[ ]{http://dx.doi.org/10.1103/PhysRevA.84.053855}{ \jr{Phys. Rev. A}
  \textbf{84}, 053855--4 (2011)}\hrefBibPDF[ ]{PRA_2011_84_53855.pdf}{PDF}.


\bibitem{Shi:2012-64006:EPL}% article
 \textsc{Z.\,W. Shi},  \textsc{H.\,G. Li},  \textsc{X.~Zhu},  and
  \textsc{X.\,J. Jiang},
Nonlocal bright spatial solitons in defocusing {K}err media supported by {PT}
  symmetric potentials,
\hrefBib[ ]{http://dx.doi.org/10.1209/0295-5075/98/64006}{ \jr{Europhys. Lett.}
  \textbf{98}, 64006--5 (2012)}\hrefBibPDF[ ]{EPL_2012_98_64006.pdf}{PDF}.


\bibitem{Hu:2011-43818:PRA}% article
 \textsc{S.\,M. Hu},  \textsc{X.\,K. Ma},  \textsc{D.\,Q. Lu},  \textsc{Z.\,J.
  Yang},  \textsc{Y.\,Z. Zheng},  and  \textsc{W.~Hu},
Solitons supported by complex {PT}-symmetric gaussian potentials,
\hrefBib[ ]{http://dx.doi.org/10.1103/PhysRevA.84.043818}{ \jr{Phys. Rev. A}
  \textbf{84}, 043818--6 (2011)}\hrefBibPDF[ ]{PRA_2011_84_43818.pdf}{PDF}.


\bibitem{Makris:2008-103904:PRL}% article
 \textsc{K.\,G. Makris},  \textsc{R.~{E}l {G}anainy},  \textsc{D.\,N.
  Christodoulides},  and  \textsc{Z.\,H. Musslimani},
Beam dynamics in {PT} symmetric optical lattices,
\hrefBib[ ]{http://dx.doi.org/10.1103/PhysRevLett.100.103904}{ \jr{Phys. Rev.
  Lett.} \textbf{100}, 103904--4 (2008)}\hrefBibPDF[
  ]{PRL_2008_100_103904.pdf}{PDF}.


\bibitem{Zhu:2011-2680:OL}% article
 \textsc{X.~Zhu},  \textsc{H.~Wang},  \textsc{L.\,X. Zheng},  \textsc{H.\,G.
  Li},  and  \textsc{Y.\,J. He},
Gap solitons in parity-time complex periodic optical lattices with the real
  part of superlattices,
\hrefBib[ ]{http://www.opticsinfobase.org/abstract.cfm?URI=ol-36-14-2680}{
  \jr{Opt. Lett.} \textbf{36}, 2680--2682 (2011)}\hrefBibPDF[
  ]{OL_2011_36_02680.pdf}{PDF}.


\bibitem{Li:2012-4543:OL}% article
 \textsc{C.\,Y. Li},  \textsc{C.\,M. Huang},  \textsc{H.\,D. Liu},  and
  \textsc{L.\,W. Dong},
Multipeaked gap solitons in {PT}-symmetric optical lattices,
\hrefBib[ ]{http://www.opticsinfobase.org/abstract.cfm?URI=ol-37-21-4543}{
  \jr{Opt. Lett.} \textbf{37}, 4543--4545 (2012)}\hrefBibPDF[
  ]{OL_2012_37_04543.pdf}{PDF}.


\bibitem{Lumer:2013-263901:PRL}% article
 \textsc{Y.~Lumer},  \textsc{Y.~Plotnik},  \textsc{M.\,C. Rechtsman},  and
  \textsc{M.~Segev},
Nonlinearly induced {PT} transition in photonic systems,
\hrefBib[ ]{http://dx.doi.org/10.1103/PhysRevLett.111.263901}{ \jr{Phys. Rev.
  Lett.} \textbf{111}, 263901--5 (2013)}\hrefBibPDF[
  ]{PRL_2013_111_263901.pdf}{PDF}.


\bibitem{Nixon:2012-4874:OL}% article
 \textsc{S.~Nixon},  \textsc{Y.~Zhu},  and  \textsc{J.\,K. Yang},
Nonlinear dynamics of wave packets in parity-time-symmetric optical lattices
  near the phase transition point,
\hrefBib[ ]{http://www.opticsinfobase.org/abstract.cfm?URI=ol-37-23-4874}{
  \jr{Opt. Lett.} \textbf{37}, 4874--4876 (2012)}\hrefBibPDF[
  ]{OL_2012_37_04874.pdf}{PDF}.


\bibitem{Dmitriev:2010-2976:OL}% article
 \textsc{S.\,V. Dmitriev},  \textsc{A.\,A. Sukhorukov},  and  \textsc{Y.\,S.
  Kivshar},
Binary parity-time-symmetric nonlinear lattices with balanced gain and loss,
\hrefBib[ ]{http://www.opticsinfobase.org/abstract.cfm?URI=ol-35-17-2976}{
  \jr{Opt. Lett.} \textbf{35}, 2976--2978 (2010)}\hrefBibPDF[
  ]{OL_2010_35_02976.pdf}{PDF}.


\bibitem{Konotop:2012-56006:EPL}% article
 \textsc{V.\,V. Konotop},  \textsc{D.\,E. Pelinovsky},  and  \textsc{D.\,A.
  Zezyulin},
Discrete solitons in {PT}-symmetric lattices,
\hrefBib[ ]{http://dx.doi.org/10.1209/0295-5075/100/56006}{ \jr{Europhys.
  Lett.} \textbf{100}, 56006--6 (2012)}\hrefBibPDF[
  ]{EPL_2012_100_56006.pdf}{PDF}.


\bibitem{Hu:2012-43826:PRA}% article
 \textsc{S.\,M. Hu},  \textsc{X.\,K. Ma},  \textsc{D.\,Q. Lu},  \textsc{Y.\,Z.
  Zheng},  and  \textsc{W.~Hu},
Defect solitons in parity-time-symmetric optical lattices with nonlocal
  nonlinearity,
\hrefBib[ ]{http://dx.doi.org/10.1103/PhysRevA.85.043826}{ \jr{Phys. Rev. A}
  \textbf{85}, 043826--7 (2012)}\hrefBibPDF[ ]{PRA_2012_85_43826.pdf}{PDF}.


\bibitem{Li:2012-23840:PRA}% article
 \textsc{H.\,G. Li},  \textsc{X.\,J. Jiang},  \textsc{X.~Zhu},  and
  \textsc{Z.\,W. Shi},
Nonlocal solitons in dual-periodic {PT}-symmetric optical lattices,
\hrefBib[ ]{http://dx.doi.org/10.1103/PhysRevA.86.023840}{ \jr{Phys. Rev. A}
  \textbf{86}, 023840--4 (2012)}\hrefBibPDF[ ]{PRA_2012_86_23840.pdf}{PDF}.


\bibitem{Hu:2012-14006:EPL}% article
 \textsc{S.\,M. Hu},  \textsc{D.\,Q. Lu},  \textsc{X.\,K. Ma},
  \textsc{Q.~Guo},  and  \textsc{W.~Hu},
Defect solitons supported by nonlocal {PT} symmetric superlattices,
\hrefBib[ ]{http://dx.doi.org/10.1209/0295-5075/98/14006}{ \jr{Europhys. Lett.}
  \textbf{98}, 14006--6 (2012)}\hrefBibPDF[ ]{EPL_2012_98_14006.pdf}{PDF}.


\bibitem{Yin:2012-19355:OE}% article
 \textsc{C.\,P. Yin},  \textsc{Y.\,J. He},  \textsc{H.\,G. Li},  and
  \textsc{J.\,N. Xie},
Solitons in parity-time symmetric potentials with spatially modulated nonlocal
  nonlinearity,
\hrefBib[ ]{http://dx.doi.org/10.1364/OE.20.019355}{ \jr{Opt. Express}
  \textbf{20}, 19355--19362 (2012)}\hrefBibPDF[ ]{OE_2012_20_19355.pdf}{PDF}.


\bibitem{Jisha:2014-13812:PRA}% article
 \textsc{C.\,P. Jisha},  \textsc{A.~Alberucci},  \textsc{V.\,A. Brazhnyi},  and
   \textsc{G.~Assanto},
Nonlocal gap solitons in {PT}-symmetric periodic potentials with defocusing
  nonlinearity,
\hrefBib[ ]{http://dx.doi.org/10.1103/PhysRevA.89.013812}{ \jr{Phys. Rev. A}
  \textbf{89}, 013812--10 (2014)}\hrefBibPDF[ ]{PRA_2014_89_13812.pdf}{PDF}.


\bibitem{Kartashov:2013-2600:OL}% article
 \textsc{Y.\,V. Kartashov},
Vector solitons in parity-time-symmetric lattices,
\hrefBib[ ]{http://dx.doi.org/10.1364/OL.38.002600}{ \jr{Opt. Lett.}
  \textbf{38}, 2600--2603 (2013)}\hrefBibPDF[ ]{OL_2013_38_02600.pdf}{PDF}.


\bibitem{Abdullaev:2011-41805:PRA}% article
 \textsc{F.\,K. Abdullaev},  \textsc{Y.\,V. Kartashov},  \textsc{V.\,V.
  Konotop},  and  \textsc{D.\,A. Zezyulin},
Solitons in {PT}-symmetric nonlinear lattices,
\hrefBib[ ]{http://dx.doi.org/10.1103/PhysRevA.83.041805}{ \jr{Phys. Rev. A}
  \textbf{83}, 041805--4 (2011)}\hrefBibPDF[ ]{PRA_2011_83_41805.pdf}{PDF}.


\bibitem{He:2012-13831:PRA}% article
 \textsc{Y.\,J. He},  \textsc{X.~Zhu},  \textsc{D.~Mihalache},  \textsc{J.\,L.
  Liu},  and  \textsc{Z.\,X. Chen},
Lattice solitons in {PT}-symmetric mixed linear-nonlinear optical lattices,
\hrefBib[ ]{http://dx.doi.org/10.1103/PhysRevA.85.013831}{ \jr{Phys. Rev. A}
  \textbf{85}, 013831--6 (2012)}\hrefBibPDF[ ]{PRA_2012_85_13831.pdf}{PDF}.


\bibitem{Abdullaev:2010-56606:PRE}% article
 \textsc{F.\,K. Abdullaev},  \textsc{V.\,V. Konotop},  \textsc{M.~Salerno},
  and  \textsc{A.\,V. Yulin},
Dissipative periodic waves, solitons, and breathers of the nonlinear
  {S}chr\"odinger equation with complex potentials,
\hrefBib[ ]{http://dx.doi.org/10.1103/PhysRevE.82.056606}{ \jr{Phys. Rev. E}
  \textbf{82}, 056606--6 (2010)}\hrefBibPDF[ ]{PRE_2010_82_56606.pdf}{PDF}.


\bibitem{Liu:2012-1934:OC}% article
 \textsc{S.~Liu},  \textsc{C.\,W. Ma},  \textsc{Y.\,Q. Zhang},  and
  \textsc{K.\,Q. Lu},
{B}ragg gap solitons in {PT} symmetric lattices with competing nonlinearity,
\hrefBib[ ]{http://dx.doi.org/10.1016/j.optcom.2011.11.065}{ \jr{Opt. Commun.}
  \textbf{285}, 1934--1939 (2012)}\hrefBibPDF[ ]{OC_2012_285_01934.pdf}{PDF}.


\bibitem{Li:2012-16823:OE}% article
 \textsc{C.\,Y. Li},  \textsc{H.\,D. Liu},  and  \textsc{L.\,W. Dong},
Multi-stable solitons in {PT}-symmetric optical lattices,
\hrefBib[ ]{http://dx.doi.org/10.1364/OE.20.016823}{ \jr{Opt. Express}
  \textbf{20}, 16823--16831 (2012)}\hrefBibPDF[ ]{OE_2012_20_16823.pdf}{PDF}.


\bibitem{Miri:2012-33801:PRA}% article
 \textsc{M.\,A. Miri},  \textsc{A.\,B. Aceves},  \textsc{T.~Kottos},
  \textsc{V.~Kovanis},  and  \textsc{D.\,N. Christodoulides},
{B}ragg solitons in nonlinear {PT}-symmetric periodic potentials,
\hrefBib[ ]{http://dx.doi.org/10.1103/PhysRevA.86.033801}{ \jr{Phys. Rev. A}
  \textbf{86}, 033801--5 (2012)}\hrefBibPDF[ ]{PRA_2012_86_33801.pdf}{PDF}.


\bibitem{Yang:2014-332:STAM}% article
 \textsc{J.\,K. Yang},
Can parity-time-symmetric potentials support families of
  non-parity-time-symmetric solitons?,
\hrefBib[ ]{http://dx.doi.org/10.1111/sapm.12032}{ \jr{Stud. Appl. Math.}
  \textbf{132}, 332--353 (2014)}\hrefBibPDF[ ]{STAM_2014_132_00332.pdf}{PDF}.


\bibitem{Zhu:2013-2723:OL}% article
 \textsc{X.~Zhu},  \textsc{H.~Wang},  \textsc{H.\,G. Li},  \textsc{W.~He},  and
   \textsc{Y.\,J. He},
Two-dimensional multipeak gap solitons supported by parity-time-symmetric
  periodic potentials,
\hrefBib[ ]{http://dx.doi.org/10.1364/OL.38.002723}{ \jr{Opt. Lett.}
  \textbf{38}, 2723--2725 (2013)}\hrefBibPDF[ ]{OL_2013_38_02723.pdf}{PDF}.


\bibitem{Nixon:2013-1933:OL}% article
 \textsc{S.~Nixon} and  \textsc{J.\,K. Yang},
Pyramid diffraction in parity-time-symmetric optical lattices,
\hrefBib[ ]{http://dx.doi.org/10.1364/OL.38.001933}{ \jr{Opt. Lett.}
  \textbf{38}, 1933--1935 (2013)}\hrefBibPDF[ ]{OL_2013_38_01933.pdf}{PDF}.


\bibitem{Jovic:2012-4455:OL}% article
 \textsc{D.\,M. Jovic},  \textsc{C.~Denz},  and  \textsc{M.\,R. Belic},
Anderson localization of light in {PT}-symmetric optical lattices,
\hrefBib[ ]{http://www.opticsinfobase.org/abstract.cfm?URI=ol-37-21-4455}{
  \jr{Opt. Lett.} \textbf{37}, 4455--4457 (2012)}\hrefBibPDF[
  ]{OL_2012_37_04455.pdf}{PDF}.


\bibitem{Yang:2014-1133:OL}% article
 \textsc{J.\,K. Yang},
Partially {PT} symmetric optical potentials with all-real spectra and soliton
  families in multidimensions,
\hrefBib[ ]{http://dx.doi.org/10.1364/OL.39.001133}{ \jr{Opt. Lett.}
  \textbf{39}, 1133--1136 (2014)}\hrefBibPDF[ ]{OL_2014_39_01133.pdf}{PDF}.


\bibitem{Chen:2014-29679:OE}% article
 \textsc{Z.\,P. Chen},  \textsc{J.\,F. Liu},  \textsc{S.\,H. Fu},
  \textsc{Y.\,Y. Li},  and  \textsc{B.\,A. Malomed},
Discrete solitons and vortices on two-dimensional lattices of {PT}-symmetric
  couplers,
\hrefBib[ ]{http://dx.doi.org/10.1364/OE.22.029679}{ \jr{Opt. Express}
  \textbf{22}, 29679--29692 (2014)}.


\bibitem{Mostafazadeh:2013-12103:PRA}% article
 \textsc{A.~Mostafazadeh},
Invisibility and {PT} symmetry,
\hrefBib[ ]{http://dx.doi.org/10.1103/PhysRevA.87.012103}{ \jr{Phys. Rev. A}
  \textbf{87}, 012103--8 (2013)}\hrefBibPDF[ ]{PRA_2013_87_12103.pdf}{PDF}.


\bibitem{Feng:2013-108:NMAT}% article
 \textsc{L.~Feng},  \textsc{Y.\,L. Xu},  \textsc{W.\,S. Fegadolli},
  \textsc{M.\,H. Lu},  \textsc{J.\,E.\,B. Oliveira},  \textsc{V.\,R. Almeida},
  \textsc{Y.\,F. Chen},  and  \textsc{A.~Scherer},
Experimental demonstration of a unidirectional reflectionless parity-time
  metamaterial at optical frequencies,
\hrefBib[ ]{http://dx.doi.org/10.1038/NMAT3495}{ \jr{Nat. Mater.} \textbf{12},
  108--113 (2013)}\hrefBibPDF[ ]{NMAT_2013_12_00108.pdf}{PDF}.


\bibitem{Mostafazadeh:2009-220402:PRL}% article
 \textsc{A.~Mostafazadeh},
Spectral singularities of complex scattering potentials and infinite reflection
  and transmission coefficients at real energies,
\hrefBib[ ]{http://dx.doi.org/10.1103/PhysRevLett.102.220402}{ \jr{Phys. Rev.
  Lett.} \textbf{102}, 220402--4 (2009)}.


\bibitem{Longhi:2011-485302:JPA}% article
 \textsc{S.~Longhi},
Invisibility in {PT}-symmetric complex crystals,
\hrefBib[ ]{http://dx.doi.org/10.1088/1751-8113/44/48/485302}{ \jr{J. Phys. A}
  \textbf{44}, 485302--16 (2011)}\hrefBibPDF[ ]{JPA_2011_44_485302.pdf}{PDF}.


\bibitem{Jones:2012-135306:JPA}% article
 \textsc{H.\,F. Jones},
Analytic results for a {PT}-symmetric optical structure,
\hrefBib[ ]{http://dx.doi.org/10.1088/1751-8113/45/13/135306}{ \jr{J. Phys. A}
  \textbf{45}, 135306--10 (2012)}\hrefBibPDF[ ]{JPA_2012_45_135306.pdf}{PDF}.


\bibitem{Suchkov:2012-33825:PRA}% article
 \textsc{S.\,V. Suchkov},  \textsc{S.\,V. Dmitriev},  \textsc{B.\,A. Malomed},
  and  \textsc{Y.\,S. Kivshar},
Wave scattering on a domain wall in a chain of {PT}-symmetric couplers,
\hrefBib[ ]{http://dx.doi.org/10.1103/PhysRevA.85.033825}{ \jr{Phys. Rev. A}
  \textbf{85}, 033825--6 (2012)}\hrefBibPDF[ ]{PRA_2012_85_33825.pdf}{PDF}.


\bibitem{Schomerus:2013-20120194:PTRSA}% article
 \textsc{H.~Schomerus},
From scattering theory to complex wave dynamics in non-hermitian {PT}-symmetric
  resonators,
\hrefBib[ ]{http://dx.doi.org/10.1098/rsta.2012.0194}{ \jr{Philos. Trans. R.
  Soc. A} \textbf{371}, 20120194--17 (2013)}.


\bibitem{Peng:2014-394:NPHYS}% article
 \textsc{B.~Peng},  \textsc{S.\,K. Ozdemir},  \textsc{F.\,C. Lei},
  \textsc{F.~Monifi},  \textsc{M.~Gianfreda},  \textsc{G.\,L. Long},
  \textsc{S.\,H. Fan},  \textsc{F.~Nori},  \textsc{C.\,M. Bender},  and
  \textsc{L.~Yang},
Parity-time-symmetric whispering-gallery microcavities,
\hrefBib[ ]{http://dx.doi.org/10.1038/NPHYS2927}{ \jr{Nature Physics}
  \textbf{10}, 394--398 (2014)}\hrefBibPDF[ ]{NPHYS_2014_10_00394.pdf}{PDF}.


\bibitem{Lin:2012-50101:PRA}% article
 \textsc{Z.~Lin},  \textsc{J.~Schindler},  \textsc{F.\,M. Ellis},  and
  \textsc{T.~Kottos},
Experimental observation of the dual behavior of {PT}-symmetric scattering,
\hrefBib[ ]{http://dx.doi.org/10.1103/PhysRevA.85.050101}{ \jr{Phys. Rev. A}
  \textbf{85}, 050101--4 (2012)}\hrefBibPDF[ ]{PRA_2012_85_50101.pdf}{PDF}.


\bibitem{Bender:2013-234101:PRL}% article
 \textsc{N.~Bender},  \textsc{S.~Factor},  \textsc{J.\,D. Bodyfelt},
  \textsc{H.~Ramezani},  \textsc{D.\,N. Christodoulides},  \textsc{F.\,M.
  Ellis},  and  \textsc{T.~Kottos},
Observation of asymmetric transport in structures with active nonlinearities,
\hrefBib[ ]{http://dx.doi.org/10.1103/PhysRevLett.110.234101}{ \jr{Phys. Rev.
  Lett.} \textbf{110}, 234101--5 (2013)}\hrefBibPDF[
  ]{PRL_2013_110_234101.pdf}{PDF}.


\bibitem{Lepri:2011-164101:PRL}% article
 \textsc{S.~Lepri} and  \textsc{G.~Casati},
Asymmetric wave propagation in nonlinear systems,
\hrefBib[ ]{http://dx.doi.org/10.1103/PhysRevLett.106.164101}{ \jr{Phys. Rev.
  Lett.} \textbf{106}, 164101--4 (2011)}.


\bibitem{Nazari:2014-9574:OE}% article
 \textsc{F.~Nazari},  \textsc{N.~Bender},  \textsc{H.~Ramezani},
  \textsc{M.\,K. Moravvej-Farshi},  \textsc{D.\,N. Christodoulides},  and
  \textsc{T.~Kottos},
Optical isolation via {PT}-symmetric nonlinear {F}ano resonances,
\hrefBib[ ]{http://dx.doi.org/10.1364/OE.22.009574}{ \jr{Opt. Express}
  \textbf{22}, 9574--9584 (2014)}\hrefBibPDF[ ]{OE_2014_22_09574.pdf}{PDF}.


\bibitem{Miroshnichenko:2010-2257:RMP}% article
 \textsc{A.\,E. Miroshnichenko},  \textsc{S.~Flach},  and  \textsc{Y.\,S.
  Kivshar},
Fano resonances in nanoscale structures,
\hrefBib[ ]{http://dx.doi.org/10.1103/RevModPhys.82.2257}{ \jr{Rev. Mod. Phys.}
  \textbf{82}, 2257--2298 (2010)}\hrefBibPDF[ ]{RMP_2010_82_02257.pdf}{PDF}.


\bibitem{Zhang:2014-13927:OE}% article
 \textsc{X.\,Y. Zhang},  \textsc{J.\,L. Chai},  \textsc{J.\,S. Huang},
  \textsc{Z.\,Q. Chen},  \textsc{Y.\,Y. Li},  and  \textsc{B.\,A. Malomed},
Discrete solitons and scattering of lattice waves in guiding arrays with a
  nonlinear p t -symmetric defect,
\hrefBib[ ]{http://dx.doi.org/10.1364/OE.22.013927}{ \jr{Opt. Express}
  \textbf{22}, 13927--13939 (2014)}\hrefBibPDF[ ]{OE_2014_22_13927.pdf}{PDF}.


\bibitem{Mostafazadeh:2013-260402:PRL}% article
 \textsc{A.~Mostafazadeh},
Nonlinear spectral singularities for confined nonlinearities,
\hrefBib[ ]{http://dx.doi.org/10.1103/PhysRevLett.110.260402}{ \jr{Phys. Rev.
  Lett.} \textbf{110}, 260402--5 (2013)}\hrefBibPDF[
  ]{PRL_2013_110_260402.pdf}{PDF}.


\bibitem{Liu:2014-13824:PRA}% article
 \textsc{X.\,L. Liu},  \textsc{S.\,D. Gupta},  and  \textsc{G.\,S. Agarwal},
Regularization of the spectral singularity in {PT}-symmetric systems by
  all-order nonlinearities: Nonreciprocity and optical isolation,
\hrefBib[ ]{http://dx.doi.org/10.1103/PhysRevA.89.013824}{ \jr{Phys. Rev. A}
  \textbf{89}, 013824--5 (2014)}\hrefBibPDF[ ]{PRA_2014_89_13824.pdf}{PDF}.


\bibitem{Dmitriev:2011-13833:PRA}% article
 \textsc{S.\,V. Dmitriev},  \textsc{S.\,V. Suchkov},  \textsc{A.\,A.
  Sukhorukov},  and  \textsc{Y.\,S. Kivshar},
Scattering of linear and nonlinear waves in a waveguide array with a
  {PT}-symmetric defect,
\hrefBib[ ]{http://dx.doi.org/10.1103/PhysRevA.84.013833}{ \jr{Phys. Rev. A}
  \textbf{84}, 013833--5 (2011)}\hrefBibPDF[ ]{PRA_2011_84_13833.pdf}{PDF}.


\bibitem{Suchkov:2012-54003:EPL}% article
 \textsc{S.\,V. Suchkov},  \textsc{A.\,A. Sukhorukov},  \textsc{S.\,V.
  Dmitriev},  and  \textsc{Y.\,S. Kivshar},
Scattering of the discrete solitons on the {PT}-symmetric defects,
\hrefBib[ ]{http://dx.doi.org/10.1209/0295-5075/100/54003}{ \jr{Europhys.
  Lett.} \textbf{100}, 54003--5 (2012)}\hrefBibPDF[
  ]{EPL_2012_100_54003.pdf}{PDF}.


\bibitem{Hu:2012-266:EPD}% article
 \textsc{S.\,M. Hu} and  \textsc{W.~Hu},
Defect solitons in optical lattices with parity-time symmetric defect,
\hrefBib[ ]{http://dx.doi.org/10.1140/epjd/e2012-30408-6}{ \jr{Eur. Phys. J. D}
  \textbf{66}, 266--5 (2012)}\hrefBibPDF[ ]{EPD_2012_66_00266.pdf}{PDF}.


\bibitem{Abdullaev:2013-43829:PRA}% article
 \textsc{F.\,K. Abdullaev},  \textsc{V.\,A. Brazhnyi},  and
  \textsc{M.~Salerno},
Scattering of gap solitons by {PT}-symmetric defects,
\hrefBib[ ]{http://dx.doi.org/10.1103/PhysRevA.88.043829}{ \jr{Phys. Rev. A}
  \textbf{88}, 043829--9 (2013)}\hrefBibPDF[ ]{PRA_2013_88_43829.pdf}{PDF}.


\bibitem{Karjanto:2015-23112:CHA}% article
 \textsc{N.~Karjanto},  \textsc{W.~Hanif},  \textsc{B.\,A. Malomed},  and
  \textsc{H.~Susanto},
Interactions of bright and dark solitons with localized {PT} - symmetric
  potentials,
\hrefBib[ ]{http://dx.doi.org/10.1063/1.4907556}{ \jr{Chaos} \textbf{25},
  023112--11 (2015)}\hrefBibPDF[ ]{CHA_2015_25_23112.pdf}{PDF}.


\bibitem{Suchkov:2011-222:RAR}% article
 \textsc{S.~Suchkov} and  \textsc{A.~Khare},
Soliton collision in discrete pt-symmetric systems without peierls-nabarro
  potential,
\hrefBib[ ]{http://lettersonmaterials.com/Upload/Journals/210/v4_222-225.pdf}{
  \jr{Lett. on Mat.} \textbf{1}, 222--4 (2011)}\hrefBibPDF[
  ]{RAR_2011_01_00222.pdf}{PDF}.


\bibitem{Bludov:2014-3382:OL}% article
 \textsc{Y.\,V. Bludov},  \textsc{C.~Hang},  \textsc{G.\,X. Huang},  and
  \textsc{V.\,V. Konotop},
{PT}-symmetric coupler with a coupling defect: soliton interaction with
  exceptional point,
\hrefBib[ ]{http://dx.doi.org/10.1364/OL.39.003382}{ \jr{Opt. Lett.}
  \textbf{39}, 3382--3385 (2014)}\hrefBibPDF[ ]{OL_2014_39_03382.pdf}{PDF}.


\bibitem{Burlak:2013-62904:PRE}% article
 \textsc{G.~Burlak} and  \textsc{B.\,A. Malomed},
Stability boundary and collisions of two-dimensional solitons in {PT}-symmetric
  couplers with the cubic-quintic nonlinearity,
\hrefBib[ ]{http://dx.doi.org/10.1103/PhysRevE.88.062904}{ \jr{Phys. Rev. E}
  \textbf{88}, 062904--8 (2013)}\hrefBibPDF[ ]{PRE_2013_88_62904.pdf}{PDF}.


\bibitem{Lin:1990-2927:PRL}% article
 \textsc{W.\,A. Lin} and  \textsc{L.\,E. Ballentine},
Quantum tunneling and chaos in a driven anharmonic-oscillator,
\hrefBib[ ]{http://dx.doi.org/10.1103/PhysRevLett.65.2927}{ \jr{Phys. Rev.
  Lett.} \textbf{65}, 2927--2930 (1990)}\hrefBibPDF[
  ]{PRL_1990_65_02927.pdf}{PDF}.


\bibitem{Grossmann:1991-516:PRL}% article
 \textsc{F.~Grossmann},  \textsc{T.~Dittrich},  \textsc{P.~Jung},  and
  \textsc{P.~Hanggi},
Coherent destruction of tunneling,
\hrefBib[ ]{http://link.aps.org/abstract/PRL/v67/p516}{ \jr{Phys. Rev. Lett.}
  \textbf{67}, 516--519 (1991)}\hrefBibPDF[ ]{PRL_1991_67_00516.pdf}{PDF}.


\bibitem{Grifoni:1998-229:PRP}% article
 \textsc{M.~Grifoni} and  \textsc{P.~Hanggi},
Driven quantum tunneling,
\hrefEmpty[ ]{Empty}{ \jr{Phys. Rep.} \textbf{304}, 229--354
  (1998)}\hrefBibPDF[ ]{PRP_1998_304_00229.pdf}{PDF}.


\bibitem{Creffield:2007-110501:PRL}% article
 \textsc{C.\,E. Creffield},
Quantum control and entanglement using periodic driving fields,
\hrefBib[ ]{http://dx.doi.org/10.1103/PhyRevLett.99.110501}{ \jr{Phys. Rev.
  Lett.} \textbf{99}, 110501--4 (2007)}.


\bibitem{Luo:2011-53847:PRA}% article
 \textsc{X.\,B. Luo},  \textsc{J.\,H. Huang},  and  \textsc{C.\,H. Lee},
Coherent destruction of tunneling in a lattice array under selective in-phase
  modulations,
\hrefBib[ ]{http://dx.doi.org/10.1103/PhysRevA.84.053847}{ \jr{Phys. Rev. A}
  \textbf{84}, 053847--7 (2011)}.


\bibitem{Moiseyev:2011-52125:PRA}% article
 \textsc{N.~Moiseyev},
Crossing rule for a {PT}-symmetric two-level time-periodic system,
\hrefBib[ ]{http://dx.doi.org/10.1103/PhysRevA.83.052125}{ \jr{Phys. Rev. A}
  \textbf{83}, 052125--5 (2011)}\hrefBibPDF[ ]{PRA_2011_83_52125.pdf}{PDF}.


\bibitem{Joglekar:2014-40101:PRA}% article
 \textsc{Y.\,N. Joglekar},  \textsc{R.~Marathe},  \textsc{P.~Durganandini},
  and  \textsc{R.\,K. Pathak},
{PT} spectroscopy of the {R}abi problem,
\hrefBib[ ]{http://dx.doi.org/10.1103/PhysRevA.90.040101}{ \jr{Phys. Rev. A}
  \textbf{90}, 040101--4 (2014)}\hrefBibPDF[ ]{PRA_2014_90_40101.pdf}{PDF}.


\bibitem{Gong:2015-42135:PRA}% article
 \textsc{J.\,B. Gong} and  \textsc{Q.\,H. Wang},
Stabilizing non-hermitian systems by periodic driving,
\hrefBib[ ]{http://dx.doi.org/10.1103/PhysRevA.91.042135}{ \jr{Phys. Rev. A}
  \textbf{91}, 042135--6 (2015)}\hrefBibPDF[ ]{PRA_2015_91_42135.pdf}{PDF}.


\bibitem{Driben:2011-51001:EPL}% article
 \textsc{R.~Driben} and  \textsc{B.\,A. Malomed},
Stabilization of solitons in {PT} models with supersymmetry by periodic
  management,
\hrefBib[ ]{http://dx.doi.org/10.1209/0295-5075/96/51001}{ \jr{Europhys. Lett.}
  \textbf{96}, 51001--5 (2011)}\hrefBibPDF[ ]{EPL_2011_96_51001.pdf}{PDF}.


\bibitem{Horne:2013-485101:JPA}% article
 \textsc{R.\,L. Horne},  \textsc{J.~Cuevas},  \textsc{P.\,G. Kevrekidis},
  \textsc{N.~Whitaker},  \textsc{F.\,K. Abdullaev},  and  \textsc{D.\,J.
  Frantzeskakis},
{PT}-symmetry management in oligomer systems,
\hrefBib[ ]{http://dx.doi.org/10.1088/1751-8113/46/48/485101}{ \jr{J. Phys. A}
  \textbf{46}, 485101--19 (2013)}\hrefBibPDF[ ]{JPA_2013_46_485101.pdf}{PDF}.


\bibitem{DAmbroise:2014-23136:CHA}% article
 \textsc{J.~D'Ambroise},  \textsc{B.\,A. Malomed},  and  \textsc{P.\,G.
  Kevrekidis},
Quasi-energies, parametric resonances, and stability limits in ac-driven
  {PT}-symmetric systems,
\hrefBib[ ]{http://dx.doi.org/10.1063/1.4883715}{ \jr{Chaos} \textbf{24},
  023136--10 (2014)}\hrefBibPDF[ ]{CHA_2014_24_23136.pdf}{PDF}.


\bibitem{Battelli:2015-353:NLD}% article
 \textsc{F.~Battelli},  \textsc{J.~Diblik},  \textsc{M.~Feckan},
  \textsc{J.~Pickton},  \textsc{M.~Pospisil},  and  \textsc{H.~Susanto},
Dynamics of generalized {PT}-symmetric dimers with time-periodic gain-loss,
\hrefBib[ ]{http://dx.doi.org/10.1007/s11071-015-1996-2}{ \jr{Nonlin. Dynam.}
  \textbf{81}, 353--371 (2015)}\hrefBibPDF[ ]{NLD_2015_81_00353.pdf}{PDF}.


\bibitem{Greenberg:2004-451:OL}% article
 \textsc{M.~Greenberg} and  \textsc{M.~Orenstein},
Irreversible coupling by use of dissipative optics,
\hrefBib[ ]{http://dx.doi.org/10.1364/OL.29.000451}{ \jr{Opt. Lett.}
  \textbf{29}, 451--453 (2004)}\hrefBibPDF[ ]{OL_2004_29_00451.pdf}{PDF}.


\bibitem{Greenberg:2005-1013:IQE}% article
 \textsc{M.~Greenberg} and  \textsc{M.~Orenstein},
Optical unidirectional devices by complex spatial single sideband perturbation,
\hrefBib[ ]{http://dx.doi.org/10.1109/JQE.2005.848948}{ \jr{IEEE J. Quantum
  Electron.} \textbf{41}, 1013--1023 (2005)}\hrefBibPDF[
  ]{IQE_2005_41_01013.pdf}{PDF}.


\bibitem{Greenberg:2004-4013:OE}% article
 \textsc{M.~Greenberg} and  \textsc{M.~Orenstein},
Unidirectional complex gratings assisted couplers,
\hrefBib[ ]{http://dx.doi.org/10.1364/OPEX.12.004013}{ \jr{Opt. Express}
  \textbf{12}, 4013--4018 (2004)}\hrefBibPDF[ ]{OE_2004_12_04013.pdf}{PDF}.


\bibitem{West:2007-8052:AOP}% article
 \textsc{B.\,R. West} and  \textsc{D.\,V. Plant},
Transfer matrix analysis of the unidirectional grating-assisted codirectional
  coupler,
\hrefBib[ ]{http://dx.doi.org/10.1364/AO.46.008052}{ \jr{Appl. Optics}
  \textbf{46}, 8052--8060 (2007)}\hrefBibPDF[ ]{AOP_2007_46_08052.pdf}{PDF}.


\bibitem{Luo:2013-243902:PRL}% article
 \textsc{X.\,B. Luo},  \textsc{J.\,H. Huang},  \textsc{H.\,H. Zhong},
  \textsc{X.\,Z. Qin},  \textsc{Q.\,T. Xie},  \textsc{Y.\,S. Kivshar},  and
  \textsc{C.\,H. Lee},
Pseudo-parity-time symmetry in optical systems,
\hrefBib[ ]{http://dx.doi.org/10.1103/PhysRevLett.110.243902}{ \jr{Phys. Rev.
  Lett.} \textbf{110}, 243902--5 (2013)}\hrefBibPDF[
  ]{PRL_2013_110_243902.pdf}{PDF}.


\bibitem{DellaValle:2013-22119:PRA}% article
 \textsc{G.~Della~Valle} and  \textsc{S.~Longhi},
Spectral and transport properties of time-periodic {PT}-symmetric tight-binding
  lattices,
\hrefBib[ ]{http://dx.doi.org/10.1103/PhysRevA.87.022119}{ \jr{Phys. Rev. A}
  \textbf{87}, 022119--6 (2013)}\hrefBibPDF[ ]{PRA_2013_87_22119.pdf}{PDF}.


\bibitem{Eichelkraut:2013-2533:NCOM}% article
 \textsc{T.~Eichelkraut},  \textsc{R.~Heilmann},  \textsc{S.~Weimann},
  \textsc{S.~Stutzer},  \textsc{F.~Dreisow},  \textsc{D.\,N. Christodoulides},
  \textsc{S.~Nolte},  and  \textsc{A.~Szameit},
Mobility transition from ballistic to diffusive transport in non-hermitian
  lattices,
\hrefBib[ ]{http://dx.doi.org/10.1038/ncomms3533}{ \jr{Nat. Commun.}
  \textbf{4}, 2533--7 (2013)}\hrefBibPDF[ ]{NCOM_2013_04_02533.pdf}{PDF}.


\bibitem{Zeuner:2015-40402:PRL}% article
 \textsc{J.\,M. Zeuner},  \textsc{M.\,C. Rechtsman},  \textsc{Y.~Plotnik},
  \textsc{Y.~Lumer},  \textsc{S.~Nolte},  \textsc{M.\,S. Rudner},
  \textsc{M.~Segev},  and  \textsc{A.~Szameit},
Observation of a topological transition in the bulk of a non-hermitian system,
\hrefBib[ ]{http://dx.doi.org/10.1103/PhysRevLett.115.040402}{ \jr{Phys. Rev.
  Lett.} \textbf{115}, 040402--5 (2015)}\hrefBibPDF[
  ]{PRL_2015_115_40402.pdf}{PDF}.


\bibitem{Meany:2015-363:LPR}% article
 \textsc{T.~Meany},  \textsc{M.~Gr{\"a}fe},  \textsc{R.~Heilmann},
  \textsc{A.~Prez-Leija},  \textsc{S.~Gross},  \textsc{M.\,J. Steel},
  \textsc{M.\,J. Withford},  and  \textsc{A.~Szameit},
Laser written circuits for quantum photonics,
\hrefEmpty[ ]{Empty}{ \jr{Laser Photon. Rev.} \textbf{9}, 363--384 (2015)}.


\bibitem{Regensburger:2012-167:NAT}% article
 \textsc{A.~Regensburger},  \textsc{C.~Bersch},  \textsc{M.\,A. Miri},
  \textsc{G.~Onishchukov},  \textsc{D.\,N. Christodoulides},  and
  \textsc{U.~Peschel},
Parity-time synthetic photonic lattices,
\hrefBib[ ]{http://dx.doi.org/10.1038/nature11298}{ \jr{Nature} \textbf{488},
  167--171 (2012)}\hrefBibPDF[ ]{NAT_2012_488_00167.pdf}{PDF}.


\bibitem{Wimmer:2015-7782:NCOM}% article
 \textsc{M.~Wimmer},  \textsc{A.~Regensburger},  \textsc{M.\,A. Miri},
  \textsc{C.~Bersch},  \textsc{D.\,N. Christodoulides},  and
  \textsc{U.~Peschel},
Observation of optical solitons in {PT}-symmetric lattices,
\hrefEmpty[ ]{Empty}{ \jr{Nat. Commun.} \textbf{6}, 7782 (2015)}\hrefBibPDF[
  ]{NCOM_2015_06_07782.pdf}{PDF}.


\bibitem{Miri:2012-23807:PRA}% article
 \textsc{M.\,A. Miri},  \textsc{A.~Regensburger},  \textsc{U.~Peschel},  and
  \textsc{D.\,N. Christodoulides},
Optical mesh lattices with {PT} symmetry,
\hrefBib[ ]{http://dx.doi.org/10.1103/PhysRevA.86.023807}{ \jr{Phys. Rev. A}
  \textbf{86}, 023807--12 (2012)}\hrefBibPDF[ ]{PRA_2012_86_23807.pdf}{PDF}.


\bibitem{Li:2013-13604:PRA}% article
 \textsc{Y.\,Y. Li},  \textsc{J.\,F. Liu},  \textsc{W.~Pang},  and
  \textsc{B.\,A. Malomed},
Symmetry breaking in dipolar matter-wave solitons in dual-core couplers,
\hrefBib[ ]{http://dx.doi.org/10.1103/PhysRevA.87.013604}{ \jr{Phys. Rev. A}
  \textbf{87}, 013604--11 (2013)}\hrefBibPDF[ ]{PRA_2013_87_13604.pdf}{PDF}.


\bibitem{Kevrekidis:2014-10102:PRA}% article
 \textsc{P.\,G. Kevrekidis},
Variational method for nonconservative field theories: Formulation and two
  {PT}-symmetric case examples,
\hrefBib[ ]{http://dx.doi.org/10.1103/PhysRevA.89.010102}{ \jr{Phys. Rev. A}
  \textbf{89}, 010102--5 (2014)}.


\bibitem{Saadatmand:2014-52902:PRE}% article
 \textsc{D.~Saadatmand},  \textsc{S.\,V. Dmitriev},  \textsc{D.\,I. Borisov},
  and  \textsc{P.\,G. Kevrekidis},
Interaction of sine-{G}ordon kinks and breathers with a parity-time-symmetric
  defect,
\hrefBib[ ]{http://dx.doi.org/10.1103/PhysRevE.90.052902}{ \jr{Phys. Rev. E}
  \textbf{90}, 052902--10 (2014)}.


\bibitem{Makris:2015-7257:NCOM}% article
 \textsc{K.\,G. Makris},  \textsc{Z.\,H. Musslimani},  \textsc{D.\,N.
  Christodoulides},  and  \textsc{S.~Rotter},
Constant-intensity waves and their modulation instability in non-hermitian
  potentials,
\hrefBib[ ]{http://dx.doi.org/10.1038/ncomms8257}{ \jr{Nat. Commun.}
  \textbf{6}, 7257--7 (2015)}\hrefBibPDF[ ]{NCOM_2015_06_07257.pdf}{PDF}.


\bibitem{Gao:1504.00978:ARXIV}% article
 \textsc{T.~Gao},  \textsc{E.~Estrecho},  \textsc{K.~Bliokh},
  \textsc{T.~Liew},  \textsc{M.~Fraser},  \textsc{S.~Brodbeck},
  \textsc{M.~Kamp},  \textsc{C.~Schneider},  \textsc{S.~Hofling},
  \textsc{Y.~Yamamoto},  \textsc{F.~Nori},  \textsc{Y.~Kivshar},
  \textsc{A.~Truscott},  \textsc{R.~Dall},  and  \textsc{E.~Ostrovskaya},
Observation of non-{H}ermitian degeneracies in a chaotic exciton-polariton
  billiard,
\hrefBib[ ]{http://arxiv.org/abs/1504.00978}{ \jr{arXiv} \textbf{\mdseries
  1504.00978} (2015)}.


\end{thebibliography}
\end{document}